\documentclass[showpacs,showkeys,prc,nofootinbib,superscriptaddress]{revtex4}

\usepackage{amsmath,amssymb,exscale}
\usepackage[dvips]{graphicx}
\usepackage{url}

\providecommand{\urlprefix}{\\}               

\usepackage[dvips,hyperfootnotes=false]{hyperref}  
\renewcommand{\boldmath}{}                         

\usepackage{dsfont,slashed}                   
\providecommand{\slashed}[1]{\not\! #1}       
\providecommand{\mathds}[1]{\boldsymbol #1}   

\newcommand{\MeV}{\,\textrm{MeV}}
\newcommand{\GeV}{\,\textrm{GeV}}
\newcommand{\fm}{\,\textrm{fm}}

\newcommand{\njlkpl}{\textrm{G}}
\newcommand{\njlcut}{\Lambda}
\newcommand{\Gamnjl}{\Gamma}
\newcommand{\Gamnjlt}{\tilde\Gamnjl}
\newcommand{\Gaml}{\Gamnjl_l}
\newcommand{\Gamlt}{\Gamnjlt_l}

\DeclareMathOperator{\density}{\rho}
\DeclareMathOperator{\Trace}{Tr}
\DeclareMathOperator{\trace}{tr}
\DeclareMathOperator{\sgn}{sgn}
\DeclareMathOperator{\prpl}{\mathcal P} 
\DeclareMathOperator{\prwd}{\mathcal W} 
\newcommand{\nf}{\mathrm{n_F}}
\newcommand{\nb}{\mathrm{n_B}}
\newcommand{\nfp}{\mathrm{n_F^+}}
\newcommand{\nfm}{\mathrm{n_F^-}}
\newcommand{\nfpm}{\mathrm{n_F^\pm}}
\newcommand{\nfmp}{\mathrm{n_F^\mp}}
\newcommand{\ordersymb}{\mathcal{O}}
\newcommand{\order}[1]{$\ordersymb(#1)$}
\renewcommand{\Re}{\textrm{Re}}
\renewcommand{\Im}{\textrm{Im}}

\newcommand{\qbq}{$q\bar q$ }
\newcommand{\ewert}[1]{\left\langle #1 \right\rangle}
\newcommand{\veq}[1]{{\vec{#1}\,}}
\newcommand{\modk}{|\veq k|}
\newcommand{\dvp}{\frac{d^4 p}{(2\pi)^4}}
\newcommand{\dvq}{\frac{d^4 q}{(2\pi)^4}}
\newcommand{\dvr}{\frac{d^4 r}{(2\pi)^4}}
\newcommand{\ddp}{\frac{d^3 p}{(2\pi)^3}}

\newcommand{\effp}{\tilde p}
\newcommand{\effk}{\tilde k}
\newcommand{\effm}{\tilde m}
\newcommand{\effmc}{{m^*}}

\newcommand{\const}{\textrm{const.}}
\newcommand{\eff}{\textrm{eff}}
\newcommand{\os}{\textrm{os}}
\newcommand{\pop}{\textrm{pop}}
\newcommand{\all}{\textrm{all}}
\newcommand{\ret}{\textrm{ret}}
\newcommand{\av}{\textrm{av}}
\newcommand{\chron}{\textrm{c}}
\newcommand{\achron}{\textrm{a}}
\newcommand{\mf}{\textrm{mf}}
\newcommand{\hart}{\textrm{H}}
\newcommand{\fock}{\textrm{F}}

\DeclareMathOperator{\qpr}{S}
\newcommand{\qpc}{\qpr^\chron}
\newcommand{\qpclo}{\qpr_\chron}
\newcommand{\qpa}{\qpr^\achron}
\newcommand{\qpret}{\qpr^\ret}
\newcommand{\qpav}{\qpr^\av}
\newcommand{\qpgr}{\qpr^{>}}
\newcommand{\qpkl}{\qpr^{<}}
\newcommand{\qpkg}{\qpr^{\lessgtr}}
\newcommand{\qpkglo}{\qpr_{\lessgtr}}
\newcommand{\qpgk}{\qpr^{\gtrless}}
\DeclareMathOperator{\qse}{\Sigma}
\newcommand{\qsemf}{\qse^\mf}
\newcommand{\qseh}{\qse^\hart}
\newcommand{\qsef}{\qse^\fock}
\newcommand{\qsec}{\qse^\chron}
\newcommand{\qseret}{\qse^\ret}
\newcommand{\qseretlo}{\qse_\ret}
\newcommand{\qseav}{\qse^\av}
\newcommand{\qsegr}{\qse^{>}}
\newcommand{\qsekl}{\qse^{<}}
\newcommand{\qsegk}{\qse^{\gtrless}}
\newcommand{\qbr}{\Gamma}
\newcommand{\qavbrall}{\langle \qbr^\all_0 \rangle}
\newcommand{\qavbrpop}{\langle \qbr^\pop_0 \rangle}
\newcommand{\qsp}{\mathcal{A}}

\DeclareMathOperator{\mespr}{D}
\newcommand{\mespc}{\mespr^\chron}
\newcommand{\mespclo}{\mespr_\chron}

\DeclareMathOperator{\mpr}{\Delta}
\newcommand{\mpc}{\mpr^\chron}
\newcommand{\mpret}{\mpr^\ret}
\newcommand{\mpav}{\mpr^\av}
\newcommand{\mpgr}{\mpr^{>}}
\newcommand{\mpkl}{\mpr^{<}}
\newcommand{\mpkg}{\mpr^{\lessgtr}}
\newcommand{\mpgk}{\mpr^{\gtrless}}
\newcommand{\mpinf}{\mpr^\infty}
\DeclareMathOperator{\empr}{\underline\Delta\,}
\newcommand{\empc}{\empr^\chron}
\newcommand{\empret}{\empr^\ret}
\newcommand{\empgk}{\empr^{\gtrless}}
\DeclareMathOperator{\mse}{\Pi}
\newcommand{\msec}{\mse^\chron}
\newcommand{\mseret}{\mse^\ret}
\newcommand{\msegr}{\mse^{>}}
\newcommand{\msekl}{\mse^{<}}
\newcommand{\msegk}{\mse^{\gtrless}}
\newcommand{\msp}{\xi}
\newcommand{\mbr}{\Gamma}

\begin{document}


\title{Short-range correlations in quark matter}
\author{F. Fr\"omel}\email{Frank.Froemel@theo.physik.uni-giessen.de}
\affiliation{Institut f\"ur Theoretische Physik, Universit\"at Giessen, Germany}
\author{S. Leupold}
\affiliation{Institut f\"ur Theoretische Physik, Universit\"at Giessen, Germany}
\affiliation{Gesellschaft f\"ur Schwerionenforschung, Darmstadt, Germany}

\begin{abstract}
We investigate the role of short-range correlations in quark matter within the framework of the SU(2) NJL model. Employing a next-to-leading order expansion in $1/N_c$ for the quark self energy we construct a fully self-consistent model that is based on the relations between spectral functions and self energies. In contrast to the usual quasiparticle approximations we take the collisional broadening of the quark spectral function consequently into account. Mesons are dynamically generated in the fashion of a random phase approximation, using full in-medium propagators in the quark loops. The results are self-consistently fed back into the quark self energy. Calculations have been performed for finite chemical potentials at zero temperature. The short-range correlations do not only generate finite widths in the spectral functions but also have influence on the chiral phase transition. 
\end{abstract}

\maketitle


\section{Introduction}

To our present knowledge, quantum chromodynamics (QCD) \cite{Peskin:1995} is the theory of the strong interaction. QCD is, however, not well suited for investigations at moderate energies. Due to the running coupling, perturbative calculations are not possible in this regime. Lattice QCD \cite{Creutz:1984mg} is technically challenging and -- in particular at finite chemical potentials -- still limited in its applications \cite{Karsch:2001cy}. Consequently, effective models have been developed to explore the low energy phenomenology of QCD. These models treat the quarks usually as quasiparticles. Even in systems with considerable quark densities, short-range correlations that lead to a collisional broadening of the spectral function are not considered.

The existence of short-range correlations in nuclear matter is well established. They are observed in $A(e,e'p)X$ and $A(e,e'pp)X$ experiments (see, e.g., \cite{deWittHuberts:1990zy,CiofidegliAtti:1990rw}) and have been the subject of many theoretical approaches, see \cite{Dickhoff:2004xx,Dickhoff:2005mb} for a review. The results of these studies agree rather well concerning the short-range effects. Even in a self-consistent calculation with a pointlike interaction \cite{Lehr:2000ua,Lehr:2001qy}  -- using only one free parameter -- good agreement with more realistic calculations and the experimental data is reached. Thus, what matters is not the detailed modeling but the overall strength of the interaction and the collisional phasespace.

The success of using pointlike interactions in nuclear matter had motivated our first approach to short-range interactions in quark matter \cite{Froemel:2001iy} on the basis of the Nambu--Jona-Lasinio (NJL) model. The NJL model  \cite{Nambu:1961tp,Nambu:1961fr,Klevansky:1992qe,Vogl:1991qt,Hatsuda:1994pi} is an effective quark model  that respects the relevant symmetries of QCD. It is frequently used to investigate phenomena related to chiral symmetry. In the last years, it has also been used to explore the effects of color superconductivity -- see, e.g., \cite{Rajagopal:2000wf,Buballa:2003qv} for recent reviews. In the present work, as a first step, we ignore color superconductivity and explore the role of short-range correlations in quark matter in the chirally broken and restored phase. Clearly, for a complete picture also the phenomenon of color superconductivity -- relevant at high chemical potentials $\mu$ and low temperatures $T$ -- should be taken into account.

At zero temperature, just above the chiral phase transition, quark densities a few times larger than normal nuclear matter density are reached. It is not unlikely that short-range correlations exist at such densities and have influence on the medium \cite{Peshier:2004bv}. A simple estimate indicates that the character of the phase transition at low $T$ changes when a small width is added to the quark propagator: 
\begin{figure}
	\centerline{
		\includegraphics[scale=.7]{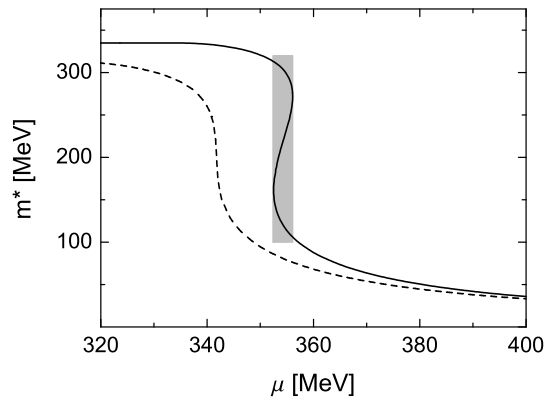}
	}
	\caption{\label{fig:naivemassgap}The quark mass as function of the chemical potential at $T=0$ in the NJL model. The solid line shows the Hartree--Fock result, the dashed line the result of a calculation with a constant quark width. The shaded area denotes a first-order phase transition. See the text for details.}
\end{figure}
The solid line in Fig.~\ref{fig:naivemassgap} shows the result of a Hartree--Fock calculation including a finite current quark mass. The 'S' shape of the curve is characteristic for a first-order phase transition \cite{Plischke:1994}. Somewhere in the shaded region the effective quark mass $\effmc$ jumps discontinuously. The dashed curve has been obtained after inserting a constant width into the quasiparticle propagator. $\effmc$ drops rapidly but continuously now -- collisional broadening has turned the phase transition into a crossover. Note that Fig.~\ref{fig:naivemassgap} is meant as an illustrative example. A more serious calculation with a self-consistently determined width will be given below.

In \cite{Froemel:2001iy} we have used the NJL interaction in a self-consistent calculation where the collisional self energy was determined by the Born diagrams, i.e., on a 2-loop level. We have found rather weak short-range effects in this approach -- more than one order of magnitude smaller than in nuclear matter. A similar approach can be found in \cite{Domitrovich:1993tz,Domitrovich:1993tq}, where the self energy was not determined self-consistently and the calculations were restricted to $\mu=0$. The results are comparable to \cite{Froemel:2001iy}.

Loop-expansions in the NJL model are questionable since the coupling is large. Expansions in the inverse number of colors, $1/N_c$, should be preferred \cite{Quack:1993ie,Dmitrasinovic:1995cb}. In this work we bring our approach from \cite{Froemel:2001iy} to a consistent level in $1/N_c$. Including all quark self energy diagrams of the next-to-leading order \order{1/N_c} in the self-consistent calculation leads to the introduction of dynamically generated mesons. In contrast to the standard Hartree+Random Phase Approximation (RPA) scheme \cite{Klevansky:1992qe}, the mesons are part of the self-consistent procedure and their properties are fed back into the quark self energy. We restrict our numerical calculations to flavor symmetric quark matter at $T=0$ here. Nonetheless, the formalism can also be used at finite $T$. An approach to flavor asymmetric systems would be technically more involved but is theoretically feasible. See \cite{Konrad:2005qm} for an approach to asymmetric nuclear matter. As already mentioned, diquark condensates and color superconductivity are presently also not considered.

$1/N_c$ extensions to the NJL model on the quasiparticle level have been discussed in \cite{Quack:1993ie,Dmitrasinovic:1995cb,Nikolov:1996jj}. It has been shown in \cite{Dmitrasinovic:1995cb,Nikolov:1996jj} that the generation of massless pions (Goldstone modes) in a perturbative expansion relies on  a careful choice of diagrams for the meson polarizations. Our self-consistent ansatz iterates diagrams up to arbitrary orders in $1/N_c$ but does not generate all contributions of any given order. Thus, the RPA pions become massive in the chirally broken phase. An approach that is self-consistent \emph{and} yields massless pions (in the chiral limit) has not been proposed so far for the NJL model. The result of our self-consistent calculation could be used as input for an additional formalism which yields realistic pion properties. However, in the approach proposed in \cite{Dmitrasinovic:1995cb} and similarly in \cite{Oertel:2000sr,vanHees:2002bv}, the Goldstone pions are not fed back into the calculation of the quark properties.

In the present work we go beyond the quasiparticle models of \cite{Dmitrasinovic:1995cb,Nikolov:1996jj}. We do not only consider the poles of the RPA spectral functions but also the continuum of off-shell states. As we will see later, these states are much more important for the quark properties than the poles. An analysis of the different on-shell contributions to the quark width will show that some of them are sensitive to the pion mass while others are not. Concerning the chirally broken phase, it will turn out that our self-consistent calculation yields a conservative estimate of the short-range correlations in quark matter.

This work is structured as follows. We introduce our self-consistent \order{1/N_c} approach in Section~\ref{ch:model}. In Section~\ref{ch:calc_selfenergies} we show how to calculate the quark self energies and RPA polarizations using full in-medium propagators. The chiral properties of the RPA pions are also discussed. In Section~\ref{ch:qq_scattering} we use quasiparticle approximations to explore the structure of the quark width qualitatively and to investigate the influence of the RPA pion mass. The technical details and the results of our numerical calculations -- where we do not use any quasiparticle approximations -- are presented in Section~\ref{ch:results}. A summary of our results and an outlook to future improvements are given in Section~\ref{ch:outlook}. The Appendices contain some of the more technical details of our approach.

\section{The formalism\label{ch:model}}

\subsection{The NJL interaction\label{sec:njl}}

The NJL model is by design the simplest effective quark interaction that resembles all relevant symmetries of QCD \cite{Klevansky:1992qe,Vogl:1991qt,Hatsuda:1994pi}. In the leading-order mean field approach -- the Hartree+RPA scheme -- a finite constituent quark mass is generated and breaks chiral symmetry dynamically at low $\mu$ and $T$, cf.~Fig.~\ref{fig:naivemassgap}. At higher $\mu$ and/or $T$ a chiral phase transition occurs. The (Hartree+)RPA pions can be identified with Goldstone modes.

The standard version of the two flavor NJL Lagrangian is given by
\begin{equation}
				\mathcal{L}_\mathrm{NJL}=\bar\psi \left( i\slashed{\partial} -m_0\right) \psi
				+\njlkpl\left[(\bar\psi\psi)^2+(\bar\psi i\gamma_5\vec\tau\psi)^2 \right] \,, \label{eq:njl_lagrangian}
\end{equation}
where $\njlkpl$ is the constant \qbq coupling strength and the $\tau_i$ are the isospin Pauli matrices. $m_0$ denotes a small current quark mass that breaks chiral symmetry explicitly. Due to the constant coupling, the model cannot be renormalized. Several regularization schemes exist \cite{Klevansky:1992qe}, the simplest one is the (three\mbox{-})momentum cutoff $\njlcut$.

$\njlkpl$, $\njlcut$ and $m_0$ are free parameters. Their values are fixed such that reasonable results are obtained for the quark condensate and the pion decay constant in vacuum -- and for the pion mass in the case $m_0\neq 0$. We list some typical mean field parameter sets in Table~\ref{tab:njlparam}.
\begin{table}
	\caption{\label{tab:njlparam} Mean field NJL parameter sets with and without (chiral limit) finite current quark mass $m_0$, using a three-momentum cutoff $\njlcut$ \protect\cite{Klevansky:1992qe}. The coupling $\njlkpl$ is fitted for Hartree calculations. The coupling $\njlkpl_\mathrm{HF}$ for the Hartree--Fock approximation is obtained from $\njlkpl$ by a rescaling factor $12/13$. $\effmc$ and $\ewert{\bar u u}$ are the effective mass and the quark condensate that are found using the given parameters.}
	\begin{ruledtabular}
			\begin{tabular}{lcccccc}
							& $m_0$		&	$\njlkpl\njlcut^2$
																&	$\njlkpl_\mathrm{HF}\njlcut^2$		
																				&	$\njlcut$	&	$\effmc$	&	$\ewert{\bar u u}^{1/3}$	\\
							& [MeV]		&				&				&	[MeV]			&	[MeV]			&	[MeV]											\\
			\hline
			Set 0		& - 			&	2.14	& 1.98	&	653				&	313				&	-250								    	\\
			Set I  	& 5.5 		&	2.19	&	2.02	&	631				&	336				&	-247											\\
		\end{tabular}
	\end{ruledtabular}
\end{table}
The simplicity of this model leads also to shortcomings like the lack of asymptotic freedom and the absence of confinement. Note that the cutoff is sometimes interpreted as a crude implementation of asymptotic freedom.

The Lagrangian (\ref{eq:njl_lagrangian}) provides scalar and pseudoscalar \qbq interaction channels.
It is also possible to construct a bosonized version of the NJL Lagrangian \cite{Klevansky:1992qe,Rehberg:1998nd},
\begin{align}
	\mathcal{L}_\mathrm{NJL}=\bar\psi \left( i\slashed{\partial} -m_0\right) \psi				
	&-\bar\psi\left( \sigma \Gamnjl_\sigma + \pi_0\Gamnjl_0 + \pi_+\Gamnjl_- + \pi_-\Gamnjl_+  \right)\psi  \notag \\
	&\quad -\frac{1}{4G}\left( \sigma^2+\pi_0^2 +2\pi_+\pi_- \right) \,, \label{eq:njl_meson}
\end{align}
where the meson fields are defined as $\sigma = -2G\bar\psi\Gamnjl_\sigma\psi$ and $\pi^{0,\pm} = -2G\bar\psi\Gamnjl_{0,\pm}\psi$, and
\begin{equation}
		\Gamnjl_\sigma = \Gamnjlt_\sigma = 1 \,, \qquad
		\Gamnjl_0 = \Gamnjlt_0 = i \gamma_5 \tau_3 \,, \qquad
		\Gamnjl_\pm = \Gamnjlt_\mp = i \gamma_5 \tau_\pm\,.
	\label{eq:gamnjl}
\end{equation}
Note that  $\tau_\pm=(\tau_1\pm i\tau_2)/{\sqrt{2}}$. $\Gamnjlt_{\sigma,0,\pm}$ are introduced here for further reference. They will be needed for the calculation of self energy and polarization diagrams. For a discussion on the quantization of (\ref{eq:njl_meson}) we refer to \cite{Rehberg:1998nd,Zhang:1992rf}. To investigate the phenomenon of color superconductivity \cite{Buballa:2003qv}, a vertex  that introduces an attractive interaction in the $qq$ channel must be added to (\ref{eq:njl_lagrangian}). We do not consider such an extension here.

\subsection{{\boldmath${1/N_c}$} expansion in the NJL model and RPA mesons\label{sec:ncexpansion}}

The QCD inspired NJL model is a strongly interacting theory. $\njlkpl\njlcut^2$ -- the relevant quantity to estimate the interaction strength -- has a value of approximately $2$, cf.~Table~\ref{tab:njlparam}. Thus, a perturbative expansion in terms of the coupling is not feasible. It is possible, however, to perform an expansion in the inverse number of colors, $1/N_c$. A simple scheme (see, e.g., \cite{Quack:1993ie,Dmitrasinovic:1995cb}) allows to determine the order in $1/N_c$ of diagrams by counting vertices and loops: A factor $\njlkpl$ is assigned to each vertex and a factor $N_c$ to each closed fermion loop. In this symbolic notation one sets $\ordersymb(G)=\ordersymb(1/N_c)$. The (free) quark propagator is of order unity. The order of a diagram with $n$ vertices and $m$ loops is then given by $\njlkpl^n N_c^m=\ordersymb(N_c^{m-n})$, i.e., the difference of loops and vertices.

\begin{figure}
	\centerline{
		\includegraphics[scale=.8]{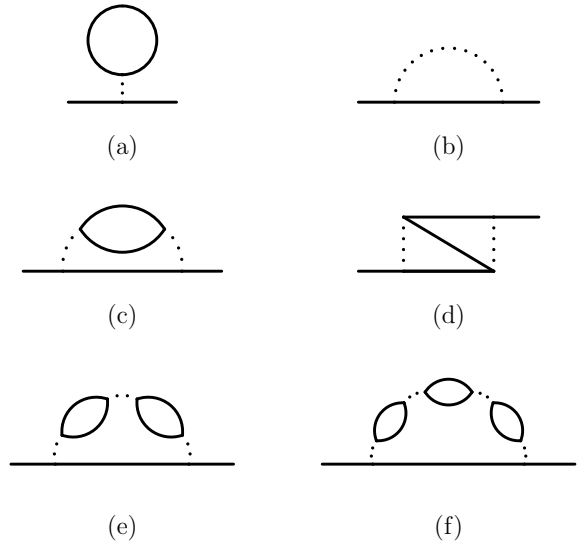}
	}
	\caption{\label{fig:hfborn}The Hartree (a) and Fock (b) self energies and the direct (c) and exchange (d) diagrams of the Born self energy. (e) and (f) are diagrams of higher order in the coupling. The pointlike \qbq vertices have been replaced by finite interaction lines to show the structure of the diagrams unambiguously.}
\end{figure}

Fig.~\ref{fig:hfborn} shows some diagrams that are of first (a,b), second (c,d), and higher orders (e,f) in the coupling. The  Hartree diagram (a) is the only contribution of order \order{1} in $1/N_c$ to the quark self energy. The Fock diagram (b) as well as the direct Born diagram (c) are of order \order{1/N_c}. The exchange Born diagram (d) is even of order \order{1/N_c^2}. 

The diagrams (e,f) are -- like (b,c) -- of order \order{1/N_c}. The two diagrams were constructed by adding pairs of loops and vertices to (c). By adding further loops in (f), diagrams of any higher order in the coupling while still of order \order{1/N_c} can be found. The \order{1/N_c} diagrams that are constructed by dressing the quark lines in the \order{1/N_c} diagrams of Fig.~\ref{fig:hfborn} with the Hartree diagram, or by dressing the Hartree loop with the \order{1/N_c} diagrams will later be generated automatically in our self-consistent approach. They are not discussed explicitly to avoid double counting.

Note that the full propagator of a self-consistent calculation can be dressed with self energies of arbitrary order. It remains \order{1} in leading order but gains contributions of higher orders in $1/N_c$. In a self-consistent approach this means that one can select Feynman diagrams by their leading order to find a consistent set of diagrams for any given order. However, every diagram will contribute in subleading orders in a possibly incomplete way. The Hartree approximation constitutes an exception. Since the Hartree self energy is of order \order{1/N_c^0} the order of a (dressed) Hartree  propagator remains exactly \order{1}.

\begin{figure*}
	\centerline{
		\includegraphics[scale=.8]{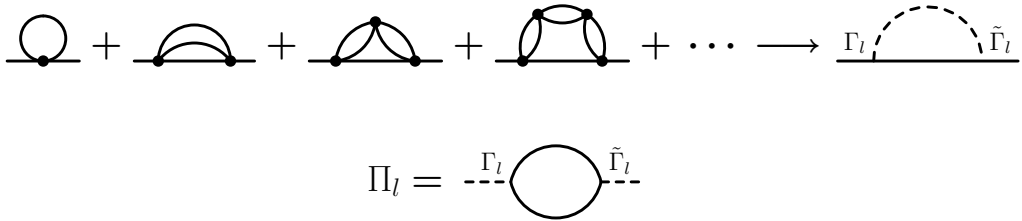}
	}
	\caption{\label{fig:mesexse}The \order{1/N_c} diagrams of the structure shown in Fig.~\ref{fig:hfborn} are added up in a random phase approximation to generate an effective meson exchange (dashed line). The first diagram is the Fock diagram, solid dots denote the NJL coupling ($-2\njlkpl$). The second line shows the RPA polarization $\mse_l$.}
\end{figure*}

Usually, all \order{1/N_c} diagrams of the structure shown in Fig.~\ref{fig:hfborn} (b,c) and (e,f) are summed up in an RPA. They can be interpreted as contributions to an effective meson exchange as shown in Fig.~\ref{fig:mesexse}.
This allows us to treat a whole class of self energy diagrams in a simple and consistent way. The meson propagator is given by  a geometric series,
\begin{align}
	\mpr_l (k) &= -2G \left[1 + (-2G)\mse_l(k) +(-2G)\mse_l(k)(-2G)\mse_l(k) +\ldots\right]  \notag \\
	&=-\frac{2 \njlkpl}{1+2\njlkpl\mse_l (k)}\,, \label{eq:mesprop}
\end{align}
where $\mse_l$ ($l=\sigma,0\pm$) is the polarization of the RPA mesons, cf.~Fig.~\ref{fig:mesexse}. Like the Lagrangian (\ref{eq:njl_meson}), the propagators have no kinetic part of the form $k^2-m^2$. The dynamics are governed by the underlying \qbq states and thus hidden in the polarizations.

The propagators have poles at $1+2\njlkpl\Re\mse_l (k)=0$. The poles correspond to bound \qbq states and are identified with the actual RPA mesons. At energies above $k_0=\mu+\effmc$ ($2\effmc$ in the vacuum), the propagators pick up contributions from unbound \qbq states. The mesons may decay into quark--antiquark pairs above this threshold. A wide range of states with large decay widths, the so-called \qbq continuum \cite{Rehberg:1998nd}, is found.

\subsection{The {\boldmath\order{1/N_c}} approach\label{sec:1Nc_approach}}

\begin{figure*}
	\centerline{
		\includegraphics[scale=.8]{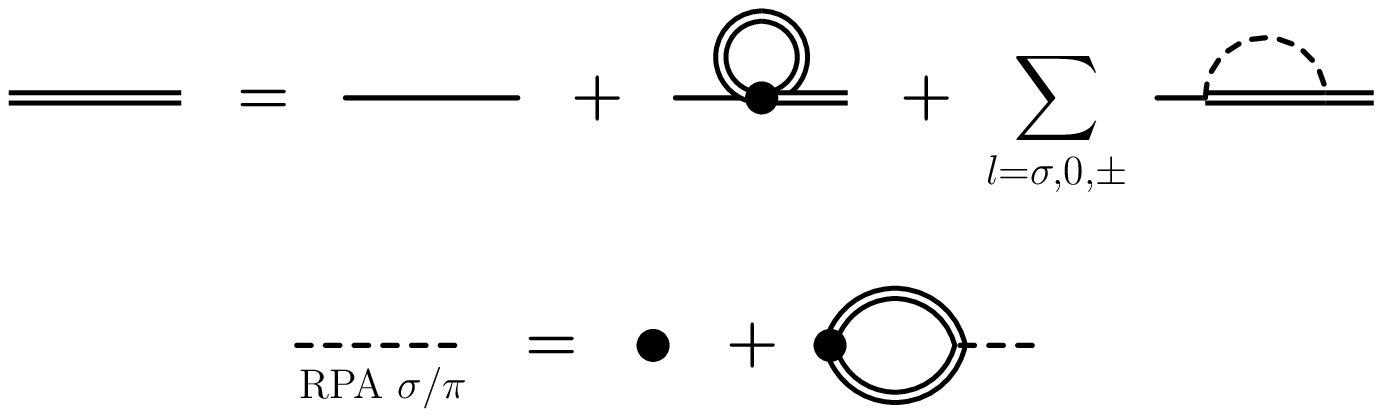}
	}
	\caption{\label{fig:dyson_full} The set of Dyson--Schwinger equations corresponding to the fully self-consistent \order{1/N_c} approach. The double lines denote dressed quark propagators, dots denote the NJL coupling ($-2\njlkpl$). The second line shows the RPA like construction of the mesons (dashed lines), cf.~(\ref{eq:mesprop}).}
\end{figure*}

The standard Hartree+RPA approach to the NJL model is of leading order in $1/N_c$. It can be summarized by the two Dyson--Schwinger equations of Fig.~\ref{fig:dyson_full}, \emph{excluding} the meson exchange diagram in the first line. The quark self energy is determined solely by the time-local Hartree diagram. Thus, the quarks remain quasiparticles and the RPA mesons have no influence on the properties of the quarks. Such a feedback can only be generated by a diagram  of higher order in $1/N_c$. Note that the first Dyson--Schwinger equation -- that determines the quark mass -- is usually called gap equation in the Hartree approximation.

In our investigation of the short-range correlations, we will determine the quark properties in next-to-leading order \order{1/N_c}. Therefore, the meson exchange diagram of Fig.~\ref{fig:mesexse} is included in the first Dyson--Schwinger equation in Fig.~\ref{fig:dyson_full}. This equation will self-consistently generate all \order{1/N_c} contributions to the self energy. In addition, it introduces a feedback of the RPA mesons\footnote{On the Hartree level (for the quark propagators) the summation of all quark loop diagrams corresponds to a (classical) random phase approximation. We follow the literature, see~e.g.~\cite{Dmitrasinovic:1995cb}, and stay with the name RPA when propagators beyond the Hartree approximation are used -- even when such RPA mesons do not satisfy the chiral theorems.} -- generated by the Dyson--Schwinger equation in the second line -- into the quark self energy. The calculation of the RPA polarizations is part of the self-consistent procedure here. We will not use a pole approximation for the mesons since the \qbq continuum plays an important role in our approach.

An analytic solution of the self-consistency problem presented in Fig.~\ref{fig:dyson_full} is not possible. A solution can only be found in an iterative, numerical calculation. By using full in-medium propagators -- including off-shell contributions -- we clearly go beyond the standard Hartree+RPA approach. The quark self energy has no longer the simple form of an effective mass. Due to the nonlocal meson exchange, it becomes complex and four-momentum dependent. The imaginary part of the self energy is generated by short-range correlations. It will be identified with the width of the quark spectral function.

To keep our model numerically simple, we will use a three-momentum cutoff $\njlcut$ \cite{Klevansky:1992qe} to regularize the divergent self energy and polarization integrals, cf.~(\ref{eq:hartree_se},\ref{eq:qsegk},\ref{eq:msegk}). Like in \cite{Froemel:2001iy} and \cite{Domitrovich:1993tz,Domitrovich:1993tq}, we use the cutoff to restrict the three-momentum of each quark to values below $\njlcut$. A meson cutoff is implicitly included since the quark loop integrals for the polarizations are only finite for meson three-momenta below $2\njlcut$. Note that the cutoff can also be implemented in other ways, see e.g. \cite{Dmitrasinovic:1995cb,Nikolov:1996jj,Oertel:2000sr}.

As long as the width does not become too large, most of the strength of the quark propagator remains in the vicinity of the on-shell peaks. The three-momentum cutoff serves then also as an energy cutoff, $|p_0|<E_\njlcut<\sqrt{\njlcut^2+{\effmc}^2}$. At higher energies, the on-shell peak will be located at three-momenta above the cutoff $\njlcut$. This is not a sharp cutoff when the peaks have a finite width, but the quark propagators at higher energies will be strongly suppressed in self energy and polarization integrals.

At this point, a remark concerning our earlier approach in \cite{Froemel:2001iy} is in order. We have not considered dynamically generated mesons there. Instead, the direct \order{1/N_c} and the exchange \order{1/N_c^2} Born diagrams (cf.~Fig.~\ref{fig:hfborn}) were used to construct the quark self energy in a loop-expansion. In comparison to nuclear matter we found rather small widths for the quarks in \cite{Froemel:2001iy}. By including an infinite number of \order{1/N_c} diagrams in the new approach, the importance of collisional broadening will increase significantly.

\subsection{Quark fields and propagators\label{sec:qprop}}

We use the real-time formalism \cite{Kadanoff:1962,Danielewicz:1982kk,Botermans:1990qi,Das:1997} for our calculations. More details on in-medium techniques can be found in \cite{Post:2003hu} and the transport theoretical approaches to quark matter \cite{Mrowczynski:1992hq,Zhang:1992rf,Rehberg:1998nd}. In the following we assume that the system is in thermal equilibrium.  For simplicity, we work only in the rest frame of the medium.

The fundamental elements of our model are the causal, the anti-causal, and the non-ordered single-particle Green's functions,
\begin{equation}
	\label{eq:qprops}
	\begin{aligned}
		\qpc_{\alpha\beta}(1,1') &= -i\ewert {\mathrm{T} [\psi_\alpha(1)\bar\psi_\beta(1')]}  \,,    \\ 
		\qpa_{\alpha\beta}(1,1') &= -i\ewert {\mathrm{\bar T} [\psi_\alpha(1)\bar\psi_\beta(1')]} \,,
	\end{aligned}
	\qquad
	\begin{aligned}
		\qpgr_{\alpha\beta}(1,1') &= -i\ewert {\psi_\alpha(1)\bar\psi_\beta(1')} \,, \\
		\qpkl_{\alpha\beta}(1,1') &= i\ewert {\bar\psi_\beta(1')\psi_\alpha(1)}  \,.
	\end{aligned}
\end{equation}
where $\alpha,\beta$ denote the spinor indices of the quark fields. Color and flavor indices have been suppressed here. $\mathrm{T}$ and  $\mathrm{\bar T}$ are the time-ordering and the anti-time-ordering operator, respectively. We also introduce the retarded and advanced propagators 
\begin{equation}
	\label{eq:qpret}
	\begin{split}
	\qpret(1,1') &= \Theta(t_1-t_{1'})[\qpgr(1,1')-\qpkl(1,1')]=\qpc(1,1')-\qpkl(1,1') \,, \\ 
	\qpav (1,1') &= \Theta(t_{1'}-t_1)[\qpkl(1,1')-\qpgr(1,1')]=\qpc(1,1')-\qpgr(1,1') \,. 
	\end{split}
\end{equation}
Note that $\qpret$ and $\qpav$ are complex conjugates of each other.

For our calculations it will be advantageous to work in momentum space. We consider infinite systems in thermal equilibrium here. Hence, the Fourier transformed propagators will depend only on the energy and the momentum of the quarks.

The relativistic quark propagators have a matrix structure in spinor space. The most general form of this Lorentz structure is found by a decomposition in terms of the 16 linear independent products of the Dirac $\gamma$ matrices (Clifford algebra). Considering the invariance under parity, time-reversal, and rotations, the Lorentz structure in the rest frame of the medium reduces to three independent components \cite{Bjorken:1965},
\begin{equation}
	\qpr(k)=\qpr_s(k) \mathds{1}_{4\times 4} +\qpr_0(k)\gamma_0 - \qpr_v(k) \vec e_k\cdot\vec\gamma
			=\qpr_s(k) +\qpr_{\mu}(k)\gamma^\mu \,,
	\label{eq:lorentzstructure}
\end{equation}
with $\vec e_k=\vec k/|\veq k|$ and $\qpr^i(k)=\qpr_v(k) e_k^i$. $\qpr_s$, $\qpr_0$, and $\qpr_v$ are scalar functions of $k_0$ and $|\veq k|$. In flavor symmetric quark matter, the functions do not have a structure in flavor space. Note that the sign convention in (\ref{eq:lorentzstructure}) differs from \cite{Froemel:2001iy} where $\qpr_v$ had the opposite sign. $\qpr_s$, $\qpr_0$, and $\qpr_v$ can always be extracted from the full propagators $\qpr_{\alpha\beta}$ using
\begin{equation}
	\qpr_s(k)=\frac{\trace \qpr(k)}{4} \,, \qquad
	\qpr_0(k)=\frac{\trace[\gamma_0 \qpr(k)]}{4} \,, \qquad
	\qpr_v(k)=\frac{\trace[\vec e_k\cdot \vec\gamma \qpr(k)]}{4} \,. \label{eq:extract_lorentz}
\end{equation}
Throughout this work we use the following convention: ``$\trace$'' denotes a trace in spinor space while ``$\Trace$'' denotes a full trace in color, flavor, and spinor space.

The spectral function $\qsp(k)$ corresponds to the density of states at a given energy $k_0$ and momentum $\vec k$. As the definitions in (\ref{eq:qprops}) show, $i\Trace\gamma_0\qpgr(1,1)$ and $-i\Trace\gamma_0\qpkl(1,1)$ are related to the density of the free and the populated states, respectively \cite{Danielewicz:1982kk}. Thus, we can introduce the spectral function in terms of these propagators,
\begin{equation}
		\qsp(k) =  i\left[\qpgr(k)-\qpkl(k)\right] = -2\Im\qpret(k) \,. \label{eq:defqsp}
\end{equation}
We use the definition $\Re F=(1/2) (F+\gamma_0 F^\dagger \gamma_0)$ and $\Im F=(1/2i) (F-\gamma_0 F^\dagger \gamma_0)$ here \cite{Froemel:2001iy,Bjorken:1965}. This definition ensures that $\Im \qpr = (\Im \qpr_s) \mathds{1}_{4\times 4}+ (\Im \qpr_\mu) \gamma^\mu$ (i.e., $\gamma_\mu=\Re\gamma_\mu$).

The spectral function inherits the Lorentz structure of the Green's functions (\ref{eq:lorentzstructure}). The  ``normalization'' of the purely real Lorentz components $\qsp_s$, $\qsp_0$ and $\qsp_\mu$ can be found by using the methods in \cite{Leupold:1999ga},
\begin{equation}
	\int_{-\infty}^{+\infty} \frac{dp_0}{2\pi}\qsp_0 (p) = 1 \,, \qquad
		\int_{-\infty}^{+\infty} \frac{dp_0}{2\pi}\qsp_s (p) = 0  \,, \qquad 
		\int_{-\infty}^{+\infty} \frac{dp_0}{2\pi}\qsp_v (p) = 0 \,. \label{eq:norm_qsp2}
\end{equation}
Both poles of a retarded propagator are located in the same complex half plane \cite{Das:1997}. When the integrand vanishes for large $|p_0|$, the integral can be closed along a contour in the other half plane. It can also be shown that
\begin{align}
	\int_{-\infty}^{+\infty} \frac{dp_0}{2\pi} \trace\qpret(p)
		=\int_{-\infty}^{+\infty} \frac{dp_0}{2\pi} \trace\left[\Gamlt \qpret(p)\Gaml\qpret(p-k)\right]=0 \,. \label{eq:sret_ints}
\end{align}

The fermionic phasespace distribution function $\nf(k)$ determines the probability that a state is populated. In thermodynamical equilibrium, $\nf(k)$  depends only on $k_0$,
\begin{equation}
	\nf(k_0) = \frac 1{e^{(k_0-\mu)/T}+1}  \quad\xrightarrow{T=0}\quad \Theta(\mu-k_0) \,. \label{eq:nf_t0}
\end{equation}
At $T=0$, all states below $\mu$ are occupied while all states above $\mu$ are free. The distribution function can be used to express $\qpgk(k)$ in terms of the spectral function \cite{Kadanoff:1962},
\begin{equation}
	\label{eq:qpgksp}
	\begin{split}
		\qpgr (k) &= -i\qsp (k)[1-\nf(k_0)] \,, \\
		\qpkl (k) &= i\qsp (k) \nf(k_0) \,.
	\end{split}
\end{equation}

In thermodynamical equilibrium, the retarded and advanced propagators in momentum space are given by
\begin{equation}
	\qpr^{\ret,\av}(k) = \frac {1}{\slashed{k} -m_0-\qse^{\ret,\av}(k)} 
	= \frac {\slashed{\effk}^{\ret,\av}+\effm^{\ret,\av}(k)}{\effk_{\ret,\av}^2-\effm^2_{\ret,\av}(k)} \,,
	\label{eq:explqpret}
\end{equation}
where we have introduced the effective masses and four-momenta,
\begin{equation}
	\label{eq:eff_momentum}
		\effm^{\ret,\av}(k) = m_0+ \qse^{\ret,\av}_s (k) \,, \qquad
		\effk^{\ret,\av}_\mu = k_\mu-\qse^{\ret,\av}_\mu(k) \,.
\end{equation}
$\qse^{\ret,\av}(k)=\qse^{\ret,\av}_s (k)+ \qse^{\ret,\av}_\mu(k)\gamma^\mu$ are the (complex) retarded and the advanced self energy, respectively. They have the same Lorentz structure as the propagators (\ref{eq:lorentzstructure}). In contrast to the time-ordered propagators $\qpr^{\chron,\achron}$ that gain extra terms \cite{Das:1997}, $\qpr^{\ret,\av}$ keep the simple form (\ref{eq:explqpret}) in the presence of a medium.

For our calculations it will be necessary to disentangle the real and imaginary parts of the propagators, e.g., to determine the spectral function. The real and the imaginary part of the denominator on the rhs.~of Eq.~(\ref{eq:explqpret}) can be identified with
\begin{align}
	\prpl(k) &= \Re\effk^2 -\Re\effm(k)^2 - \left[\Im \qseret_\mu(k)\Im \qseretlo^\mu(k)-\Im \qseret_s(k)^2\right] \,, 
			\label{eq:denom_prop_real}\\
	\prwd(k) &= -2 \left[\Re\effk_\mu\Im\qseretlo^\mu(k) +\Re\effm(k)\Im\qseret_s(k)\right]  
			\label{eq:denom_prop_imag}
\end{align}
for the retarded propagator. Using $\qseretlo^* (k)=\qseav (k)$, we find $\prpl^\av(k) = \prpl(k)$, $\prwd^\av(k) = -\prwd(k)$
for the advanced propagator. Note that $\Re\effm=\Re\effm^{\ret,\av}$ and $\Re\effk=\Re\effk^{\ret,\av}$. The expansion with the complex conjugates of the denominators yields \cite{Henning:1992dv}
\begin{equation}
	\qpr^{\ret,\av}(k) = \frac {\left(\slashed{\effk}^{\ret,\av} +\effm^{\ret,\av}(k)\right)\left(\prpl(k) \mp i\prwd(k)\right)}
	{\prpl^2(k) + \prwd^2(k)} \,.
	\label{eq:cmplxqpret}
\end{equation}
The upper sign in the numerator refers to the retarded and the lower sign to the advanced propagator, respectively. In both cases the functions $\prpl$ and $\prwd$ of (\ref{eq:denom_prop_real},\ref{eq:denom_prop_imag})  are used.

The zeros of $\prpl(k)$ determine the poles of the propagators.  $\prwd(k)$ generates the width. In the following, we will refer to the broadened poles as on-shell peaks. The on-shell energy $k_0^\os(\veq k)$ and the on-shell momentum $\vec k_\os(k_0)$ are defined by
\begin{equation}
	\prpl\left(k_0^\os(\veq k),\veq k\right)=0 \, ,  \qquad 
	\prpl\left(k_0,\vec k_\os(k_0)\right)=0 \,. \label{eq:os_def}
\end{equation}
We note that $\prpl$ and $\prwd$ both depend on the full self energy, not only the real or imaginary part. As we will see later, $\Im\qseret$ does not become too large in the on-shell regions. Hence, the influence of the quadratic terms in (\ref{eq:denom_prop_real}) is limited.

The real and the imaginary part of the propagators can be easily extracted from Eq.~(\ref{eq:cmplxqpret}). We identify the Lorentz components of the spectral function with
\begin{equation}
	\label{eq:qspectral}
		\qsp_s(k) = 2\frac{\Re\effm\prwd-\Im\qseret_s\prpl}{\prpl^2 + \prwd^2} \,,  \qquad
		\qsp_\mu(k) = 2 \frac{\Re\effk_\mu\prwd+\Im\qseret_\mu\prpl}{\prpl^2 + \prwd^2} \,.
\end{equation}
In the absence of a current quark mass, the Lorentz scalar components $\qpr_s$ and $\qse_s$ vanish when chiral symmetry is restored. The Lorentz structure is then simplified even further. In case of a finite $m_0$, $\qpr_s$ and $\qse_s$ will become small but do not vanish entirely.

\subsection{Meson propagators and spectral functions\label{sec:mprop}}

The $\sigma$ and $\pi^{0,\pm}$ Green's functions in terms of the meson fields \cite{Rehberg:1998nd} can be defined in analogy to the quark Green's functions (\ref{eq:qprops},\ref{eq:qpret}). Like for the quarks (\ref{eq:defqsp}), the $\sigma$ and $\pi^{0,\pm}$ spectral functions are introduced via the non-ordered propagators $\mpgk$,
\begin{equation}
		\msp(k) =  i\left[\mpgr(k)-\mpkl(k)\right] 
		= -2\Im\mpret(k) \,. \label{eq:defmsp}
\end{equation}
We have dropped the indices $\sigma,0,\pm$ since the definition is the same for all mesons. Note that ${\mpret}^*=\mpav$.
In thermodynamical equilibrium, the bosonic distribution function
\begin{equation}
	\nb(k_0) = \frac 1{e^{k_0/T}-1}  \quad\xrightarrow{T=0}\quad -\Theta(-k_0) \,, \label{eq:nb_t0}
\end{equation}
is -- like $\nf$ -- a simple step function at zero temperature. We can use $\nb$ to express the non-ordered propagators $\mpgk$ in terms of the spectral functions \cite{Kadanoff:1962},
\begin{equation}
	\label{eq:mpgksp}
	\begin{split}
		\mpgr (k) &= -i \msp (k)[1+\nb(k_0)]\,, \\
		\mpkl (k) &= -i \msp (k) \nb (k_0) \,. 
	\end{split}
\end{equation}
Note the subtle differences to the relations for the quarks. The distributions $\nf$  in (\ref{eq:qpgksp}) correspond to $-\nb$ in (\ref{eq:mpgksp}), i.e., Pauli blocking is replaced by Bose enhancement.

In Section~\ref{sec:ncexpansion}, the RPA meson propagators have been determined in terms of the NJL coupling and the polarizations $\mse$, cf.~Eq.~(\ref{eq:mesprop}),
\begin{equation}
	\mpret_l (k) = -\frac{2 \njlkpl}{1+2\njlkpl\mseret_l (k)} \,, \label{eq:mesprop_ret}
\end{equation}
where $l=\sigma,0,\pm$ denotes the type of the meson. We can easily disentangle the real and the imaginary part of the retarded propagators. The spectral function $\msp=-2\Im\mpret$ is identified with
\begin{equation}
	\msp_l(k) = \frac{-2 \Im\mseret_l(k)}
					{\left(\frac{1}{2\njlkpl}+ \Re\mseret_l(k)\right)^2 + \Im\mseret_l(k)^2} \,. \label{eq:mspectral}
\end{equation}

The term $\frac{1}{2\njlkpl}+ \Re\mseret_l$ determines the pole structure of $\msp_l$. The imaginary parts of the polarizations generate the width. The meson spectral functions are not normalized to a certain value. There are no explicit coupling factors that enter the quark--meson interaction vertices, cf.~Eq.~(\ref{eq:njl_meson}). Hence, the RPA spectral functions can be interpreted as the products of normalized spectral functions and the effective quark--meson couplings.

\subsection{Self energy, polarizations and widths\label{sec:se_pol_width}}

In analogy to the single-particle Green's functions (\ref{eq:qprops}), we can define causal, anti-causal, and collisional (non-ordered) self energies $\qsegk$ \cite{Danielewicz:1982kk,Botermans:1990qi,Das:1997}. We have already encountered the retarded and advanced self energies $\qse^{\ret,\av}$. They are defined as 
\begin{equation}
	\label{eq:qse_decomp}
	\begin{split}
		\qseret(1,1')&=\qsemf(1,1')+\Theta(t_1-t_{1'})\left[\qsegr (1,1')-\qsekl (1,1')\right] \,,  \\
		\qseav(1,1')&=\qsemf(1,1')+\Theta(t_{1'}-t_1)\left[\qsekl (1,1')-\qsegr (1,1')\right] \,. 
	\end{split}
\end{equation}
In our \order{1/N_c} approach, the mean field self energy $\qsemf$ is identified with the Hartree self energy $\qseh$, cf.~Fig.~\ref{fig:dyson_full}. It is mainly responsible for the generation of a constituent quark mass. The collisional self energies $\qsegk$ correspond to the meson exchange diagram. They generate the short-range effects that arise from decays and collisions in the medium.

The Fourier transformed  $\qseret(k)$ enters the spectral function $\qsp(k)=-2\Im\qpret(k)$ (\ref{eq:qspectral}). As we can see in (\ref{eq:denom_prop_imag},\ref{eq:cmplxqpret}), the width of the spectral function is determined by  $-2\Im\qseret(k)$. In momentum space we find $-2\Im \qseret(k)=i\qsegr(k)-i\qsekl(k)$ \cite{Danielewicz:1982kk}, a relation similar to (\ref{eq:defqsp}). Therefore, we define the quark width $\qbr$ by the sum of the collisional self energies,
\begin{equation}
	\qbr(k) = i\qsegr(k)-i\qsekl(k) = -2\Im \qseret(k) \label{eq:def_qbr}\,.
\end{equation}

Note that $-i\qsekl(k)$ and $i\qsegr(k)$ are identical to the total collision rates for scattering into (gain rate) and out of (loss rate) the configuration $(k_0,\veq k)$, respectively \cite{Kadanoff:1962,Froemel:2001iy}. The relation on the rhs.~of (\ref{eq:def_qbr}) is just the optical theorem \cite{Peskin:1995,Das:1997}. It relates the imaginary part of a Feynman amplitude ($\Im \qseret$) to a total cross section or collision rate ($i\qsegr-i\qsekl$). Cutting the meson exchange diagram in Fig.~\ref{fig:mesexse} yields processes like $q\rightarrow q \pi$ and $q \bar q\rightarrow \sigma$. We will examine these processes in more detail in Section~\ref{ch:qq_scattering}.

Like the quark propagators (\ref{eq:lorentzstructure}) and self energies, the width consists of three Lorentz components $\qbr_{s,0,v}$ which are functions of $k_0$ and $|\veq k|$. Using the definition of the on-shell energy and momentum (\ref{eq:os_def}), we can introduce on-shell self energies and an on-shell width,
\begin{equation}
	\begin{aligned}
		\qse_\os(k_0)    &= \qse(k_0,\vec k_\os(k_0)) \,, \\ \qse_\os(\veq k) &= \qse(k_0^\os(\veq k),\veq k) \,,		 
	\end{aligned}
	\quad
	\begin{aligned}
		\qbr_\os(k_0)    &= \qbr(k_0,\vec k_\os(k_0))\,, \\	\qbr_\os(\veq k) &= \qbr(k_0^\os(\veq k),\veq k)\,.
	\end{aligned}
	\label{eq:qseos}
\end{equation}

The RPA polarizations do not include a time-local contribution. Thus, the retarded polarization  $\mseret_l$ can be expressed solely in terms of the collisional polarizations $\msegk_l$,
\begin{align}
	\mseret(1,1')&=\Theta(t_1-t_{1'})\left[\msegr (1,1')-\msekl (1,1')\right] \,. 
\end{align}
Like $\qsegk$, the collisional polarizations $i\msegr$ and $-i\msekl$ can be identified with total collision rates. We find processes like $\pi\rightarrow q\bar q$, $\sigma q\rightarrow q$, etc., when cutting the polarization diagram in Fig.~\ref{fig:mesexse}. Thus, we introduce the RPA meson widths by
\begin{equation}
	\mbr_l (k) = [i\msegr_l(k)-i\msekl_l(k)]/(2 k_0) = -\Im \mseret_l (k)/k_0 \label{eq:def_mbr} \,.
\end{equation}
The additional factor $1/k_0$ in comparison to (\ref{eq:def_qbr}) is required to get the dimension of the widths right. The factor $1/2$ is conventional. It ensures that the nonrelativistic limit of this definition is identical to the widths from nonrelativistic models \cite{Leupold:2000ma,Yao:2006px}.

\section{Calculation of self energies and polarizations\label{ch:calc_selfenergies}}

\subsection{Mean field self energy and quark width\label{sec:mf_se}}

In our \order{1/N_c} approach of Fig.~\ref{fig:dyson_full}, the Hartree diagram takes the role of the mean field self energy $\qsemf$.  It is calculated using a full in-medium propagator that includes a finite width,
\begin{align}
	-i\qsemf=-i\qseh 
	        = 2 \njlkpl\int \dvp  \Trace \qpc(p) \label{eq:hartree_se}
	        = 8 i \njlkpl N_f N_c \int \dvp \qsp_s (p) \nf(p_0) \,.
\end{align}
Due to the trace, only the isospin scalar ($l=\sigma$) NJL vertex contributes to $\qseh$. On the rhs.~we have used Eqs.~(\ref{eq:qpret},\ref{eq:lorentzstructure},\ref{eq:sret_ints}). The Lorentz structure of $\qseh$ is simple -- it has only a Lorentz scalar component. Note that the integral must be regularized by a cutoff.

The definitions of the quark condensate and the quark density are closely related to $\qsemf$. For the quark condensate we find in isospin symmetric quark matter, cf.~(\ref{eq:hartree_se}),
\begin{align}
\label{eq:quark_cond}
	\ewert{\bar u u} = \ewert{\bar d d} = \frac 12\ewert{\bar\psi\psi}=-\frac i2 \int\dvp \Trace \qpc (p)
	= -\frac 1{4\njlkpl}\qseh \,.
\end{align}
The quark number density $\density(\mu,T)$ is related to $\ewert{\psi_\alpha^\dagger \psi_\alpha} =  -i\Trace \gamma_0 \qpkl(1,1)$. Using (\ref{eq:qpgksp}) the density can be calculated by integrating the spectral function weighted with the occupation propability over all states with a positive effective energy $\Re \effp_0$ (\ref{eq:eff_momentum}),
\begin{align}
		\density(\mu) 
		&= 4 N_f N_c \int_\njlcut \ddp \int_{Re \effp_0=0}^\infty \frac{dp_0}{2\pi} \qsp_0(p) \nf(p_0) \,. \label{eq:qrk_density}
\end{align}

$\qsp_0$ is the only Lorentz component of $\qsp$ that remains finite in the nonrelativistic limit. In a quasiparticle approximation, the peaks of $\qsp_0$ become $\delta$-functions, see (\ref{eq:mf_spectral}). At zero temperature the integral can then be solved analytically,
\begin{align}
	\density_\mathrm{qp}(\mu)= N_f N_c/(3\pi^2) k_F^3 \,, \label{eq:qp_density}
\end{align} 
where $k_F=\sqrt{\mu^2-\effmc^2}$ is the quasiparticle Fermi momentum. Note that the Fermi momentum of the \order{1/N_c} approach is defined by $\prpl(\mu,k_F)=0$, cf.~(\ref{eq:denom_prop_real},\ref{eq:os_def}).

\subsection{Collisional self energies and quark widths\label{sec:qwidth}}

The collisional self energies $\qsegk$ are calculated from the meson exchange diagram in Fig.~\ref{fig:mesexse}, using full in-medium propagators. The Feynman rules of the real-time formalism \cite{Danielewicz:1982kk} yield
\begin{align}
	-i\qsegk (k) = \sum_l \int \dvp\dvr (2\pi)^4 \delta^4 (k-p+r)  \Gaml  \qpgk(p) \Gamlt  \mpkg_l (r) \,. \label{eq:qsegk}
\end{align}
The contributions from the different RPA pions are equivalent up to the different propagators $\mpgk_0$, $\mpgk_\pm$ and the isospin factors in $\Gaml,\Gamlt$ (\ref{eq:gamnjl}). In flavor symmetric quark matter, the isospin matrices $\tau_l$ commute with the quark propagators. Moreover, we have $\mpr_0=\mpr_+=\mpr_-$ since the polarizations $\mse_{0,\pm}$ are equal, cf.~(\ref{eq:msegk_sigmapi}). Thus, it is not necessary to treat the different pions separately here. After introducing universal pion propagators $\mpgk_\pi$, their contributions can be added up,
\begin{align}
	- i\qsegk(k) = \int \dvp &\left[ 
		\qpgk_s(p) \left\{\mpkg_{\sigma}(p-k) - 3\mpkg_{\pi}(p-k)\right\} \right. \notag \\
		&\quad + \left. \qpgk_{\mu}(p)\gamma^\mu \left\{\mpkg_{\sigma}(p-k) + 3\mpkg_{\pi}(p-k)\right\} \right] \,. 
		\label{eq:qsegkexpl}
\end{align}

To determine $\Im\qseret$ and the quark width, it is convenient to replace the propagators in (\ref{eq:qsegkexpl}) by spectral functions according to Eqs.~(\ref{eq:qpgksp},\ref{eq:mpgksp}). The width is given by the sum of the collisional self energies as shown in (\ref{eq:def_qbr}),
\begin{align}
	\qbr_{s,\mu}(k)= \int\dvp [\nf(p_0)+\nb(p_0-k_0)] 
		\qsp_{s,\mu}(p) \left[\msp_{\sigma}(p-k) \mp 3\msp_{\pi}(p-k)\right] \,, \label{eq:qwidth_sp}
\end{align}
where the upper sign refers to the Lorentz scalar and the lower sign to the Lorentz vector component. $\qsp_i (p)$ should be read as $\qsp_v (p) \cos \vartheta e_k^i$ when determining $\qbr_v$.

For $T=0$, $\nf$ and $\nb$ become simple step functions (\ref{eq:nf_t0},\ref{eq:nb_t0}). This can be used to combine the distribution functions with the limits of the $p_0$ integration,
\begin{align}
	\int^{+\infty}_{-\infty}dp_0 [\nf(p_0)+\nb(p_0-k_0)]\dotsi 
		\quad \xrightarrow{T=0} \quad 
			\int^\mu_{k_0}dp_0 \dotsi \,.	\label{eq:qwidth_mu}
\end{align}
The width is zero at $k_0=\mu$ in this case: States at the Fermi energy are stable (quasiparticles). Due to Pauli blocking it is not possible to scatter into our out of those states. For finite temperature this restriction is less strict. At low temperatures, however, the width will remain small for energies in the vicinity of the chemical potential.

\subsection{Collisional polarizations and RPA meson widths\label{sec:mes_width}}

For the collisional polarizations $\msegk_l(k)$, cf.~Fig.~\ref{fig:mesexse}, we  find
\begin{align}
 -i\msegk_l (k) 
	= -\int \dvp\dvq (2\pi)^4 \delta^4 (k-p+q) \Trace \left[i \Gamlt i \qpgk(p)i \Gaml i \qpkg(q)\right]  \,. \label{eq:msegk}
\end{align}
Working out the trace yields in flavor symmetric quark matter, where the isospin matrices $\tau_l$ commute with the quark propagators,
\begin{align}
	-i\msegk_{\sigma,\pi} (k) = - 4 N_f N_c \int \dvp \left[\pm\qpgk_s(p)\qpkg_s(p-k) + \qpgk_{\mu}(p)\qpkglo^{\mu}(p-k)  \right] \,.
	\label{eq:msegk_sigmapi}
\end{align}
The upper sign in the integrand refers to $\sigma$ and the lower sign to the $\pi$ case. Since $\tau_\mp\tau_\pm=1\pm \tau_3$ and $\Trace\tau_3=0$, the same result is found for all pions.

From Eq.~(\ref{eq:mesprop}) and $\msegk_0 =\msegk_+ =\msegk_-$ follows that $\mpgk_0 (k)= \mpgk_+ (k)= \mpgk_- (k)$. Thus, we can introduce a universal pion propagator $\mpr_\pi$, as already done in (\ref{eq:qsegkexpl}). This simplification is not possible in flavor asymmetric matter. There, the quark propagators have a nontrivial structure in flavor space and will not commute with $\tau_l$.

The widths of the RPA mesons are found with the help of Eq.~(\ref{eq:def_mbr}). We replace the quark propagators in (\ref{eq:msegk_sigmapi}) by spectral functions and distribution functions $\nf$,
\begin{align}
	\mbr_{\sigma,\pi}(k) = \frac{2 N_f N_c}{k_0} &\int \dvp [\nf(p_0-k_0)-\nf(p_0)]  \notag \\
		&\quad\times \left[\pm\qsp_s(p)\qsp_s(p-k) + \qsp_{\mu}(p)\qsp^{\mu}(p-k)\right]	\,.  \label{eq:mbr_sigma_pi}           
\end{align}
Note that the integral is antisymmetric in $k$. To see that, one has to perform the substitution $p\rightarrow p-\frac 12 k$. $\nf$ becomes a simple step function (\ref{eq:nf_t0}) at zero temperature. This can be used to rewrite the $p_0$ integration,
\begin{align}
	\int^{+\infty}_{-\infty}dp_0 [\nf(p_0-k_0)-\nf(p_0)]\dotsi 
	\quad\xrightarrow{T=0}\quad \int^{\mu+k_0}_\mu dp_0 \dotsi \,. \label{eq:mwidth_t0}
\end{align}

\subsection{Dispersion integrals\label{sec:disprel}}

In the NJL model, cutoffs are used to regularize the self energy and polarization integrals. Such cutoffs yield a violation of analyticity when we calculate both, the real \emph{and} the imaginary parts of $\qseret$ and $\mseret_l$, using Feynman rules. Some properties of the spectral functions -- like the normalization of $\qsp(k)$ (\ref{eq:norm_qsp2}) -- would be lost.

Our main interest in the present work is to investigate the collisional broadening of the spectral functions. For the self-consistent approach it is more reasonable to preserve analyticity than to calculate the real parts directly. Therefore, we will calculate only $\Im\qseret$ ($\qbr_{s,\mu}$) and $\Im\mseret_l$ ($\qbr_{\sigma,\pi}$) from the Feynman rules as shown in Sections~\ref{sec:qwidth} and \ref{sec:mes_width}. The real parts will be determined by the dispersion relations
\begin{equation}
	\label{eq:disp_rel}
	\begin{split}
		\Re\qseret (k_0,\veq k) &= \qsemf +\frac 1 {2\pi} \mathcal{P}\int^{+\infty}_{-\infty} d p_0
					\frac{\qbr(p_0,\veq k)}{k_0-p_0} +\const \,, \\
		\Re\mseret_l(k_0,\veq k) &= \frac 1 {\pi} \mathcal{P}\int^{\infty}_0 d p_0^2
					\frac{p_0\mbr_l(p_0,\veq k)}{k_0^2-p_0^2} +\const
	\end{split}
\end{equation}
In the second relation, we have used the antisymmetry of $p_0\mbr_l$ that has been mentioned after Eq.~(\ref{eq:mbr_sigma_pi}). Constant -- or at least energy independent -- contributions to the real parts do not interfere with analyticity and cannot be calculated from dispersion integrals. Thus, the (purely real) mean field self energy must be calculated separately.

As indicated in (\ref{eq:disp_rel}), there exist further energy independent contributions to $\Re\qseret$ and $\Re\mseret_l$. They should not be ignored. For example, the RPA pion mass depends strongly on the inclusion of such shifts. Identifying constant terms -- like the Fock diagram that is part of the meson exchange in Fig.~\ref{fig:mesexse} -- requires an investigation of the real parts that are given by the Feynman rules. The details of this rather technical exercise can be found in the Appendices \ref{app:re_mespol} and \ref{app:re_qse}. The full dispersion relations including all energy independent shifts are given in (\ref{eq:qrk_disp}) for $\Re\qseret$ and in (\ref{eq:remseret_full}) for $\Re\mseret_l$.

\subsection{Masses of the RPA mesons\label{sec:meson_masses}}

As discussed in Section~\ref{sec:ncexpansion}, we use the RPA mesons mainly as a tool to generate the next-to-leading order quark--quark interactions. Nonetheless, it is interesting to explore the properties of the mesons. The quasiparticle approaches of \cite{Dmitrasinovic:1995cb,Nikolov:1996jj,Oertel:2000sr} have shown that the RPA pions, i.e. the bound \qbq states, may loose their Goldstone boson character in $1/N_c$ extensions of the NJL model. In Appendix~\ref{app:hfrpa}, that is briefly summarized below, we demonstrate how this effect arises on the mean field level.

\begin{figure}
\centerline{
		\includegraphics[scale=.7]{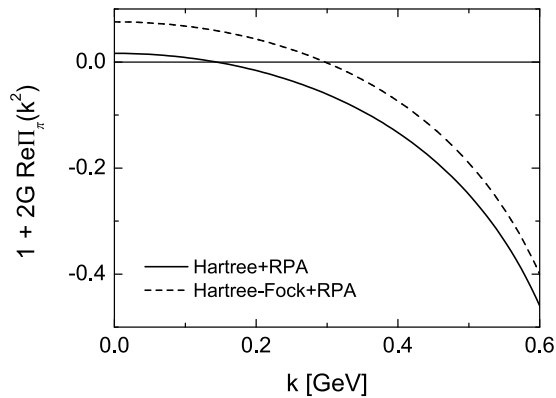}
	}
	\caption{\label{fig:1_re_Pi_pi} The real part of the denominator of the RPA pion propagator in the Hartree+RPA and Hartree--Fock+RPA approximations, using parameter set I from Table~\ref{tab:njlparam}. The zeros determine $m_\pi$, see the text (and Appendix~\ref{app:hfrpa}) for details.}
\end{figure}

The poles of the retarded RPA propagators are given by the zeros of $1+2\njlkpl \Re\mseret_l(k)$, cf.~(\ref{eq:mesprop_ret}). In the Hartree+RPA approximation (see (\ref{eq:hfproppi_denom},\ref{eq:hfpropsigma_denom}) in Appendix~\ref{app:hfrpa} below), the term $1+2\njlkpl \mse_{\mathrm{n}} = (\effmc-\qsemf)/\effmc$ becomes zero in the chiral limit ($m_0=0$), cf.~(\ref{eq:dshift_m0_hartree}). The poles are determined by $\Re\mseret_{\mathrm{d},\pi}\sim k^2 \Im\mathrm{I}(k)$ and $\Re\mseret_{\mathrm{d},\sigma}\sim (k^2-4\effmc^2) \Im\mathrm{I}(k)$. Hence, the RPA pions are massless and the RPA sigma has a mass of $2\effmc$. For finite $m_0$, $1+2\njlkpl \mse_{\mathrm{n}}$ will generate only a small shift and the RPA pions remain light. The effect is shown in Fig.~\ref{fig:1_re_Pi_pi}.

In the \order{1/N_c} approach, $\Re\mseret_l$ is given by (\ref{eq:remseret_full}) and (\ref{eq:mse_nondisp2},\ref{eq:remseret_d}). As numerical checks confirm, the properties of $\mseret_{\mathrm{d} \sigma,\pi} \sim [ k^2 \mathrm{I}^\ret- \mathrm{J}^\ret_{\sigma,\pi}]$ do not change too much. $\mathrm{J}^\ret_{\pi}$ remains very small and $\mathrm{J}^\ret_{\sigma}$ is approximately given by $4 \bar M^2 \mathrm{I}^\ret(k)$, where $\bar M=m_0+\Re\qseret_{\os,s}(\vec k=0)$ is the constant (real) part of $M$ from (\ref{eq:num_pol_sigmapi}). However, the term $1+2\njlkpl \mse_{\mathrm{n}}$ will -- like in the Hartree--Fock+RPA approach (\ref{eq:fock_shift}) -- not vanish for $m_0=0$. Eq.~(\ref{eq:mse_nondisp2}) yields
\begin{align}	
	1+2\njlkpl \mse_{\mathrm{n}} &=	 \frac{m_0+\qse^s_\const-\qsemf}{m_0+\qse^s_\const}  \notag \\
	&\quad +\frac{16\njlkpl N_f N_c}{m_0+\qse^s_\const} 
			\int\dvp \Im \frac{\qseret_s(p)-\qse^s_\const}{\effp^2_\ret-\effm^2_\ret} \nf(p_0) \,. \label{eq:meson_poles_full}
\end{align}
The first term corresponds to the mean field result (\ref{eq:hfproppi_denom},\ref{eq:hfpropsigma_denom})  with $\effmc=m_0+\qse^s_\const$. The second term appears only in the full calculation, where $\qseret$ becomes complex and four-momentum dependent.

It is reasonable to chose $\qse^s_\const$ as large as possible (see the discussion below (\ref{eq:qsemf_split})) to minimize the second term. Due to next-to-leading order contributions, $\qse^s_\const$ will then not be equal to $\qsemf$. Using a numerical result of our approach, $\qsemf \approx 0.75\,\qse^s_\const$, leads to values of $0.15-0.2$ for the first term while the second term remains small. Hence, the RPA pions are massive even in the chiral limit. Qualitatively, this resembles the Hartree--Fock+RPA approximation discussed in Appendix~\ref{app:hfrpa}, see Fig.~\ref{fig:1_re_Pi_pi}. Quantitatively, the shift exceeds the Hartree--Fock+RPA value of $0.08$.

The large shift raises the question which influence the RPA pion mass has on the quark properties in our self-consistent calculation. We will address this problem in Section~\ref{ch:qq_scattering}. There, it will be found that the off-shell states of the pion propagator -- and not the bound \qbq states -- have the largest influence on the quark self energy. The \qbq continuum states in the timelike region are not generated by poles of the pion propagator but by large decay widths. They are largely independent of the RPA pion mass and differ within reasonable limits in the \order{1/N_c} and the Hartree+RPA approaches. The large RPA pion mass may lead, however, to a suppression of the contributions from the spacelike region of the propagator to the quark width. Our analysis in Section~\ref{ch:qq_scattering} indicates that our approach will -- on average -- underestimate the short-range effects in the chirally broken phase.

Let us conclude this section by reviewing why the chiral properties of our model are broken and how they may be restored. 
It has been shown in \cite{Dmitrasinovic:1995cb,Nikolov:1996jj} that the chiral properties of dynamically generated pions in a perturbative expansion depend on a careful choice of polarization diagrams. Only when certain contributions cancel each other, the pions become massless in the chiral limit. In $1/N_c$ extended schemes, the quark propagators are not purely of order \order{1}, cf.~Section~\ref{sec:ncexpansion}. When the RPA polarizations are calculated using these propagators, contributions in subleading orders will be automatically generated -- in an incomplete way. Thus, the cancellation effect also remains incomplete  beyond leading order and the RPA pions become massive.

\begin{figure}
	\centerline{
		\includegraphics[scale=.8]{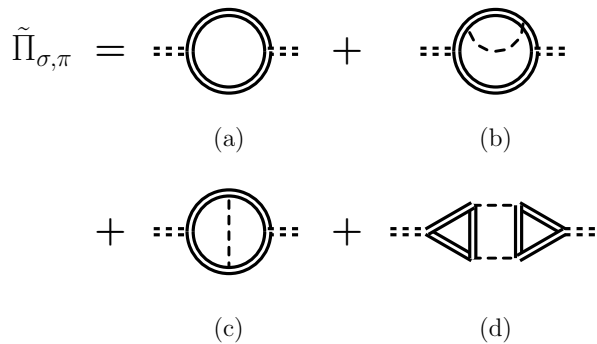}
	}
	\caption{\label{fig:corr_dmitra}The $1/N_c$ extended polarizations of \protect\cite{Dmitrasinovic:1995cb}. Double solid and single dashed lines indicate self-consistently calculated quark and RPA meson propagators, respectively. Double dashed lines indicate $1/N_c$ extended mesons. See the text for details.}
\end{figure}

A $1/N_c$ extension to the NJL model that restores the chiral theorems in next-lo-leading order can be found in \cite{Dmitrasinovic:1995cb}. The (quasiparticle) approach is based on Dyson--Schwinger equations similar to those in Fig.~\ref{fig:dyson_full} (see Figs.~1 and 2 in \cite{Dmitrasinovic:1995cb}). These equations, that yield massive RPA pions, are solved self-consistently as a first step. The physical mesons are found in a second step, where the $1/N_c$ extended polarizations shown in Fig.~\ref{fig:corr_dmitra} 
-- including next-to-leading order corrections\footnote{Note that a restoration of the chiral properties in higher orders may be possible by using Bethe-Salpeter equations that generate vertex corrections -- like those in Fig.~\ref{fig:corr_dmitra}(c,d) -- in all orders \cite{vanHees:2002bv}.} -- are calculated using the self-consistent quark and RPA propagators. The $1/N_c$ improved pions are not further modified, i.e., they are not reinserted into a self-consistent calculation. This would destroy the chiral properties again.

Following the approach of \cite{Dmitrasinovic:1995cb}, we could determine the Goldstone modes of our model (the diagram in Fig.~\ref{fig:corr_dmitra}(b) is generated automatically). However, this would have no influence on the properties of the quarks -- the main interest of the present work. They would still be calculated from the Dyson--Schwinger equations of Fig.~\ref{fig:dyson_full} where the massive RPA pions enter. Hence, such an extension is of limited use for us.

A fully self-consistent approach that yields massless pions has not yet been proposed for the NJL model. (In \cite{Leupold:2006bp} such a scheme has been proposed which, however, does not work in the vicinity of the phase transition.) Adding the vertex corrections of Fig.~\ref{fig:corr_dmitra}(c,d) to the second Dyson--Schwinger equation in Fig.~\ref{fig:dyson_full} should improve -- but not fix -- the chiral properties of the pions and generate a feedback on the quarks. However, the numerical efforts would increase significantly by involving two- and three-loop diagrams that must be calculated in every iteration. Since the numerical calculations are already rather involved we do not pursue this strategy any further at the present stage. Accepting that our approach yields a conservative estimate of the short-range correlations in the chirally broken phase, we use the (uncorrected) RPA mesons in our calculations and ignore the vertex corrections for the time being.

\section{Quark and meson scattering\label{ch:qq_scattering}}

The self energies  $-i\qsekl(k)$ and $i\qsegr(k)$ are identical to the total collision rates for scattering into and out of a quark state with four-momentum $k$ \cite{Kadanoff:1962}. Due to the variety of contributing scattering and decay processes, $\qsegk$ -- and thus the width $\qbr=i\qsegr(k)-i\qsekl(k)$ -- have a complicated structure and a non-trivial energy dependence \cite{Weldon:1983jn}. For a better understanding of this structure, we will investigate the collision rates on the level of the Hartree+RPA approximation at zero temperature. The results will be helpful later when the results of the full calculation are discussed. In addition, they allows us to estimate the influence of a too large RPA pion mass on the quark properties.

\subsection{Mean field spectral functions}

In the Hartree approximation, the quark spectral function $\qsp$ consists of two quasiparticle peaks,
\begin{align}
	\qsp(k)=2\pi(\slashed{k}+\effmc) \frac{1}{2E_k}[\delta(k_0-E_k)-\delta(k_0+E_k)]\,, \label{eq:mf_spectral}
\end{align}
where $E_k=(\vec k^2 +\effmc^2)^{1/2}$ and $\effmc=m_0+\qseh$. We interpret quark states $k=(k_0,\veq k)$ at negative $k_0$ as antiquark states $\bar k=(\bar k_0,\veq k)$ that have a positive energy $\bar k_0=-k_0$. The two peaks in (\ref{eq:mf_spectral}) can then be identified with a quark peak at $k_0=E_k$ and an antiquark peak at $\bar k_0=E_k$. An antiquark corresponds to a hole in the populated quark states. Since the chemical potential is positive in our calculations, there are no holes in the Dirac sea. Consequently, there are no antiquarks in the medium at $T=0$.

Each of the meson spectral functions $\msp_{\sigma,\pi}$ (\ref{eq:mspectral}) consists of two components that we will treat separately. Bound \qbq states generate the poles of the RPA propagators and show up as peaks in $\msp_{\sigma,\pi}$. In addition, the spectral functions include broad ranges of unbound \qbq states, cf.~Section~\ref{sec:ncexpansion}: The \qbq continuum is located at $|k_0|>\mu+\effmc$, i.e., above the threshold for the decay  into \qbq pairs \cite{Rehberg:1998nd}. Landau damping ($\pi q\rightarrow q$, etc.) generates further contributions at spacelike four-momenta \cite{Kapusta:2006}. In those regions, the spectral functions are characterized by large widths $\mbr_{\sigma,\pi}$ and not by poles.

The RPA pion peaks are located below the \qbq decay threshold. Hence, they generate pronounced contributions to the collision rates that are well separated from the continuum contributions. For the quasiparticle component of the RPA pion spectral function we can use an expression similar to (\ref{eq:mf_spectral}),
\begin{align}
	\msp_\pi(k) \sim \delta(k_0-E^\pi_k)-\delta(k_0+E^\pi_k) \,, \label{eq:qqbound_spec}
\end{align}
where $E^\pi_k=(\vec k^2+m_\pi^2)^{1/2}$. $m_\pi$ is the effective pion mass that is determined by the poles of the RPA propagator. We will refer to a meson with negative energy $k_0$ as a meson with positive energy $\bar k_0=-k_0$ in the following. Since we are only interested in energy thresholds, it is not necessary to specify $\msp_\pi$ in more detail. To investigate the off-shell contributions to the quark width, we will later replace the incoming and outgoing mesons in (\ref{eq:qsegk}) by \qbq pairs. The RPA sigma peaks are located above the \qbq decay threshold and thus are rather broad. For our qualitative investigation, it is sufficient to investigate the \qbq continuum contributions in this case.

\subsection{Thresholds\label{seq:scatt_thresholds}}

The collisional self energies in Eq.~(\ref{eq:qsegk}) can be immediately identified with total collision rates: The propagators $-i\qpkl$ and $i\qpgr$ (\ref{eq:qpgksp}) are just the densities of the populated and the free quark states. The meson propagators can be interpreted accordingly, cf.~(\ref{eq:mpgksp}). Hence, the loss rate $i\qsegr(k)$ corresponds to a quark with four-momentum $k$ that is added to the system and scatters off a meson into a free quark state. The gain rate $-i\qsekl(k)$, on the other hand, corresponds to decays of a quark from the medium into a quark with four-momentum $k$ and a meson.

Note that incoming quarks with negative energy correspond to outgoing antiquarks (and vice versa) \cite{Weldon:1983jn}. The $\delta$-function in (\ref{eq:qsegk}) yields $k_0+r_0=p_0$, corresponding to the processes $kr\rightleftarrows p$. Since the energy of an antiquark is given by $\bar k_0=-k_0$, the condition should be reinterpreted as $r_0=\bar k_0+ p_0$ when $k_0$ is negative. This corresponds to the processes $r\rightleftarrows \bar k p$. Accordingly,  an incoming meson with negative $r_0$ is reinterpreted as an outgoing meson with positive energy $\bar r_0=-r_0$.

\begin{figure}
	\begin{center}
		\includegraphics[scale=.8]{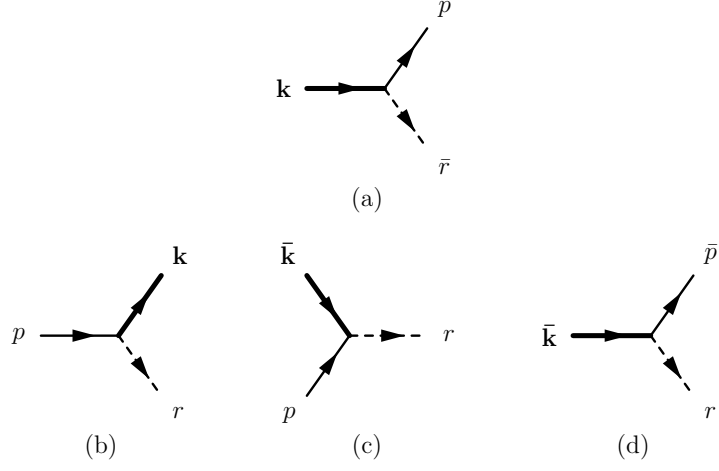}
	\end{center}
\caption{\label{fig:qmscattering}Scattering and decay processes corresponding to $\qsegk(k)$ (\ref{eq:qsegk}). The thick lines carry the external four-momentum $k$. See the text for details.}
\end{figure}

To investigate the contributions from the RPA pion peaks to the collision rates, we replace the propagators in (\ref{eq:qsegk}) by the spectral functions (\ref{eq:mf_spectral},\ref{eq:qqbound_spec}) and the appropriate distribution functions, cf.~(\ref{eq:qpgksp},\ref{eq:mpgksp}). This generates four terms with different combinations of the quasiparticle peaks for each of the two self energies $\qsegk$. At zero temperature, some terms are ruled out by the combination of $\delta$- and step functions in the propagators. In total we find
\begin{align}
	 i\qsegr_0(k) 	&\sim	\int_\njlcut d^3p  ~ \delta(k_0-E_p-E^\pi_{p-k}) \, [1-\nf(E_p)] \, \nb(E_p-k_0)
	 								\label{eq:scattgr_qm} \,, \\
	-i\qsekl_0(k)	&\sim	\int_\njlcut d^3p  ~ \delta(k_0-E_p+E^\pi_{p-k}) \, \nf( E_p) \, [1+\nb(E_p-k_0)]  \notag \\
									&\quad+			\int_\njlcut d^3p  ~ \delta(k_0+E_p+E^\pi_{p-k}) \,  \, [1+\nb(-E_p-k_0)] \,.
									\label{eq:scattkl_qm}
\end{align}
The corresponding processes are shown diagrammatically in Fig.~\ref{fig:qmscattering}. There is one process (a) that contributes to the loss rate $i\qsegr$ and -- taking into account that $k_0$ can be positive or negative in the first term of (\ref{eq:scattkl_qm}) -- three processes (b-d) that contribute to the gain rate $-i\qsekl$ (antiquark loss rate). In the last term of (\ref{eq:scattkl_qm}), the Bose distribution demands $k_0<-E_p$. Hence, the integral will only be finite for negative $k_0$. 

We can now determine the energy thresholds of the processes in Fig.~\ref{fig:qmscattering}. Besides four-momentum conservation, the integrands of (\ref{eq:scattgr_qm},\ref{eq:scattkl_qm}) include the following constraints: Incoming quarks from the medium must have a positive energy below the chemical potential. Outgoing quarks must have an energy higher than $\mu$ (Pauli blocking). Outgoing antiquarks and mesons must have an energy that exceeds their mass. In addition, the (anti\mbox{-})quark three-momenta are regularized by the cutoff $\njlcut$ of the model. Since (\ref{eq:mf_spectral}) provides a sharp relation between the energy and the momentum of an (anti\mbox{-})quark, the energies $p_0$ and $\bar p_0$ are restricted to values below $E_\njlcut=\sqrt{\njlcut^2+\effmc^2}$.

In Table~\ref{tab:qm_thresholds} we summarize the thresholds that can be found for the processes of Fig.~\ref{fig:qmscattering}.
\begin{table*}
	\caption{\label{tab:qm_thresholds}Energy thresholds of the processes in Fig.~\ref{fig:qmscattering}, see the text for details. Note that $E^\pi_{\njlcut-k} = [(\njlcut+|\veq k|)^2+m_\pi^2]^{1/2}$, where $m_\pi$ is the pion mass that is determined by the poles of the RPA propagator.}
\begin{ruledtabular}
		\begin{tabular}{l @{\hspace{3em}} r @{\hspace{1.5em}}ccc@{\hspace{1.5em}} l}
			Process &  \multicolumn{5}{l}{Thresholds}\\
			\hline
			\\[-1ex]
			\mbox{\renewcommand{\arraystretch}{1.2}$\begin{array}{l}\mbox{(a)} \end{array}$}		
								& \mbox{\renewcommand{\arraystretch}{1}
										$\left.\begin{matrix}
												\mu+m_\pi \\[2ex] 
												\sqrt{\vec k^2+(\effmc+m_\pi)^2}
										\end{matrix}\right\}$}			
								&	$<$ & $k_0$ & $<$
								&	$E_\njlcut+E^\pi_{\njlcut-k}$
			\\[5ex]
			\mbox{\renewcommand{\arraystretch}{1.2}$\begin{array}{l}\mbox{(b),(c)}\\ \mbox{\small($\effmc>m_\pi$)} \end{array}$}
								& $\mu-\sqrt{(|\veq k|+k_F)^2+m_\pi^2}	$	
								&	$<$ & $k_0$ & $<$
								&	\mbox{\renewcommand{\arraystretch}{1}
									$\left\{\begin{matrix}
											\mu-m_\pi \\[2ex] 
											\sqrt{\vec k^2+(\effmc-m_\pi)^2}
									\end{matrix}\right.$}
			\\[6ex]
			\mbox{\renewcommand{\arraystretch}{1.2}$\begin{array}{l}\mbox{(b),(c)}\\ \mbox{\small($\effmc<m_\pi$)} \end{array}$}		
								&	$\sqrt{(|\veq k|-k_F)^2+m_\pi^2}-\mu$ 
								&	$<$ & $\bar k_0$ & $<$
								& \mbox{\renewcommand{\arraystretch}{1}
									$\left\{\begin{matrix}
										E_{\njlcut+k_F}^\pi-\mu \\[2ex]
										\sqrt{\vec k^2 +(\effmc-m_\pi)^2}
									\end{matrix}\right.$}	
			\\[5ex]
			\mbox{\renewcommand{\arraystretch}{1.2}$\begin{array}{l}\mbox{(d)} \end{array}$}
								& $\sqrt{\vec k^2+(\effmc+m_\pi)^2}$						
								&	$<$ & $\bar k_0$ & $<$
								&	$E_\njlcut+E^\pi_{\njlcut-k}$ 
			\\[2ex]
		\end{tabular}
	\end{ruledtabular}
\end{table*}
Where two values are given for the same threshold, the first one follows from Pauli blocking or the upper energy limit for quarks from the medium and the second one is a kinematical constraint. These limits are complementary. Which one is stricter depends on $\modk$. The thresholds in Table~\ref{tab:qm_thresholds} are not always the strictest limits that can be found. However, to identify the contributions from individual processes to the quark width they are perfectly suitable.

\begin{figure}
	\centerline{
		\includegraphics[scale=.8]{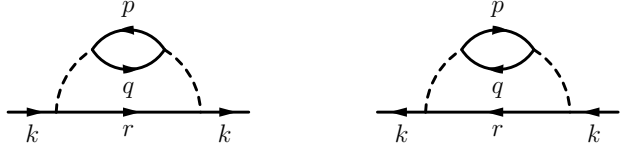}
	}
	\caption{\label{fig:mod_quark_se}Self energy diagrams corresponding to $-i\qsegr$ (left) and $-i\qsekl$ (right) in the rewritten form of Eq.~(\ref{eq:mod_qse}).}
\end{figure}

The continuum of off-shell states in the RPA spectral functions cannot be represented by quasiparticles. To investigate their contributions to the collision rates in a similar way as the contributions from the bound states, we rewrite the integrals in Eq.~(\ref{eq:qsegk}). Thereby we replace the outgoing mesons by (quasiparticle) \qbq pairs. Using the relation $\mpgk_l =\mpret_l \msegk_l \mpav_l$ \cite{Zhang:1992rf,Greiner:1998vd,Chou:1984es} in  Eq.~(\ref{eq:qsegk}) we find
\begin{align}
	-i\qsegk (k) = -i\sum_l & \int \dvp\dvq\dvr (2\pi)^4\delta^4(k+p-q-r)  \label{eq:mod_qse}\\
 	&\quad\times \Gaml \qpgk(r) \Gamlt \mpret_l(r-k) 
 			\Trace \left[ \Gamlt \qpkg(p) \Gaml \qpgk(q)\right] \mpav_l(r-k) \,. \notag
\end{align}
Diagrammatically, the self energies (\ref{eq:mod_qse}) have the form shown in Fig.~\ref{fig:mod_quark_se}. 
Eq.~(\ref{eq:mod_qse}) and the diagrams in Fig.~\ref{fig:mod_quark_se} resemble the approach of \cite{Froemel:2001iy}, where we have calculated the collisional self energies from the Born diagrams in Fig.~\ref{fig:hfborn}. Here, however, the bare NJL coupling $\njlkpl^2$ is replaced by the propagators of the dynamically generated mesons, $\mpret \mpav={|\mpret|}^2$.

The thresholds are now generated by the incoming and outgoing (anti\mbox{-})quarks that are explicitly given by the propagators $\qpgk$. Hence, we can ignore the off-shell structure of the meson propagators in (\ref{eq:mod_qse}). It is also not necessary to distinguish between $\sigma$ and $\pi$ contributions -- they differ only in magnitude but not in the thresholds.

\begin{figure}
	\centerline{
		\includegraphics[scale=.8]{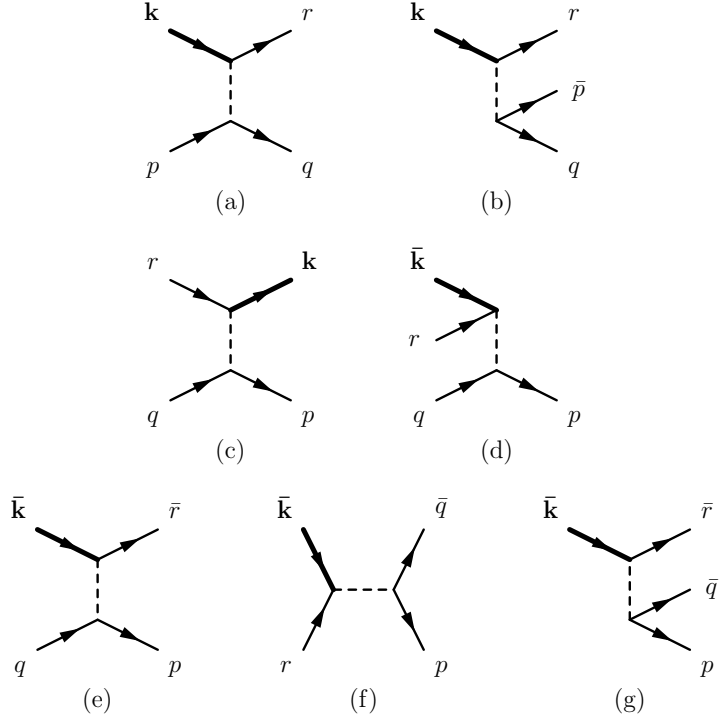}
	}
	\caption{\label{fig:qqscattering}Scattering and decay processes corresponding to $\qsegk(k)$ in the form of Eq.~(\ref{eq:mod_qse}). Thick lines carry the external four-momentum $k$. See the text for details.}
\end{figure}

Replacing the quark propagators in (\ref{eq:mod_qse}) by mean field spectral functions (\ref{eq:mf_spectral}) and Fermi distributions, we find the processes shown in Fig.~\ref{fig:qqscattering}.
Many of the formally possible processes are ruled out by the distribution functions (no incoming antiquarks). For simplicity, we do not display u-channel processes in Fig.~\ref{fig:qqscattering}. They have the same thresholds as the shown t-channel processes (the full calculation includes both channels).

The processes (a) and (b) in Fig.~\ref{fig:qqscattering} are contributions to the loss rate $i\qsegr(k)$. (c) and (d) are two of the contributions to the gain rate $-i\qsekl$. Note that they differ only in the sign of $k_0$. The processes (e)-(g) are the remaining contributions to $-i\qsekl$ for negative $k_0$. 
The thresholds for these processes are found by applying the same constraints to the incoming and outgoing quarks as in the analysis of the processes in Fig.~\ref{fig:qmscattering}. We have summarized all thresholds in Table \ref{tab:qq_thresholds}.

\begin{table*}
	\caption{\label{tab:qq_thresholds}Energy thresholds of the processes in Fig.~\ref{fig:qqscattering}, see the text for details.}
	\begin{ruledtabular}
	\begin{tabular}{l @{\hspace{3em}} r @{\hspace{1.5em}}ccc@{\hspace{1.5em}} l}
		Process & \multicolumn{5}{l}{Thresholds} \\
		\hline
		\\[-1ex]
		(a) 			& $\mu$
							&	$<$ & $k_0$ & $<$
							&	$2E_\njlcut-\effmc$
		\\[3ex]
		(b) 			& \mbox{\renewcommand{\arraystretch}{1}
									$\left.\begin{matrix}
											2\mu+\effmc	 \\[2ex] 
											\sqrt{\vec k^2+(3\effmc)^2}
									\end{matrix}\right\}$}			
							&	$<$ & $k_0$ & $<$
							&	$3E_\njlcut$ 
		\\[6ex]				
		(c),(d)	& $2\mu-\sqrt{(|\veq k|+2k_F)^2+\effmc^2}$			
							&	$<$ & $k_0$ & $<$
							&	$\mu$ 
		\\[3ex]							
		(e),(f)	& $\effmc$																		
							&	$<$ & $\bar k_0$ & $<$
							&	$2 E_\njlcut-\effmc$
		\\[3ex]
		(g) 			& \mbox{\renewcommand{\arraystretch}{1}
									$\left.\begin{matrix}
											\mu+2\effmc	 \\[2ex] 
											\sqrt{\vec k^2+(3\effmc)^2}
									\end{matrix}\right\}$}			
							&	$<$ & $\bar k_0$ & $<$
							&	$3E_\njlcut$ 
		\\[5ex]	
	\end{tabular}
	\end{ruledtabular}
\end{table*}

\subsection{Density dependence\label{sec:densitydepend}}

Eqs.~(\ref{eq:qsegk}) and (\ref{eq:mod_qse}) allow us to make some estimates on the density dependence of the processes in Figs.~\ref{fig:qmscattering} and \ref{fig:qqscattering}. Each incoming quark from the medium can be identified with an integral over $\qpkl$ for positive energies, cf.~(\ref{eq:qrk_density}), and thus introduces a linear density dependence into the corresponding collision rate. Each outgoing quark -- not counting quarks with the external four-momentum $k$ -- corresponds to an integral over $\qpgr$, i.e., the density of the free states. Due to Pauli blocking and the (fixed) cutoff, this density decreases when $\mu$ rises.

It follows from these considerations that the growth of the collision rates -- and the short-range correlations -- will be limited. We have already investigated this effect for nucleons in nuclear matter \cite{Froemel:2003dv}. There, Pauli blocking leads to a saturation of the correlations at a density of $2-3$ times normal nuclear matter density.

Pauli blocking becomes effective as soon as a medium is present. Its influence increases continuously as $\mu$ rises. The cutoff acts only on outgoing states with large energy. When the total incoming energy is low, such states cannot be reached. Depending on the considered process, the cutoff may be only relevant if $\mu$ and/or $k_0$ are large.

We use process (c) of Fig.~\ref{fig:qqscattering} as an example to illustrate the role of the cutoff: In this process the cutoff becomes relevant for  $p_0>E_\njlcut$, i.e., when the incoming energy exceeds $E_\njlcut+k_0$. The maximum energy of the two incoming quarks is $2\mu$. Hence, the cutoff suppresses only contributions to the collision rates when $2\mu>E_\njlcut+k_0\,(>E_\njlcut)$. For lower $\mu$, final states with energies above $E_\njlcut$ cannot be reached kinematically.

Note that $E_\njlcut=(\njlcut^2+\effmc^2)^{1/2}$ is -- due to $\effmc$ -- density dependent and jumps at the phase transition. A numerical check shows that $\mu\approx E_\njlcut/2$ in the chirally broken phase. Above the phase transition, however, $\mu$ is well beyond this limit. Therefore, we can expect that the density dependence of process (c) in Fig.~\ref{fig:qqscattering} and some of the other processes changes significantly at the phase transition. All these considerations will help us to interpret our numerical results presented below in Section~\ref{ch:results}.

\subsection{On-shell width\label{sec:os_processes}}

The on-shell peaks dominate the structure of the quark spectral function. Hence, processes that contribute to the on-shell width have the largest influence on the properties of the medium.  This is of particular interest for the present \order{1/N_c} approach: The thresholds of the processes involving bound \qbq states depend on $m_\pi$, cf.~Table \ref{tab:qm_thresholds}. As we already know, the mass of the RPA pions in our approach will be larger than that of the physical pions in the chirally broken phase -- and thus shift the thresholds. Furthermore, we will see below that the magnitude of the contributions from some of the processes in Fig.~\ref{fig:qqscattering} depends on the pion mass. This will have some impact on the on-shell width and the properties of the medium.

A process contributes to the on-shell width when the on-shell energy $E_k=(\vec k^2+\effmc^2)^{1/2}$ is located within its thresholds. As Table~\ref{tab:qq_thresholds} shows, the processes (a), (c), and (e,f) of Fig.~\ref{fig:qqscattering} contribute to the on-shell width of all free quark states, all populated quark states, and all antiquark states, respectively. The processes (b,d,g) do not reach the on-shell regions when $\effmc$ is finite.

For massless quarks, the upper threshold of process (d) and the lower kinematical limits of (b) and (g) are located at $E_k$. However, phasespace just opens at this point and thus contributions to the on-shell width are not generated. This picture will not change substantially when collisional broadening is taken into account. If the (on-shell) quark width does not become too large -- and this will be found numerically, see Fig.~\ref{fig:avg_width} -- the thresholds are still rather sharp.
For (b,g) we have also to consider that the constraints from Pauli blocking are stricter than the kinematical ones at moderate $\modk$.

The processes (a,b,d) of Fig.~\ref{fig:qmscattering} behave similar to (b,g) of Fig.~\ref{fig:qqscattering}. When $m_\pi$ is finite, the lower kinematical thresholds of (a) and (d) and the upper kinematical threshold of (b) keep the processes off-shell. Pauli blocking prevents (a) and (b) even from entering the off-shell regions near the Fermi energy. When $m_\pi$ vanishes, the thresholds of (a,b,d) will be located at $E_k$. Due to the same arguments as before, there will be no contributions to the on-shell width on the quasiparticle level. Small contributions are possible when collisional broadening is considered.

Process (c) of Fig.~\ref{fig:qmscattering} is the only process including a bound \qbq state that may contribute to the (antiquark) on-shell width. For this to happen, $m_\pi$ must be larger than $2\effmc$. The RPA pions of the \order{1/N_c} approach as well as the physical pions will exceed this limit only in the chirally restored phase. Since the masses of \order{1/N_c} RPA pions and Hartree+RPA pions are much closer to each other in the chirally restored phase (we will discuss the numerical results for the RPA pion mass in Section~\ref{sec:rpa_results}), such on-shell contributions are no artifact of the present approach. Note that the limit $\bar k_0<E_{\njlcut+k_F}^\pi-\mu$ in Table \ref{tab:qm_thresholds} is a simplification of $\bar k_0<\mu-[(\modk+k_F)^2 +m_\pi^2]^{1/2}$ (for $\modk>\effmc/(m_\pi-\effmc)k_F$). This condition forbids on-shell contributions above a certain value of $\modk$. An interesting observation holds for all populated on- and off-shell quark states: The contributions to the width -- process (b) of Fig.~\ref{fig:qmscattering} and (c) of Fig.~\ref{fig:qqscattering} -- do not include antiquarks. Even when process (c) of Fig.~\ref{fig:qmscattering} contributes to the antiquark on-shell width, this will have no \emph{direct} influence on the quark states.

The analysis of the thresholds shows that the on-shell quark width in the chirally broken phase is determined by the processes of Fig.~\ref{fig:qqscattering}. All those processes do not involve bound \qbq states (treating the broad RPA sigma peak as part of the \qbq continuum). The thresholds of the processes are independent of the RPA pion mass, cf.~Table~\ref{tab:qq_thresholds}. The processes involving bound \qbq states do not contribute to the collision rates in the on-shell region, even when the pions have a reasonable mass. The gap between the on-shell energy and the thresholds of these processes is on the order of $m_\pi$. Considering a realistic pion mass of $140\MeV$, this gap is much larger than the typical width of the on-shell peaks that we find numerically.

So far, we have  not considered the influence of the pion mass on the magnitude of the various contributions to the on-shell width. In the processes of Fig.~\ref{fig:qqscattering}, we do not probe the on-shell region of the RPA pion propagator (i.e., the bound \qbq states). However, the propagator may also depend on the pion mass in its off-shell regions.  Thus, it is not unlikely that the pion mass has influence on the on-shell quark width even though the width is generated by the ``correct'' processes.

\begin{figure}
	\centerline{
		\includegraphics[scale=0.7]{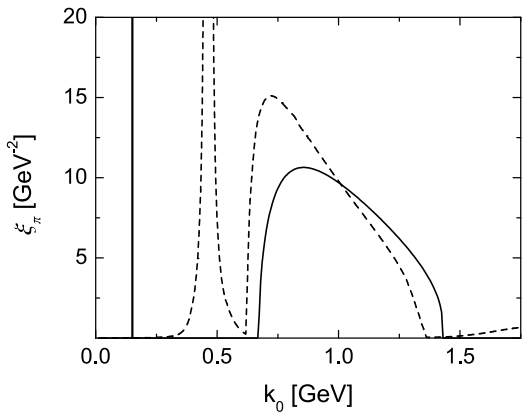}
	}
	\caption{\label{fig:cont_comp}The RPA pion spectral function at a constant three-momentum of $50\MeV$. The solid line shows the result of the Hartree+RPA approximation in vacuum ($\effmc=336\MeV$), the dashed line the result of the \order{1/N_c} approach at a density of $0.05\rho_0$ ($\effmc=326\MeV$). Parameter set I of Tables~\ref{tab:njlparam} and \ref{tab:fullnjlparam} has been used.} 
\end{figure}

Let us split up the relevant processes of Fig.~\ref{fig:qqscattering} into the t-channel processes (a,c,e) and the s-channel process (f). In the s-channel process -- like in the decays (b,d,g) that do not contribute to the on-shell width -- the exchanged meson carries a timelike four-momentum that must exceed $k_\pi^2>(\effmc+\mu)^2>(2\effmc)^2$. In other words, the magnitude of the scattering rate is sensitive to the \qbq continuum of the RPA spectral functions. We show numerical results for the \qbq continuum of the RPA pion spectral function in the Hartree+RPA approximation and in the \order{1/N_c} approach in Fig.~\ref{fig:cont_comp}. 
The differences in size and shape of the continuum are within reasonable limits -- we are comparing a leading order to a next-to-leading order scheme here. A significant $m_\pi$ dependence of the \qbq continuum cannot be observed. This means that the magnitude of the contributions from process (f) to the on-shell quark width is -- like the thresholds -- independent of the large RPA pion mass (in the chirally broken phase) of our \order{1/N_c} approach. We note that the same is true for the contributions of the decays (b,d,g) to the off-shell quark width.

In the t-channel processes (a,c,e), the mesons must carry a spacelike four-momentum $k_\pi^2<0$ \cite{Yao:2006px}. Typically, one would expect that the range of the meson-exchange interaction in these processes scales with the inverse mass of the exchanged meson, i.e. $R\sim 1/m_\pi$. The simple estimate $\qbr\sim\rho\sigma v$, $\sigma\sim \pi R^2$ (where $\sigma$ is the total quark--quark cross section and $v$ the relative velocity of two colliding quarks) suggests a quadratic $m_\pi$ dependence of the quark width, $\qbr\sim 1/m_\pi^2$. Since the RPA pions of our approach are almost three times heavier than physical pions in the chirally broken phase, this means that the contributions from the processes (a,c,e) to the on-shell quark width could be suppressed.

The off-shell properties of our dynamically generated mesons differ from those of regular mesons -- we come back to this point in Appendix~\ref{app:re_qse}. Hence, we cannot directly transfer the above estimate to our approach. A numerical check of the term $|\mpret_\pi(k_\pi)|^2$ in (\ref{eq:mod_qse}), that we do not discuss in detail here, shows that the real suppression effect is $k_\pi^2$ dependent (we integrate over $k_\pi=r-k$ in the total collision rates) and ranges -- on average -- well below one order of magnitude. The naive estimate $\qbr\sim 1/m_\pi^2$ yields only an upper limit for the suppression and overestimates the influence of the large pion mass on the on-shell width.

This brings us to the end of our investigation of the $m_\pi$ dependence of the quark width. We have found that our approach generates the on-shell width from the ``correct'' processes, i.e. from the same processes as an approach with a more reasonable pion mass -- this is important for the structure of the width and its density dependence. However, the contributions from several processes are suppressed by the too large RPA pion mass. In summary, this means that our approach represents a conservative approximation of the short-range effects in quark matter. A calculation with a more realistic pion mass should lead to stronger short-range correlations. Let us clarify once again that these considerations affect only our results in the chirally broken phase. Above the phase transition, the RPA pion mass of our approach will be close to the pion mass of the Hartree+RPA approximation. Thus, there will be no suppression of the quark width in the chirally restored phase.

For completeness, we will consider the correctional diagrams (c,d) of Fig.~\ref{fig:corr_dmitra}. A self-consistent inclusion of these diagrams will fix the pion mass in next-to-leading-order and shift the limits of the processes in Fig.~\ref{fig:qmscattering} closer to the Hartree+RPA values in the chirally broken phase. More important than this shift, the contributions from the t-channel processes of Fig.~\ref{fig:qqscattering} will rise. Since the chiral properties of the pions remain disturbed in higher orders, the RPA pions will still be heavier than physical pions. The correctional diagrams will also modify Eq.~(\ref{eq:mod_qse}) and thus the \qbq continuum entering Fig.~\ref{fig:qqscattering}. These corrections are, however, only next-to-leading order contributions to the quark width and thus suppressed in $1/N_c$.

We note that the $m_\pi$ dependence of the thresholds in Table~\ref{tab:qm_thresholds} may lead to differences between calculations with massive and with massless pions in the chirally broken phase. In contrast to the density dependent (on-shell) processes (a,c,e,f) of Fig.~\ref{fig:qqscattering}, the decays (a,d) of Fig.~\ref{fig:qmscattering} occur even in vacuum. Thus, their small contributions to the width in the on-shell regions -- for $m_\pi=0$ -- may be larger than those of the actual on-shell processes at low densities.

When moving close to the on-shell regions, the large off-shell contributions of the decays may also have influence on the results, even when the on-shell width does not change too much. The left panel of Fig.~\ref{fig:specwidthresig} shows a similar effect: In the \order{1/N_c} approach, the contributions from process (c) of Fig.~\ref{fig:qmscattering} come rather close to the antiquark on-shell region in the chirally broken phase. They generate the small peak in the width at low negative $k_0$. This peak is reflected in the spectral function by a significant structure next to the antiquark on-shell peak.

The processes (a) and (d) of Fig.~\ref{fig:qmscattering} are less density dependent than (c). Thus, they generate larger off-shell contributions at low densities -- they can be identified with the shoulders on the large bumps in the left panel of Fig.~\ref{fig:specwidthresig}. When those processes move closer  to the on-shell regions -- which happens when $m_\pi$ becomes small compared to the quark width -- the impact will surely be more extreme than that of process (c) in Fig.~\ref{fig:specwidthresig}. It is important to stress that the pions must be (almost) massless to observe the just discussed effect. The mass of the physical pions, $m_\pi\approx 140\MeV$, is considerably larger than the typical on-shell quark width in our approach, cf.~Fig.~\ref{fig:os_qse0}.

We come to the end of our quasiparticle analysis of the scattering processes. The investigation of the thresholds and the density dependence of the individual processes in Sections~\ref{seq:scatt_thresholds} and \ref{sec:densitydepend}, respectively, will be very helpful for the understanding of the results of our full \order{1/N_c} calculations.

\section{Numerics and results\label{ch:results}}

\subsection{Details of the calculation}

Using the present \order{1/N_c} approach, we have performed numerical calculations for flavor symmetric quark matter in thermodynamical equilibrium. We have investigated a wide range of chemical potentials $\mu$, allowing for quark matter both, in the chirally broken and restored phase. The calculations have been restricted to zero temperature.

In the Hartree+RPA approximation, the NJL coupling $\njlkpl$ and the cutoff $\njlcut$ are fitted to the quark condensate and the pion decay constant. If a finite current quark mass $m_0$ is desired, it is used to adjust $m_\pi$. Presently, we do not calculate the $1/N_c$ corrected physical pions (Goldstone modes) and have no access to their mass and the pion decay constant. Hence, we proceeded in a way analogous to the transition from the Hartree to the Hartree--Fock approximation in \cite{Klevansky:1992qe} to adjust the parameter sets of Table~\ref{tab:njlparam} to our approach: We kept $\njlcut$ and $m_0$ fixed and lowered the (Hartree) coupling $\njlkpl$ by $22\%$ so that a reasonable value for the quark condensate (\ref{eq:quark_cond}) was obtained. Note that the too large RPA pion mass cannot be fixed by adjusting $m_0$, cf.~(\ref{eq:meson_poles_full},\ref{eq:fock_shift}). The readjusted \order{1/N_c} NJL parameters are shown in Table \ref{tab:fullnjlparam}.
\begin{table}
	\caption{\label{tab:fullnjlparam} Parameter sets for the self-consistent \order{1/N_c} approach,  based on the Hartree parameter sets in Table \ref{tab:njlparam}. $\njlcut$ is a three-momentum cutoff. The couplings were reduced by $22\%$ to readjust the quark condensate. $\effmc$ and $\ewert{\bar u u}$ are the effective mass (\ref{eq:eff_mass_const}) and the quark condensate (\ref{eq:quark_cond}) that are found with those parameters in vacuum.}
	\begin{ruledtabular}
	\begin{tabular}{lccccc}
						& $m_0$					&		$\njlkpl\njlcut^2$		&		$\njlcut$ 		&		$\effmc$ 				&		$\ewert{\bar u u}^{1/3}$ 	\\
						& [MeV]					&													&		[MeV]					&		[MeV]						&		[MeV]											\\
		\hline
		Set 0 	& - 						&  	1.68									&		653						&		308							&		-248								    	\\
		Set I  	& 5.5 					&		1.72									&		631						&		326							&		-244											\\
	\end{tabular}
	\end{ruledtabular}
\end{table}

As discussed in Section~\ref{ch:qq_scattering}, we identify quark states at negative $k_0$ with antiquark states $\bar k=(\bar k_0,\veq k)$ at positive $\bar k_0$. Note that, strictly speaking, the quark-antiquark border is not located at $k_0=0$ but at $\Re \tilde k_0=k_0-\Re\qseret_0(k_0,\veq k)=0$. $\Re\qseret_0$ shifts  the energy scale by less than $5\MeV$ in the chirally broken and less than $50\MeV$ in the chirally restored phase. In the calculations we use the correct border, e.g., when determining the quark density and the momentum distribution. In the discussion, however, we just refer to positive and negative energy states to denote quarks and antiquarks, respectively.

For the discussion of the results (not for the numerics) we also introduce an effective quark mass that is 
real and constant. Like in \cite{Domitrovich:1993tz,Domitrovich:1993tq}, we define $\effmc$ in terms of the real part of the on-shell self energy, cf.~Eqs.~(\ref{eq:os_def},\ref{eq:qseos}), for a quark at rest,
\begin{align}
	\effmc= m_0+\Re \qseret_{s,\os} (\vec k=0)=\Re\effm^{\ret,\av}(k_0^\os(0),0) \,. \label{eq:eff_mass_const}
\end{align}
The numerical results show that the momentum dependence of $\Re\effm(k)$ ranges below $5\%$ for the on-shell states. Hence, $\effmc$ yields a good classification for the states in the vicinity of the on-shell peak at all momenta.

\subsection{Numerical implementation}

The self-consistency problem of Fig.~\ref{fig:dyson_full} can be solved iteratively. Starting from a Hartree propagator that is modified by adding a small width, we calculate the Dyson--Schwinger equations consecutively (beginning with the lower one) until the results converge. In the calculations we keep $\mu$ fixed while the quark mass and the density may change.

\begin{figure*}
	\centerline{
		\includegraphics[scale=1]{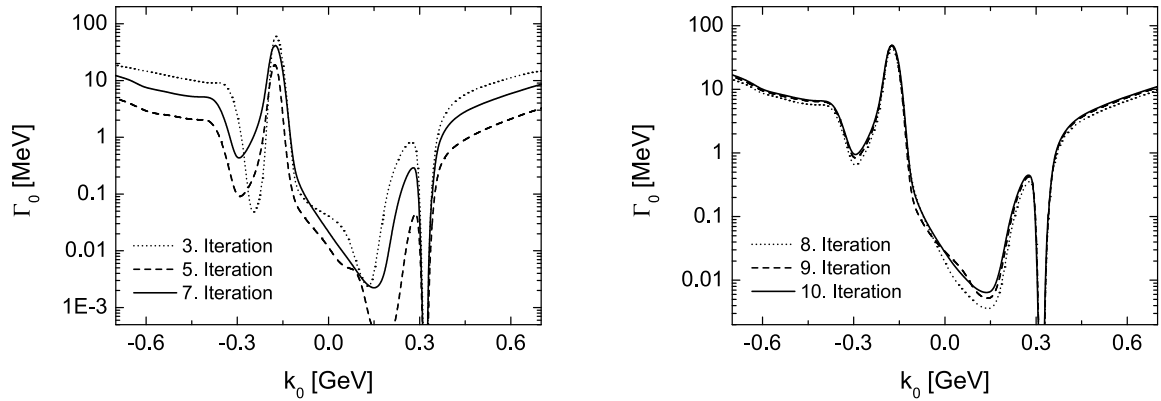}
	}
	\caption{\label{fig:conv_lomu}The quark width in an iterative calculation at a chemical potential of $\mu=323\MeV$ ($\rho=0.5\rho_0$), using parameter set I from Table \ref{tab:fullnjlparam}. The calculation has been initialized with a constant width of $\qbr_0=5\MeV$, $\qbr_{s,v}=0$ and $\effmc=325\MeV$.}
\end{figure*}

Fig.~\ref{fig:conv_lomu} shows how the quark width converges in a calculation in the chirally broken phase.
The strong variations in the first iterations are related to fluctuations of the quark mass.  The quasiparticle relation $\rho\sim (\mu^2-\effmc^2)^{3/2}$, cf.~(\ref{eq:qp_density}), indicates that small changes of $\effmc$ have significant impact on the density  -- and thus the width, cf.~Section~\ref{sec:densitydepend} -- when $\effmc$ and $\mu$ are of comparable size. Self-consistency is reached after $10$ iterations.

Calculations in the chirally restored phase, where $\effmc$ is much smaller than $\mu$, converge faster (4-5 iterations). Close to the phase transition, the gap equation may have several solutions. More iterations can then be necessary to find one of the stable solutions.

Solving the width integrals (\ref{eq:qwidth_sp},\ref{eq:mbr_sigma_pi}) is the most time consuming part of the numerical calculation -- the integrands consist of products of spectral functions with pronounced on-shell peaks. We use the package CUBPACK \cite{Cools:2003} for this task. Since it handles vector integrands we can calculate all Lorentz components of the quark width in one turn (per iteration). To support the numerical routines, we use a substitution of the general form $\vec p^{\,2}\rightarrow z=\arctan[\vec p^{\,2}-\vec p^{\,2}_\os(p_0)]$ for the quark three-momenta. The slope of $\arctan(x)$ is largest at $x=0$. Hence, the on-shell peaks are smeared out in the coordinate $z$.

The quark self energy and the meson polarizations depend separately on the energy and the modulus of the three-momentum. Thus, the Dyson--Schwinger equations must be solved on numerical grids. For the quarks and mesons we cover ranges of $|k_0|\leq 3.5\GeV$, $|\veq k| \leq \njlcut$ and $|k_0|\leq 3\GeV$, $|\veq k| \leq 2\njlcut$, respectively.  Using grids with $300 \times 100$ (quarks) and $300\times 200$ (mesons) mesh points we achieve a resolution of $d k_0 \approx 20\MeV$ and  $d |\veq k|\approx 6\MeV$. This choice is justified by the results -- the structure of  $\qseret$ and $\mseret_l$ is smooth with respect to the mesh size. The spectral functions (\ref{eq:qspectral},\ref{eq:mspectral}) are not discretized on the grid since the on-shell peaks would be lost.

Note that the limits for $\modk$ are fixed by the cutoff scheme, cf.~Section~\ref{sec:1Nc_approach}. In comparison to these limits, the energy limits are chosen rather high. $\Re\qseret$ and $\Re\mseret$ are calculated from dispersion integrals that range to infinitely high energies. Hence, $\qbr_{s,\mu}$ and $\mbr_{\sigma,\pi}$ must vanish at the grid borders. We have to go beyond the thresholds of Section~\ref{ch:qq_scattering} here since the quark width generates (small) contributions above the quasiparticle thresholds. The quark spectral function at such high energies will not sizably influence the widths (\ref{eq:qwidth_sp},\ref{eq:mbr_sigma_pi}). At higher energies, the on-shell peaks are located at three-momenta above the cutoff. Only the much smaller off-shell contributions enter the integrals.

\subsection{Collisional broadening\label{sec:res_collbroad}}

\begin{figure*}
	\centerline{
		\includegraphics[scale=1]{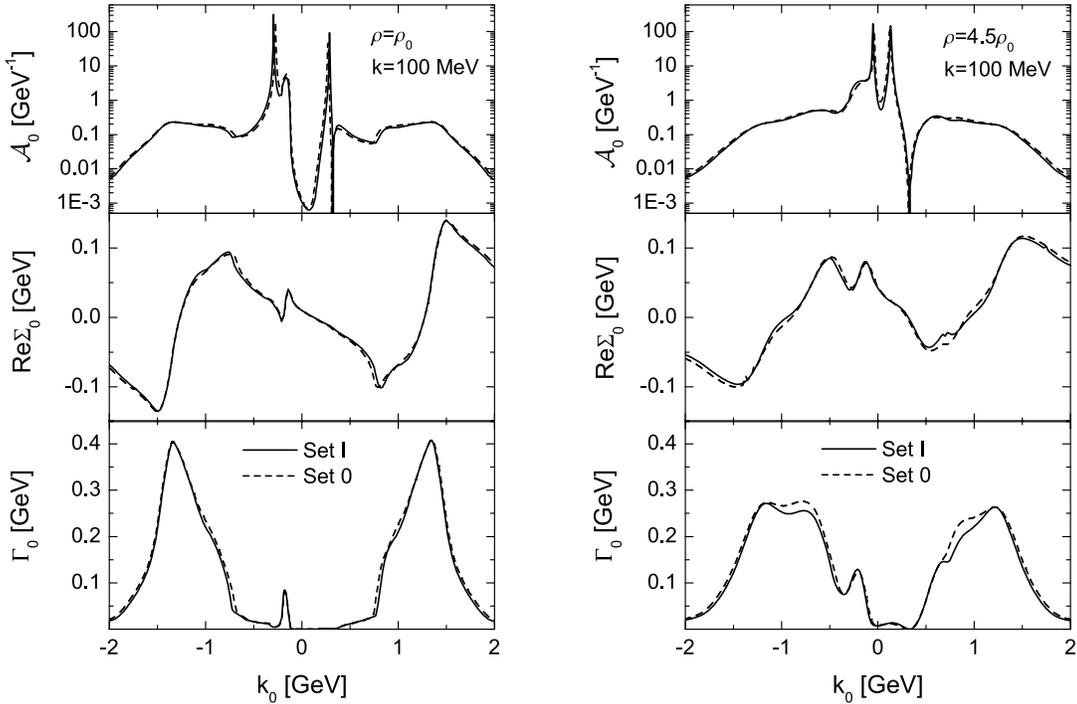}
	}
	\caption{\label{fig:specwidthresig}The spectral function $\qsp_0$, the real part $\Re\qseret_0$, and the width $\qbr_0$ at fixed three-momentum in the chirally broken (left) and restored (right) phases, using the parameter sets of Table \ref{tab:fullnjlparam}. See Figs.~\ref{fig:specwidthcuts_ld} and \ref{fig:specwidthcuts} for details.}
\end{figure*}

We begin with a qualitative overview. Fig.~\ref{fig:specwidthresig} shows the quark spectral function $\qsp_0$, the corresponding width $\qbr_0$, and $\Re\qseret_0$ in the chirally broken and restored phases as cuts at a constant three-momentum. 
The $\gamma_0$-components correspond to the norelativistic limits of $\qsp$ and $\qseret$ and show all relevant features. Furthermore, $\qbr_0$ yields the largest contribution to the width $\mathcal{W}$ (\ref{eq:denom_prop_imag}) and the quark density (\ref{eq:qrk_density}) is determined by $\qsp_0$.

Note that we use $\rho_0=0.17\fm^{-3}$ to normalize the quark density. The numerical value of $\rho_0$ corresponds to the nucleon density in normal nuclear matter $\rho_\textrm{nm}$. Nonetheless, $\rho$ is here a quark density and not a nucleon density. Assuming that each nucleon consists of three constituent quarks, quark matter with a density of $\rho_0$ should be compared to nuclear matter at $\rho_\textrm{nm}/3$.

The spectral functions in Fig.~\ref{fig:specwidthresig} are dominated by the on-shell peaks. In the upper left panel of Fig.~\ref{fig:specwidthresig}, we find the antiquark peak at $k_0\approx -290 \MeV$ and the quark peak at $k_0\approx 290\MeV$, cf.~(\ref{eq:os_def}). In the right panel, the peaks are located at lower energies since the mass of the quarks drops significantly in the chirally restored phase. The peak-like structure right next to the antiquark peak ($k_0\approx-180\MeV$) in the left panel (it turns into the shoulder of the antiquark peak in the right panel) is generated by a pronounced contribution to the quark width -- see the lower panel. In contrast to the on-shell peaks, it does not correspond to a pole of the quark propagator. We will come back to this structure in Section~\ref{sec:res_width_scatt}.

Most of the strength of the spectral functions in Fig.~\ref{fig:specwidthresig} is located close to the on-shell peaks, note the $\log$ scale for $\qsp_0$. The broad underground of off-shell states is generated by the quark width $\qbr_0$. A detailed discussion of the structure of the width -- that also explains the off-shell structure of the spectral functions -- can be found below. Here, we only note that $\qbr_0$ is strongly energy dependent, far off-shell it rises up to $400\MeV$. When the density increases, $\qbr_0$ becomes larger at low $|k_0|$ while it drops at large $|k_0|$. As we will see, this is in agreement with the estimates of Section~\ref{sec:densitydepend}.

The impact of the cutoff at large $|k_0|$  can  be clearly observed in Fig.~\ref{fig:specwidthresig}. $\qbr_0$ and $\qsp_0$ drop rapidly and will eventually vanish. It can be checked that $\Re\qseret_0$ approaches $\qsef_{\eff,0}$, cf.~Appendix~\ref{app:re_qse}. The influence of the cutoff -- that has been introduced for technical reasons -- is not entirely unphysical. The combination of a constant coupling with a cutoff roughly resembles the running coupling of QCD \cite{Peskin:1995}: Quarks with large momenta are free and do not participate in interactions.  Hence, the cutoff can be understood as a crude approximation of asymptotic freedom. In Fig.~\ref{fig:specwidthresig}, the width breaks down for the asymptotically free states at large $|k_0|$.

\begin{figure*}
	\centerline{
		\includegraphics[scale=.8]{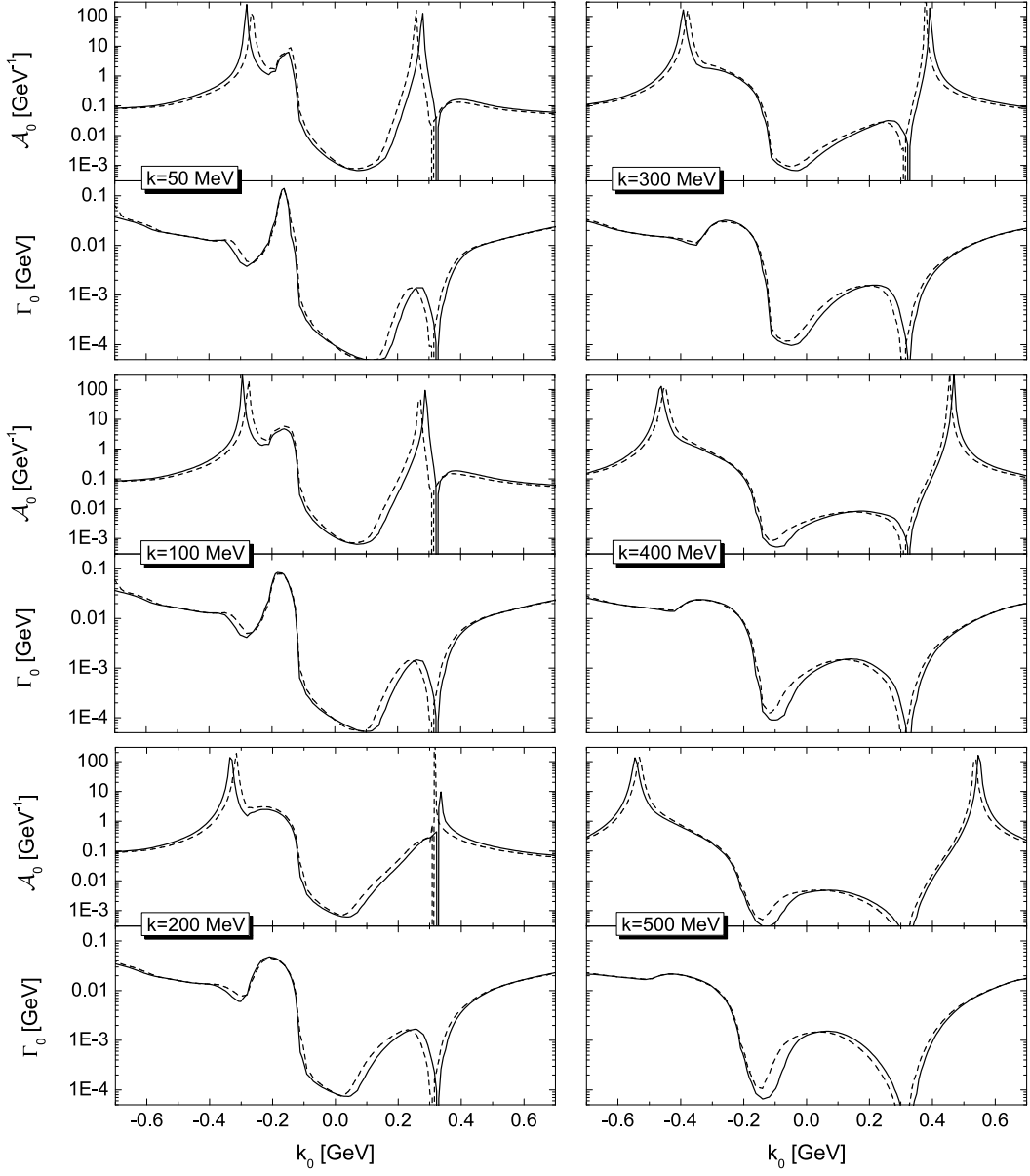}	
	}
	\caption{\label{fig:specwidthcuts_ld}Cuts of the quark spectral function $\qsp_0$ and the width $\qbr_0$ in the chirally broken phase ($\rho=\rho_0$) at several three-momenta, using parameter sets 0 (dashed lines, $\mu=308.5\MeV$, $\effmc=263\MeV$, $k_F=188\MeV$) and I (solid lines, $\mu=325\MeV$, $\effmc=284\MeV$, $k_F=186\MeV$). Note the different scale in $k_0$ as compared to Fig.~\ref{fig:specwidthresig}.}
\end{figure*}

\begin{figure*}
	\centerline{
		\includegraphics[scale=.8]{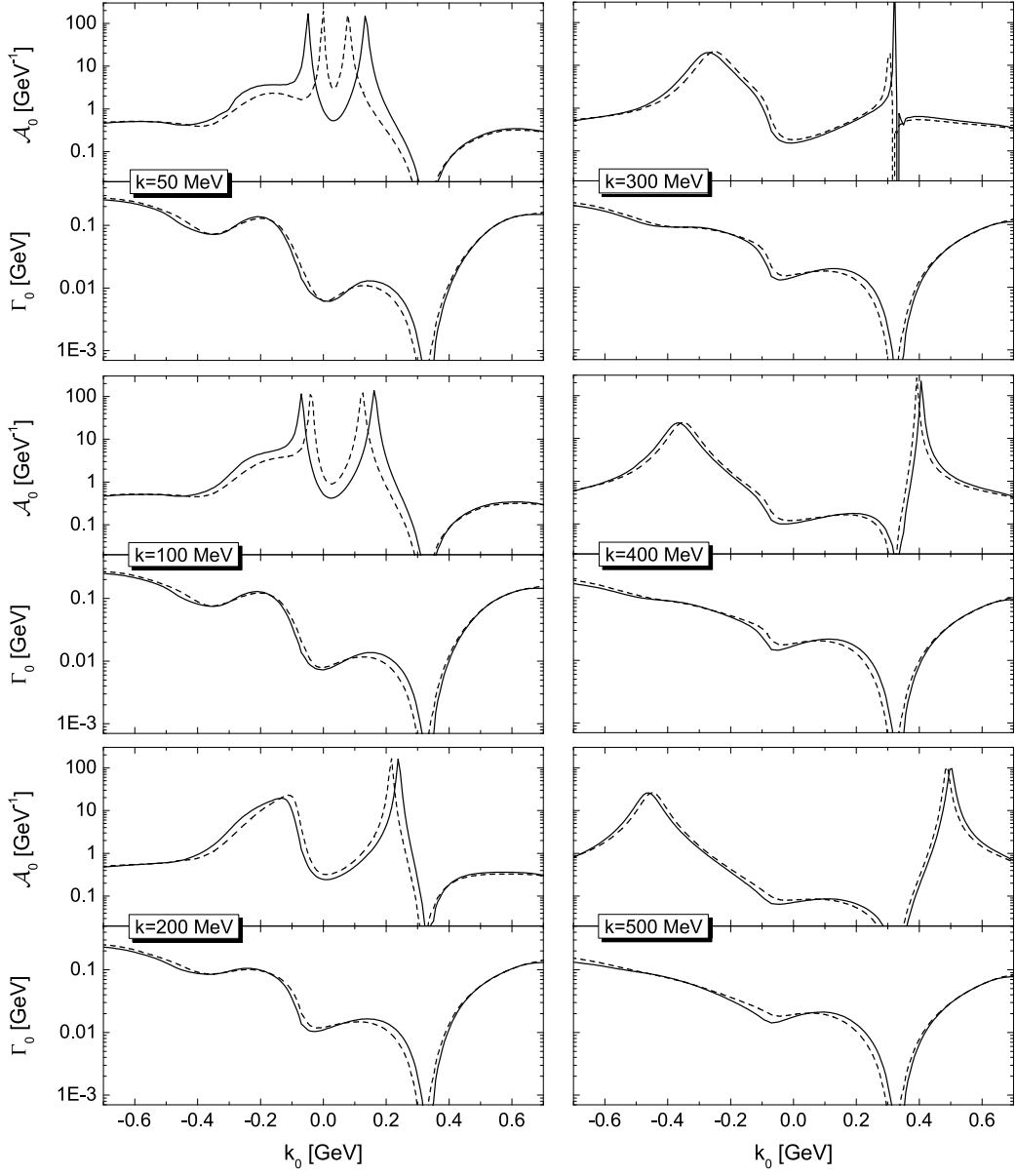}
	}
	\caption{\label{fig:specwidthcuts}Cuts of the quark spectral function $\qsp_0$ and the width $\qbr_0$ in the chirally restored phase ($\rho=4.5\rho_0$) at several three-momenta, using parameter sets 0 (dashed lines, $\mu=318\MeV$, $\effmc=0$, $k_F=312\MeV$) and I (solid lines, $\mu=332\MeV$, $\effmc=100\MeV$, $k_F=315\MeV$).}
\end{figure*}

Figs.~\ref{fig:specwidthcuts_ld} and \ref{fig:specwidthcuts} show $\qsp_0$ and $\qbr_0$ in more detail for  $|k_0|<E_\njlcut=k^\os_0(\njlcut)\approx\sqrt{\njlcut^2+{\effmc}^2}$. This is the region where the on-shell peaks are not suppressed by the cutoff in the integrands of (\ref{eq:qwidth_sp},\ref{eq:mbr_sigma_pi}), cf.~Section~\ref{sec:1Nc_approach}. Cuts for several three-momenta are shown, at the same densities as in Fig.~\ref{fig:specwidthresig}.

A noticeable feature of the width and the spectral function, are the zeros at $k_0=\mu$. Due to Pauli blocking, the states at the Fermi energy are stable for $T=0$, cf.~Section~\ref{sec:qwidth}. Close to the Fermi momentum, the quark peak of $\qsp_0$ turns almost into a sharp quasiparticle peak -- see the $200\MeV$ cut of Fig.~\ref{fig:specwidthcuts_ld} ($k_F=188\MeV$) and the $300\MeV$ cut of Fig.~\ref{fig:specwidthcuts} ($k_F=312\MeV$). Note that $\qsp_0$ and $\qbr_0$ would not vanish at $k_0=\mu$ at finite temperatures. In vacuum, $\qsp_0$ and $\qbr_0$ would be symmetric in $k_0$ while $\Re\qseret_0$ would be antisymmetric. This symmetry is induced by the invariance of the vacuum ground state under charge conjugation, see Appendix~E in \cite{Post:2004phd}. Figs.~\ref{fig:specwidthresig}-\ref{fig:specwidthcuts} show that a finite quark density breaks the symmetry between quarks and antiquarks -- in particular for $|k_0|<\njlcut$.

The results for parameter sets 0 and I differ only weakly in the chirally broken phase. $\njlkpl$ and $\njlcut$ are of similar size, the value of $m_0$ is negligible when $\effmc$ is large. The differences are slightly larger in the chirally restored phase since the effective masses differ significantly: A finite $m_0$ breaks chiral symmetry explicitly. Hence, the quarks have a mass of $100\MeV$ for parameter set I (see also Fig.~\ref{fig:qmass_c1}) while they become massless for set 0. Consequently, the on-shell peaks are shifted, cf.~(\ref{eq:os_def}). Since we compare results at the same density -- not at the same chemical potential -- the zeros at $k_0=\mu$ are also not exactly in the same position, cf.~(\ref{eq:qp_density}).

\begin{figure*}
	\centerline{
		\includegraphics[scale=1]{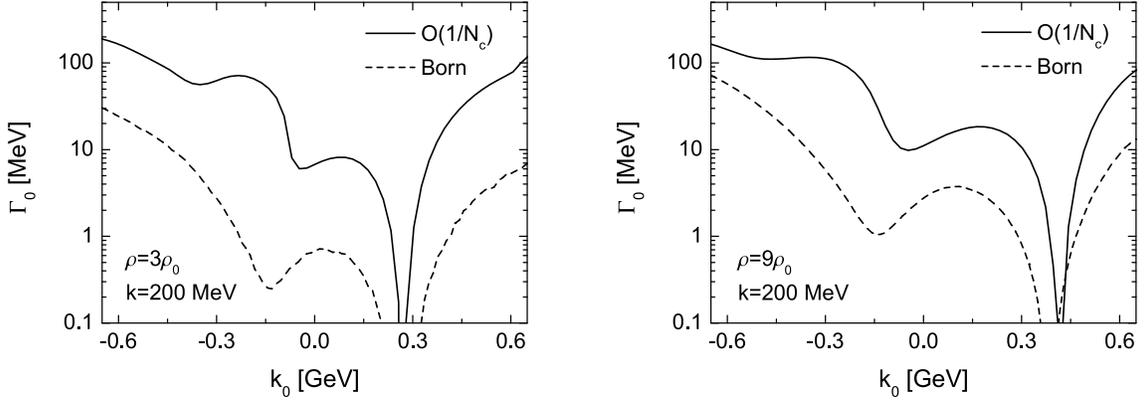}
	}
	\caption{\label{fig:quarkwidth_oldnew}Cuts of the quark width $\qbr_0$ in the chirally restored phase at constant three-momentum. Solid lines show the \order{1/N_c} results using parameter set 0 from Table~\ref{tab:fullnjlparam}, dashed lines the results from \protect\cite{Froemel:2001iy} using parameter set 0 (Hartree) from Table~\ref{tab:njlparam}.}
\end{figure*}

Fig.~\ref{fig:quarkwidth_oldnew} shows a comparison between the present work and our earlier approach in \cite{Froemel:2001iy}.
In the loop-expansion of \cite{Froemel:2001iy} we have calculated the quark width from the Born diagrams shown in Fig.~\ref{fig:hfborn}(c,d). Since we found rather small widths in \cite{Froemel:2001iy}, it was justified to use the Hartree parameter set of Table \ref{tab:njlparam} there. Note that the chemical potentials are not equal since the (density dependent) real part of the quark self energy -- in particular $\Re\qseret_0$ -- were ignored in \cite{Froemel:2001iy}.

Qualitatively, there is some agreement between the results in Fig.~\ref{fig:quarkwidth_oldnew}. We did not introduce RPA mesons in \cite{Froemel:2001iy}, processes involving bound \qbq states did not contribute to the width there. Hence, the bump at negative energies is only found in the \order{1/N_c} results (see the discussion below for details). Quantitatively, the results differ by one order of magnitude. The simple expansion in terms of the coupling constant has clearly missed large contributions to the short-range correlations by ignoring the terms that also are of next-to-leading order in $1/N_c$ but of higher orders in the coupling.

\subsection{Correspondence to scattering and decay processes\label{sec:res_width_scatt}}

To understand the full structure of $\qsp_0$ in Figs.~\ref{fig:specwidthresig}-\ref{fig:specwidthcuts} we have to analyze the structure of $\qbr_0$. Recall our discussion from Section~\ref{ch:qq_scattering} -- all processes that contribute to the width are shown in Figs.~\ref{fig:qmscattering} and \ref{fig:qqscattering}. We will refer to them as \ref{fig:qmscattering}(a), \ref{fig:qqscattering}(a), etc.~in the following. The energy thresholds of the processes are given -- in terms of $\effmc$ and $m_\pi$ -- in Tables \ref{tab:qm_thresholds} and \ref{tab:qq_thresholds}. In Table \ref{tab:full_thresholds} we list the numerical values that are found by inserting the masses of the present approach. 
\begin{table*}
	\caption{\label{tab:full_thresholds}The thresholds of Tables \ref{tab:qm_thresholds} and \ref{tab:qq_thresholds}, adjusted to the results for $|\veq k|=100\MeV$ and parameter set I in Figs.~\ref{fig:specwidthresig}-\ref{fig:specwidthcuts}. The left column corresponds to $\mu=325\MeV$ ($\effmc=284\MeV$, $m_\pi=444\MeV$), the right column to $\mu=332\MeV$ ($\effmc=100\MeV$, $m_\pi=416\MeV$). See the text for details.}
	\begin{ruledtabular}
		\begin{tabular}{ll@{\hspace{3em}}rcccr@{\hspace{4em}}rcccr}
		\multicolumn{2}{l}{Process} 					& \multicolumn{10}{l}{Thresholds [MeV]} \\		
		\cline{3-12}
		&& \multicolumn{5}{l}{ch.~broken phase} & \multicolumn{5}{l}{ch.~restored phase} \\
		\hline
			 													&(a) 			& $769$		&$<$&$k_0$&$<$&				$1547$		&	$748$		&$<$&$k_0$&$<$&				$1480$\\
		Fig.~\ref{fig:qmscattering} &(b),(c)	& $127$		&$<$&$\bar k_0$&$<$&	$189$			&	$136$		&$<$&$\bar k_0$&$<$&	$331$\\
			 													&(d)			& $735$		&$<$&$\bar k_0$&$<$&	$1547$		&	$526$		&$<$&$\bar k_0$&$<$&	$1480$\\
		\hline
																&(a) 			& $325$		&$<$&$k_0$&$<$&				$1100$		&	$332$		&$<$&$k_0$&$<$&				$1178$\\
																&(b)			& $934$		&$<$&$k_0$&$<$&				$2076$		&	$764$		&$<$&$k_0$&$<$&				$1917$\\
		Fig.~\ref{fig:qqscattering} &(c),(d)	& $99$		&$<$&$k_0$&$<$&				$325$			&	$-73$		&$<$&$k_0$&$<$&				$332$\\
																&(e),(f) 	& $284$		&$<$&$\bar k_0$&$<$&	$1100$		&	$100$		&$<$&$\bar k_0$&$<$&	$1178$\\
																&(g)			& $893$		&$<$&$\bar k_0$&$<$&	$2076$		&	$532$		&$<$&$\bar k_0$&$<$&	$1917$\\
	\end{tabular}
	\end{ruledtabular}
\end{table*}
The real parts of $\qseret_\mu$ have not been taken into account. Hence, the thresholds are slightly shifted with respect to the curves in Figs.~\ref{fig:specwidthresig}-\ref{fig:specwidthcuts}. Note that the strict quasiparticle thresholds are softened by the finite quark width.

The processes \ref{fig:qqscattering}(b) and (g) generate the huge bumps of $\qbr_0$ at large $|k_0|$ in Fig.~\ref{fig:specwidthresig}. Such decays can occur at zero density since no partner from the medium is required. The bumps decrease at higher $\rho$ since the outgoing quarks are subject to Pauli blocking. The smaller bump at low negative $k_0$ is generated by \ref{fig:qmscattering}(c). Since the RPA pion is rather heavy, the thresholds for \ref{fig:qmscattering}(b,c) are located at negative $k_0$ and \ref{fig:qmscattering}(b) is suppressed.

In the chirally broken phase, the decays \ref{fig:qmscattering}(a,d) can be identified with the shoulders on the inner slopes of the big bumps in Fig.~\ref{fig:specwidthresig}. \ref{fig:qmscattering}(a) remains visible in the chirally restored phase while \ref{fig:qmscattering}(d) is hidden below the increased contributions from \ref{fig:qqscattering}(e,f). \ref{fig:qqscattering}(e,f) contribute to $\qbr_0$ between the large and the small bump in the antiquark sector. The contributions from \ref{fig:qqscattering}(a) and \ref{fig:qqscattering}(c,d) are located right above and below the Fermi energy, respectively. Since \ref{fig:qqscattering}(a,c-f) depend on quarks from the medium, their influence on $\qbr_0$ increases substantially in the chirally restored phase.

Note that \ref{fig:qmscattering}(c) generates a structure in $\qsp_0$ that looks like a second antiquark peak in the chirally broken phase at $k_0\approx -180\MeV$, cf.~Fig.~\ref{fig:specwidthcuts_ld}. This is an artifact of the present approach. For a more reasonable $m_\pi$ in the chirally broken phase, the contributions from \ref{fig:qmscattering}(c) would be located further off-shell. $\qsp_0$ at positive energies is free from such artifacts. A process comparable to \ref{fig:qmscattering}(c) ,i.e. $k\bar q\rightarrow \pi$, would require an antiquark from the medium. Fig.~\ref{fig:specwidthcuts} shows that process \ref{fig:qmscattering}(c) has considerable influence on the antiquark on-shell width in the chirally restored phase. There, however, the Hartree+RPA pion acquires a considerable mass, too. Thus, the contributions of \ref{fig:qmscattering}(c) should be located approximately in the right position.

The quarks of the medium populate all states at positive energies below the chemical potential.
In the cuts for $\modk>k_F$ in Figs.~\ref{fig:specwidthcuts_ld} and \ref{fig:specwidthcuts}, the quark on-shell peak is located above the chemical potential. Nonetheless, the spectral functions do not entirely vanish in the range $0<k_0<\mu$. Populated off-shell states exist due to the width generated by \ref{fig:qqscattering}(c). Those states are important for the momentum distribution of the medium that will be investigated in Section~\ref{sec:momdist}.

We close the discussion on the structure of $\qbr_0$ and $\qsp_0$ with an important observation: The numerical results confirm our quasiparticle considerations of Section~\ref{ch:qq_scattering}. The width in the range $|k_0|<E_\njlcut\,(\approx630-700\MeV)$ -- where the on-shell peaks are not suppressed -- is mostly generated by the processes of Fig.~\ref{fig:qqscattering}. Those processes do not involve bound \qbq states. It is interesting to note that the contributions of the collisional processes to the overall width in the chirally broken phase are small in comparison to the huge off-shell contributions from decays. This can be readily explained by the strong density dependence of the collisions with partners from the medium. In addition, we have found in Section~\ref{sec:os_processes} that the contributions from the t-channel (a,c,e) processes of Fig.~\ref{fig:qqscattering} depend -- in contrast to the decays -- on the RPA pion mass. Thus, we may underestimate these processes below the chiral phase transition.

\subsection{RPA mesons in the {\boldmath\order{1/N_c}} approach\label{sec:rpa_results}}

\begin{figure*}
	\centerline{
		\includegraphics[scale=1]{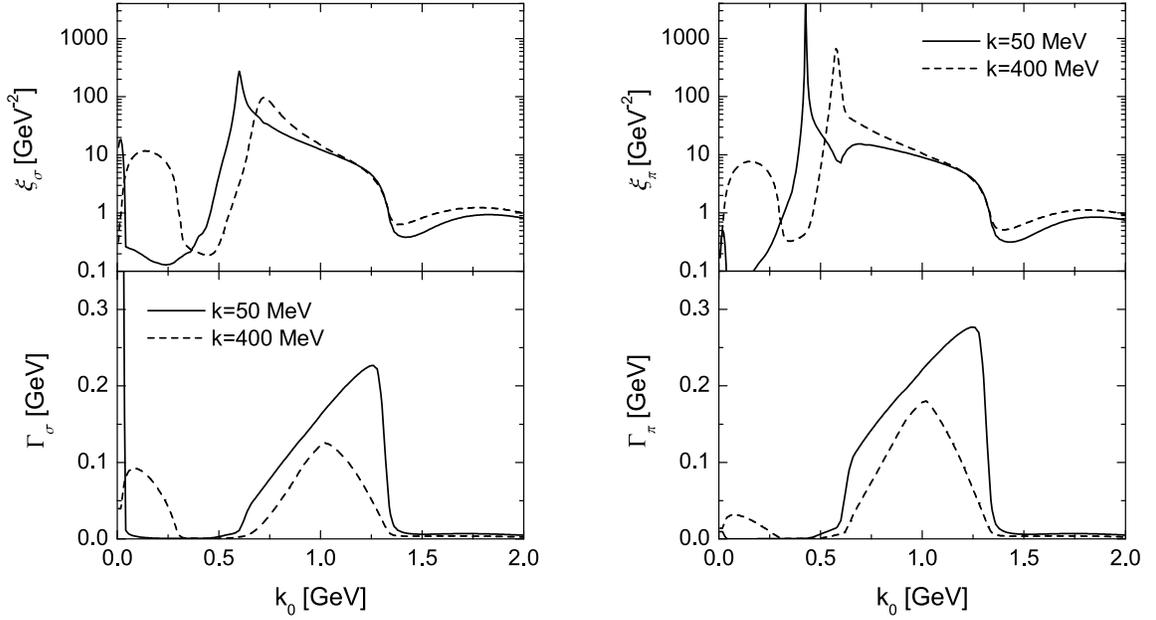}
	}
	\caption{\label{fig:meson_specwidth}Cuts of the RPA meson spectral functions and widths at constant three-momenta. The calculation was performed in the chirally broken phase at $\mu=308.5\MeV$ ($\rho=\rho_0$, $\effmc=263\MeV$), using parameter set 0.}
\end{figure*}

The RPA mesons that we discuss here cannot be identified with the physical $\sigma$ and $\pi$ -- the $1/N_c$ corrections of \cite{Dmitrasinovic:1995cb}, cf.~Section~\ref{sec:meson_masses}, are presently not calculated. Fig.~\ref{fig:meson_specwidth} shows the spectral functions and widths of the RPA mesons in the chirally broken phase.
Since parameter set 0 is used (chiral limit), the physical pion should be massless.  In the \order{1/N_c} calculation, however, we find $m_\sigma=603\MeV$ and $m_\pi=424\MeV$. As expected, the RPA pions are no Goldstone bosons.

The widths in Fig.~\ref{fig:meson_specwidth} are dominated by the \qbq continuum. The quasiparticle threshold for \qbq decays is $k_0=\mu+\effmc$, above $2E_\njlcut=2(\njlcut^2+{\effmc}^2)^{1/2}$ the decays are suppressed by the cutoff. We find small but finite widths beyond these limits in the \order{1/N_c} calculation. The magnitude of the \qbq continuum drops for larger $\modk$. For $|\veq k|>2\njlcut$, at least one quark propagator in (\ref{eq:msegk}) must have a momentum above the cutoff. Consequently, the integral becomes zero and the \qbq continuum vanishes.

At spacelike four-momenta, Landau damping ($q\pi\rightarrow q$, $q\sigma\rightarrow q$) \cite{Kapusta:2006} generates another structure. Since the widths are given by $\mbr_{\sigma,\pi}(k)=-\Im\mseret_{\sigma,\pi}(k)/k_0$ (\ref{eq:def_mbr}), they can become rather large at small $k_0$. However, they do not diverge for $k_0\rightarrow 0$.

The continuum parts of the spectral functions $\msp_{\sigma,\pi}$ in Fig.~\ref{fig:meson_specwidth} are in good agreement with the established mean field approaches. They should generate reasonable contributions to the quark width. The on-shell peak of the RPA pion --  generated by bound \qbq states -- is located below the \qbq continuum. Thus it is rather sharp in comparison to the sigma peak.

\begin{figure*}
	\centerline{
		\includegraphics[scale=1]{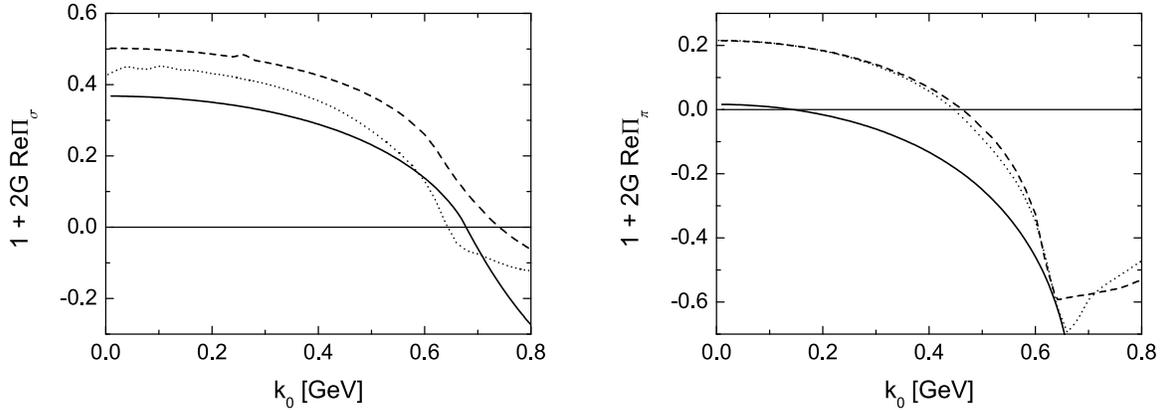}
	}
	\caption{\label{fig:meson_real_full_vs_mf}The real parts of the denominators of $\mpret_{\sigma,\pi}$ at $\modk=50\MeV$ in the chirally broken phase. The solid lines show Hartree+RPA results at zero density, the dashed and the dotted lines results from \order{1/N_c} calculations at $\rho=0.05\rho_0$ and $\rho=\rho_0$, respectively. Parameter set I from Tables \ref{tab:njlparam} and \ref{tab:fullnjlparam} has been used.}
\end{figure*}

In Fig.~\ref{fig:meson_real_full_vs_mf} we take a closer look at the real parts of the denominators of $\mpret_{\sigma,\pi}$. The zeros of $1+2\njlkpl\Re\mseret_{\sigma,\pi}$ determine $m_{\sigma,\pi}$.
At small $k_0$, the \order{1/N_c} pion curves are shifted by $\approx 0.2$ in comparison to the Hartree+RPA result. This is in agreement with the estimate from Section~\ref{sec:meson_masses}: About $50\%$ of the shift can be attributed to the Fock self energy, cf.~(\ref{eq:fock_shift}). Dispersive contributions to $\Re\mseret$ generate the remaining $50\%$. $m_\pi$ increases from $142\MeV$ (Hartree+RPA) to $460\MeV$. Note that the RPA pion is not only heavier than the physical pion but also broader. The on-shell peak is shifted closer to the \qbq continuum, where the width is larger.

In the case of the RPA sigma, the shift of the real parts is smaller. Furthermore, the zero is located in a region where the shift has -- due to the slope of the curve -- less influence on $m_\sigma$. The \order{1/N_c} effects are only on the order of $10\%$. Note that Fig.~\ref{fig:meson_real_full_vs_mf} also shows that $m_\sigma$ has a stronger density dependence than $m_\pi$.

\subsection{Chiral phase transition\label{sec:phasetrans}}

\begin{figure*}
	\centerline{
		\includegraphics[scale=1]{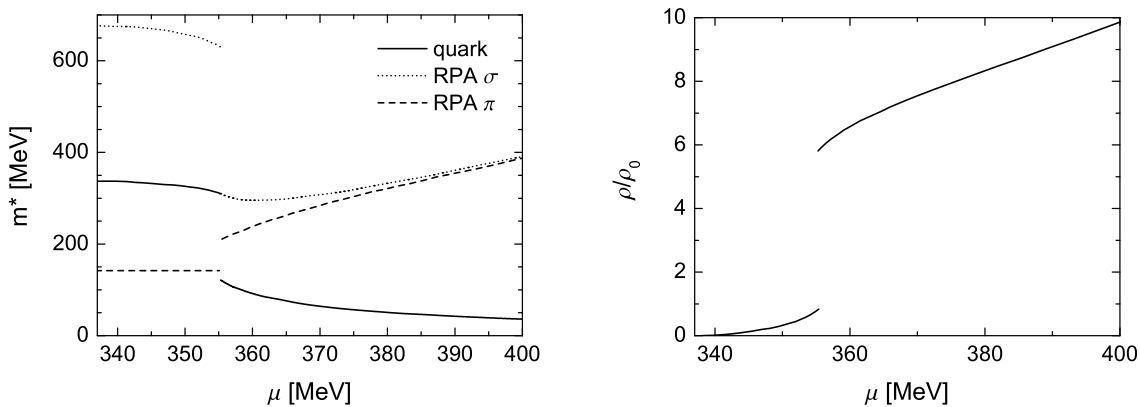}
	}
	\caption{\label{fig:hrpa_masses}The effective quark and RPA meson masses (left) and the quark density (right) in the Hartree+RPA approximation, using parameter set I from Table \ref{tab:njlparam}. A first-order chiral phase transition occurs at $\mu=355\MeV$. (Data taken from \protect\cite{Klevansky:1992qe,Asakawa:1989bq}.)}
\end{figure*}

To examine the influence of the short-range correlations on the chiral phase transition, we investigate the effective quark mass $\effmc$ (\ref{eq:eff_mass_const}) as a function of the chemical potential. Before discussing the \order{1/N_c} results, we first take a closer look at the Hartree(\mbox{--}Fock) results in Figs.~\ref{fig:naivemassgap} (solid line) and \ref{fig:hrpa_masses}.

In the shaded region of Fig.~\ref{fig:naivemassgap}, the gap equation (cf.~Section~\ref{sec:1Nc_approach}) has three solutions $\effmc(\mu)$ for a given value of $\mu$. The (inverted) 'S' shape of the curve is characteristic for the order parameter of a first-order phase transition \cite{Plischke:1994}. The upper and the lower solution (chirally broken and restored phase) correspond to minima of the thermodynamical potential \cite{Schwarz:1999dj,Klevansky:1992qe} and thus are stable. The intermediate solution corresponds to a maximum of the potential and is metastable. Note that $\effmc$ remains finite in the chirally restored phase since a small $m_0$ breaks chiral symmetry explicitly.

The energetically favored solution for $\effmc$ corresponds to the lowest minimum of the thermodynamical potential. Depending on $\mu$, this can be either the upper or the lower solution. The phase transition is located where both minima have the same value. At this point, $\effmc$ drops discontinuously while $\rho$, cf.~(\ref{eq:qp_density}), increases. The intermediate range of (metastable) masses and densities is not realized. The quark density $\rho=3\rho_0$ -- corresponding to nuclear matter at normal density -- lies within this gap in Fig.~\ref{fig:hrpa_masses}. This is a known shortcoming of the NJL model on the quasiparticle level \cite{Klevansky:1992qe}.

\begin{figure*}
	\centerline{
		\includegraphics[scale=1]{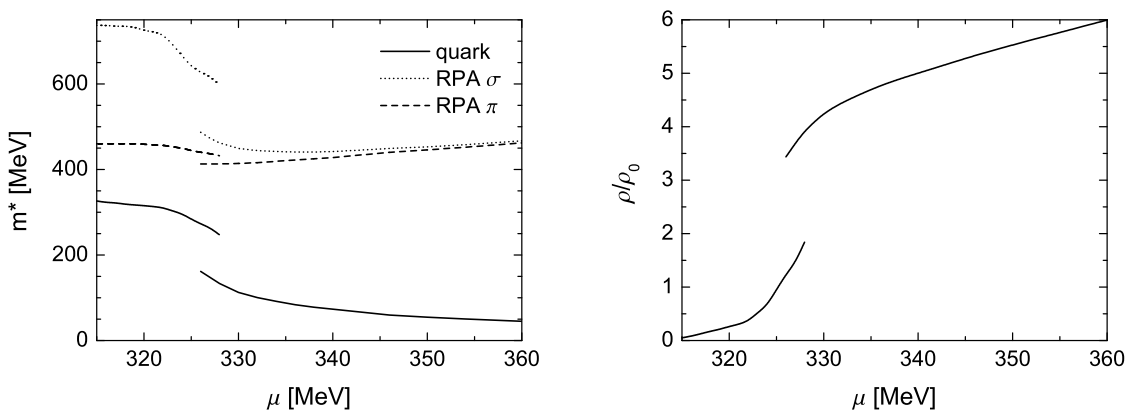}
	}
	\caption{\label{fig:qmass_c1}The effective quark and RPA meson masses (left) and the quark density (right) in the \order{1/N_c} approach, using parameter set I. In the range $326\MeV<\mu<328\MeV$ a first-order phase transition occurs.}
\end{figure*}

Fig.~\ref{fig:qmass_c1} shows the effective quark mass that is found in the \order{1/N_c} calculation, using parameter set I.
In contrast to the simple estimate in Fig.~\ref{fig:naivemassgap} (dashed line) that ignores the energy and density dependence of the short-range correlations, the phase transition is still of first order. The 'S' shape in Fig.~\ref{fig:qmass_c1} is interrupted since the iterative calculations will not converge for the metastable solutions in the vicinity of the phase transition.

The short-range correlations are strongly density dependent since they are generated by interactions with the medium. Only when $\mu$ -- and thus $\rho$ -- increases, deviations from the mean field models can be observed in Fig.~\ref{fig:qmass_c1}. The right panel of Fig.~\ref{fig:qmass_c1} shows the strong $\mu$ dependence  of the quark density. Recall the quasiparticle relation $\rho\sim(\mu^2-{\effmc}^2)^{3/2}$, cf.~(\ref{eq:qp_density}). The explicit cubic $\mu$ dependence of $\rho$ is enhanced where $\effmc(\mu)$ changes rapidly, i.e., near the phase transition.

The region of chemical potentials for which several solutions for $\effmc$ exist shrinks from $4\MeV$ (Hartree) to only $2\MeV$ in the \order{1/N_c} approach (we will not determine the exact location of the phase transition from the thermodynamical potential). For the gaps between the upper and the lower solutions for $\effmc$ and $\rho$ we find $\Delta\effmc=111-116\MeV$ and $\Delta\rho=2.1-2.2\,\rho_0$. The Hartree approximation, on the other hand, yields $\Delta\effmc=188\MeV$, $\Delta\rho=4.4\rho_0$ \cite{Klevansky:1992qe}.

We conclude that the short-range correlations are not strong enough to turn the first-order phase transition into a smooth crossover at zero temperature. However, the gaps at the phase transition have decreased by $40\%-50\%$. The quark density of $3\rho_0$ is still located within the gap. Note that the smaller $\effmc$ in the chirally broken phase and the \order{1/N_c} term $\qsef_{\eff,0}$, cf.~(\ref{eq:eff_fock_comp},\ref{eq:qrk_disp}), shift the phase transition to a lower $\mu$.

Figs.~\ref{fig:hrpa_masses} and \ref{fig:qmass_c1} also show the RPA meson masses. The RPA pion of the \order{1/N_c} approach has a considerable mass in the chirally broken phase. The RPA sigma mass is closer to the Hartree+RPA result. At the phase transition, $m_\sigma$ does not drop as far as in the mean field calculation. $m_\pi$ drops instead of increasing, thus moving towards the mean field result. At higher $\mu$, $m_\sigma$ and $m_\pi$ resemble the Hartree+RPA result qualitatively and quantitatively. $m_\sigma$ and $m_\pi$ converge at large $\mu$ (chiral partners) -- restored chiral symmetry is still a good symmetry in the \order{1/N_c} approach.

We do not show the results of a calculation using parameter set 0 explicitly here. In the chiral limit ($m_0=0$), the results for the chirally broken phase -- including $m_\pi$ -- are very similar to those in Fig.~\ref{fig:qmass_c1}. $G$ and $\njlcut$ are almost equal in both parameter sets, $m_0$ is irrelevant when $\effmc$ is large. Since $\qseret_s$, and thus $\effmc$, vanishes above the phase transition in the chiral limit -- chiral symmetry is fully restored -- the 'S' shape of the $\effmc$ curve in  Fig.~\ref{fig:qmass_c1} turns into a '2' shape. Note that the solution $\effmc=0$ of the gap equation exists for all $\mu$ in the chiral limit. At large $\mu$, where $\effmc$ becomes very small in Fig.~\ref{fig:qmass_c1}, the results for the two parameter sets are close to each other again.

\begin{figure}
	\centerline{
		\includegraphics[scale=0.7]{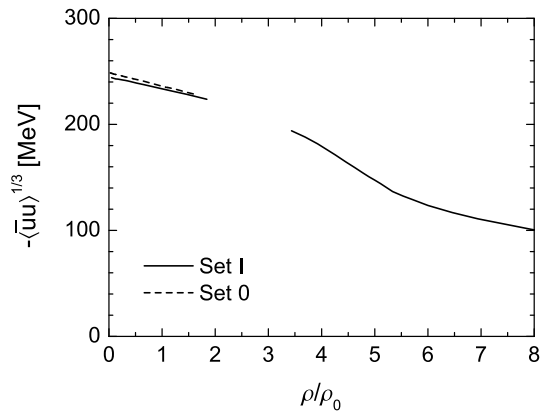}
	}
	\caption{\label{fig:qcond_rho}The quark condensate as a function of the density. The gap in the curves corresponds to the range of densities that is skipped at the phase transition, cf.~Fig.~\ref{fig:qmass_c1}.}
\end{figure}

Fig.~\ref{fig:qcond_rho} shows the quark condensate $\ewert{\bar u u}$. 
Like $\effmc$, $\ewert{\bar u u}$ is large at low densities. The results in the chirally broken phase are almost equal for both parameter sets. When parameter set 0 is used, the quark condensate drops to zero above the phase transition. It drops but remains finite when chiral symmetry is explicitly broken.

\subsection{On-shell self energy\label{sec:os_se}}

For a quantitative investigation of the collisional self energy we will restrict ourselves to the on-shell states (\ref{eq:os_def}). Fig.~\ref{fig:specwidthresig} shows that most of the strength of the spectral function remains in the vicinity of the on-shell peaks. The off-shell structure of the self energy has only limited influence on the properties of the medium.

Our calculations show that the energy and momentum dependence of $\Re\effm_\os$ -- the real part of $\effm^\ret(k)$ (\ref{eq:eff_momentum}) on the mass shell -- is weak. Due to the large constant Hartree self energy, the effective masses at $\vec k=0$ and  $|\veq k|=\njlcut$ differ by only $4-5\%$. The Fock self energy $\qsef_s = \qseh/({2 N_f N_c})$ that is part of the meson exchange provides another constant contribution. Adding up $m_0+\qseh+\qsef_s$, we find that $80-90\%$ of $\Re\effm_\os$ are determined on the mean field level. Only the remaining $10-20\%$ are generated by short-range correlations. Recall that the coupling -- and thus $\qseh$ (\ref{eq:hartree_se}) -- has been lowered by $22\%$ in the  \order{1/N_c} approach. Hence, the collisional self energy just replaces the missing mean field contributions without generating new effects.

\begin{figure*}
	\centerline{
		\includegraphics[scale=1]{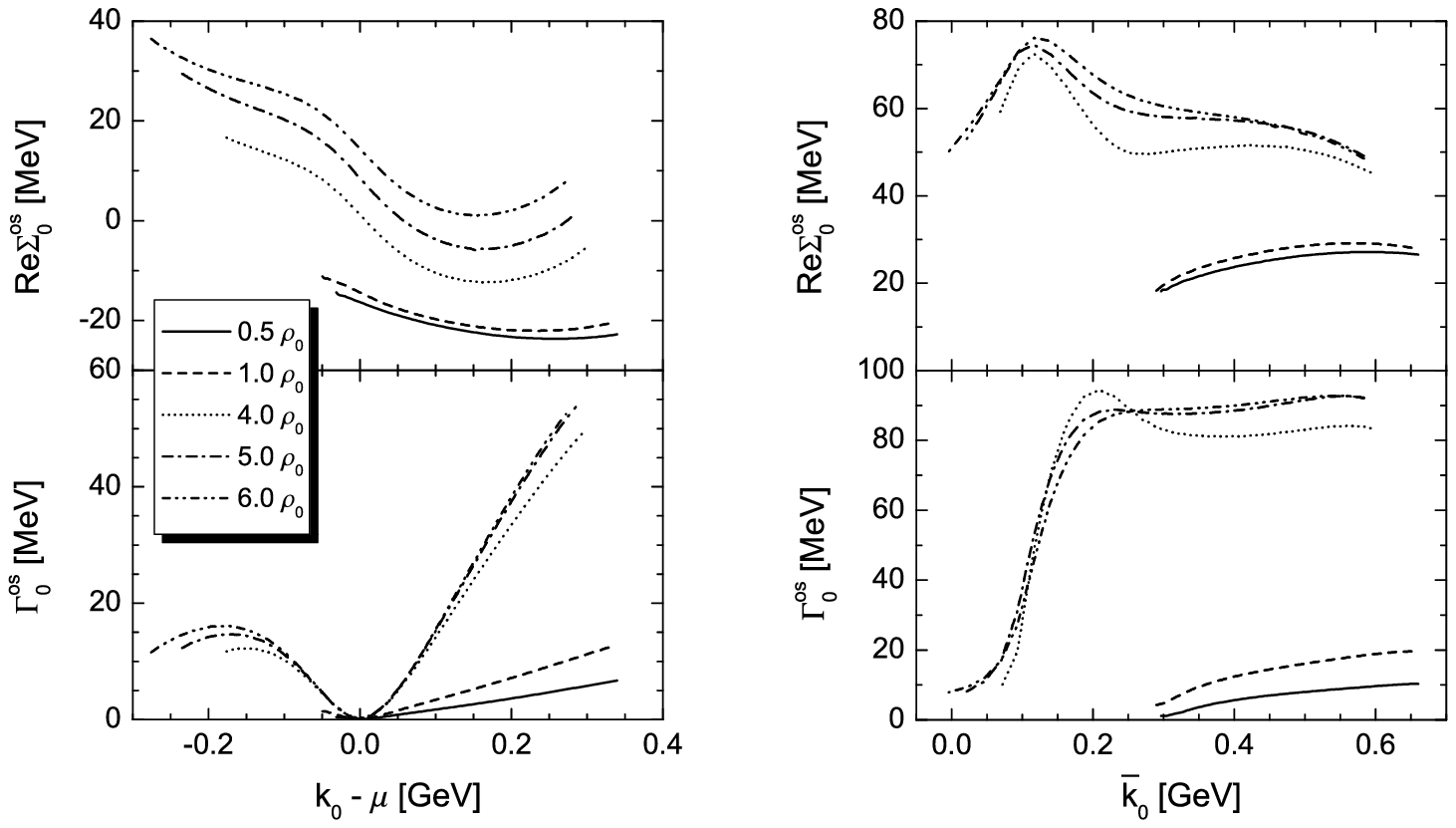}
	}
	\caption{\label{fig:os_qse0}On-shell real parts of $\qseret_0$ and the on-shell width $\qbr^\os_0$ for the quark and antiquark states at five different densities in the chirally broken ($0.5\rho_0,\rho_0$) and restored phase ($4\rho_0,5\rho_0,6\rho_0$), using parameter set I. The curves start at $k^\os_0(\modk=0)$ and end at $k^\os_0 (\modk=\njlcut)$. Note the different scales.}
\end{figure*}

The $\gamma_\mu$-components  of $\Re\qseret_\os$ shift the peaks of the spectral function, cf.~(\ref{eq:denom_prop_real}). We will not discuss $\Re\qseret_{v,\os}$ here since it is small in comparison to the other components and to $\modk$. $\Re\qseret_{0,\os}$ is shown in the upper panels of Fig.~\ref{fig:os_qse0}.
In the chirally broken phase, $\Re\qseret_{0,\os}$ is moderately density and energy dependent. Up to a small shift ($\qsef_{\eff,0}$ is small at low $\mu$, cf.~Appendix~\ref{app:re_qse}) it is approximately antisymmetric in $k_0$. 
It cannot be absorbed into $\mu$ like the constant $\Re\qseret_0=\qsef_0$ of the Hartree--Fock approach.

$\Re\qseret_{0,\os}$ becomes larger in the chirally restored phase. The energy dependence increases and the antisymmetry is lost. The non-dispersive $\qsef_{\eff,0}$ is on the order of $10-25\MeV$. At low energies, $\Re\qseret_{0,\os}$ has the same sign for quarks and antiquarks. Thus, both peaks of the spectral function are shifted to higher energies, see the cuts for small $\modk$ in Fig.~\ref{fig:specwidthcuts}. Note that the density dependence has grown for the quarks but is still moderate for the antiquarks.

The bump at $\bar k_0\approx 120\MeV$ (right panel of Fig.~\ref{fig:os_qse0}) is generated by process \ref{fig:qmscattering}(c). Above the phase transition, this process moves into the antiquark on-shell region. As discussed before, the processes involving bound \qbq states have no influence on the on-shell self energy in the quark sector.

$\Re\qseret_{0,\os}$ clearly modifies the properties of the medium in comparison to the mean field approaches where $\Re\qseret_{0,\os}$ is zero or constant. Shifting the quark peak of $\qsp$ with respect to the (fixed) chemical potential has, e.g., influence on $\effmc$ and $\rho$, cf.~(\ref{eq:hartree_se},\ref{eq:qrk_density}). Since $\Re\qseret_{0,\os}$ increases at large $\rho$, the impact is stronger in the chirally restored phase.

We turn now to the on-shell width. The lower panels of Fig.~\ref{fig:os_qse0} show $\qbr_0^{\os}$ -- the largest Lorentz component of $\qbr_\os$ -- for the quark and antiquark states. Like in the cuts of Figs.~\ref{fig:specwidthresig}-\ref{fig:specwidthcuts}, the width vanishes at $k_0=\mu$. In the vicinity of $\mu$ we find a quadratic energy dependence. Below the phase transition, the populated quark states are located close to the Fermi energy. Their width -- generated by process \ref{fig:qqscattering}(c) -- remains small. Above the Fermi energy and in the antiquark sector, $\qbr_0^{\os}$ shows a linear density dependence -- the relevant processes \ref{fig:qqscattering}(a) and \ref{fig:qqscattering}(e) both involve one quark from the medium.

In the chirally restored phase, i.e. for the densities $\ge 4\rho_0$ in Fig.~\ref{fig:os_qse0}, the region of populated states becomes larger and $\qbr_0^{\os}$ increases up to $10-20\MeV$. It drops at low $k_0$ since the threshold for process \ref{fig:qqscattering}(c) is approached, cf.~Section~\ref{seq:scatt_thresholds}. The widths of the free quark states and of the antiquark states are again larger. The density dependence of $\qbr_0^{\os}$ weakens drastically at higher $\rho$. We can interpret that as a -- Pauli blocking and cutoff induced -- saturation of the short-range effects, cf.~Section~\ref{sec:densitydepend}. We have observed the same effect for nucleons in nuclear matter at \emph{nucleon} densities of a few times $\rho_0$ \cite{Froemel:2003dv}.

For a better understanding of the density dependence of the width, we can investigate the average width of the populated quark states $\qavbrpop$ and of all quark states $\qavbrall$ (excluding antiquarks). We define $\qavbrpop$ as
\begin{align}
	\qavbrpop=\frac{\int_0^\njlcut dk k^2\int_{\Re\effk_0=0}^\infty dk_0 \qbr_0 (k) \qsp_0(k)\nf(k_0)}
											{\int_0^\njlcut dk k^2\int_{\Re\effk_0=0}^\infty dk_0 \qsp_0(k)\nf(k_0)} \,, \label{eq:avg_width}
\end{align}
using again the three-momentum cutoff $\njlcut$. Note that the denominator is just the density (\ref{eq:qrk_density}).  $\qavbrall$ is found by removing the distribution functions in (\ref{eq:avg_width}). Since we weight $\qbr_0$ with the spectral function, the on-shell width will dominate the average widths.

\begin{figure*}
	\centerline{
		\includegraphics[scale=1]{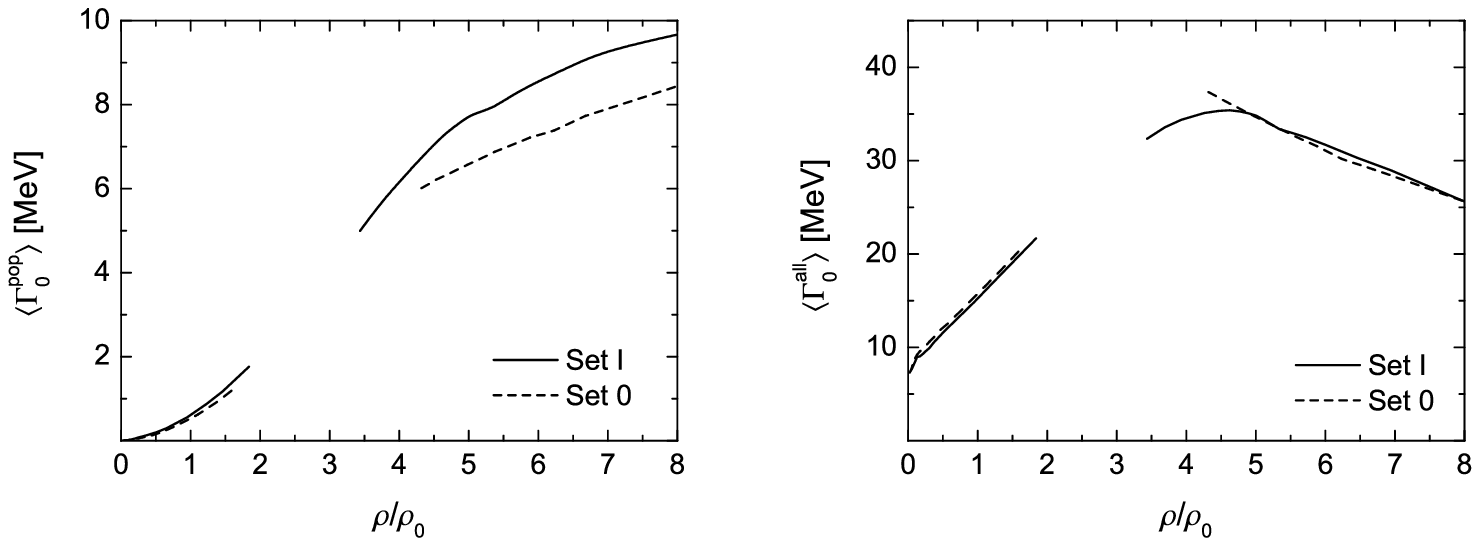}
	}
	\caption{\label{fig:avg_width}The average widths  $\qavbrpop$ and $\qavbrall$. The gap in the curves corresponds to the range of densities that is skipped at the chiral phase transition, cf.~Fig.~\ref{fig:qmass_c1}.}
\end{figure*}

Fig.~\ref{fig:avg_width} shows $\qavbrpop$ and $\qavbrall$  from calculations with both parameter sets.
$\qavbrpop$  remains smaller than $2\MeV$ in the chirally restored phase. Above the phase transition, we can observe the expected saturation effect -- $\qavbrpop$ does not exceed $10\MeV$. $\qavbrall$ grows up to $20\MeV$ below the phase transition. In the chirally restored phase, however, it does not only saturate but starts to decrease. Note that the structure of the curves in Fig.~\ref{fig:avg_width} can be readily understood from the considerations in Section~\ref{ch:qq_scattering}. We will not discuss it here in more detail.

The results for the average widths appear rather small in comparison to Figs.~\ref{fig:specwidthcuts_ld}, \ref{fig:specwidthcuts}, and \ref{fig:os_qse0}. They are, however, comparable to the nuclear matter results in \cite{Froemel:2003dv}. At $T=0$ and $\rho_\textrm{nm}=\rho_0$, the average width of the populated nucleon states is $1.5\MeV$. At $3\rho_0$, a width of $6\MeV$ is found. Of course, nucleon and quark widths should not be compared directly. Nonetheless, the quantitative agreement indicates that the short-range correlations are of similar importance in both systems.

In Section~\ref{sec:phasetrans} we have found that the chiral phase transition does not turn into a smooth crossover like in the naive calculation of Fig.~\ref{fig:naivemassgap}. In the extended Hartree approach, we have used a constant width of $24\MeV$. Below the phase transition, the realistic value for $\qavbrpop$ is one order of magnitude smaller. Hence, it is not surprising that we still find a first-order phase transition. Recall that we consider a conservative estimate of the short-range effects in the chirally broken phase here, cf.\ Section~\ref{sec:os_processes}. A calculation with a more realistic pion mass should yield an even smaller mass gap or turn the phase transition into a crossover. We note that the nucleon width is strongly temperature dependent \cite{Froemel:2003dv}. A similar behavior can be expected for the quark width. This should lead to a stronger influence of the short-range correlations on the phase transition at finite $T$. Such a study is, however, beyond the scope of the present work.

\subsection{Momentum distribution\label{sec:momdist}}

To estimate the importance of the off-shell states further away from the on-shell peaks, cf.~Figs.~\ref{fig:specwidthresig}-\ref{fig:specwidthcuts}, we investigate the momentum distribution of the quarks in the medium. The momentum distribution is closely related to the density (\ref{eq:qrk_density}). It is given by
\begin{align}
	n(|\veq k|)=\frac 1\pi \int_{\Re\effk_0=0}^\infty dk_0 \qsp_0 (k)\nf(k_0)  \,.
\end{align}
The factor $1/\pi$ has been chosen to normalize the momentum distribution, cf.~(\ref{eq:norm_qsp2}).

In a quasiparticle approach (\ref{eq:mf_spectral}), we would find the momentum distribution of a free Fermi gas that turns into a step function $\Theta(k_F-|\veq k|)$ at $T=0$. The finite quark width modifies this simple picture. The off-shell states shift some strength away from the peaks in $\qsp_0$. For $\modk<k_F$ this leads to a depletion of the momentum distribution since states above the chemical potential are not populated. For $\modk>k_F$, the off-shell states shift some strength back below the chemical potential. Those populated states yield -- even at $T=0$ -- a so-called high-momentum tail in the momentum distribution. The depletion and the high-momentum tail can be interpreted as a universal measure for the short-range correlations.

\begin{figure*}
	\centerline{
		\includegraphics[scale=1]{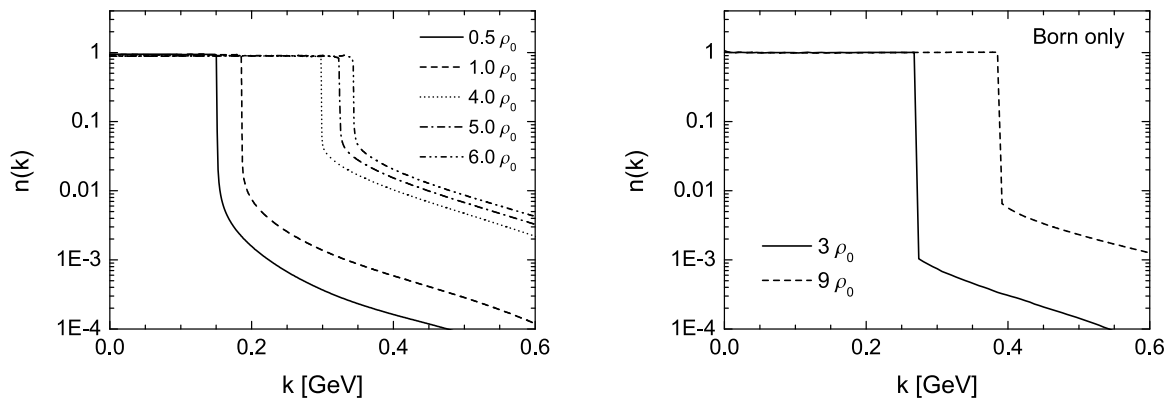}
	}
	\caption{\label{fig:momdist}The momentum distribution of quarks in quark matter. The left panel shows \order{1/N_c} results, using parameter set I from Table~\ref{tab:fullnjlparam}. The solid and the dashed line correspond to quark matter in the chirally broken phase, the other lines to quark matter in the chirally restored phase. The right panel shows results from \protect\cite{Froemel:2001iy}, using parameter set 0 from Table~\ref{tab:njlparam} and restoring chiral symmetry by hand.}
\end{figure*}

The left panel of Fig.~\ref{fig:momdist} shows the momentum distribution at several densities.
As expected for an infinite system at $T=0$, we find a sharp step at $k_F$ \cite{Dickhoff:2005mb}. Recall that the width in the range $0<k_0<\mu$ is generated by process \ref{fig:qqscattering}(c). The density dependence of this process saturates above the phase transition, cf.~Fig.~\ref{fig:avg_width} (left panel). This behavior is reflected in $n(|\veq k|)$. While the high-momentum tails at the lower densities differ significantly in size, they are very close to each other at the higher densities.

At low $\modk$, the momentum distribution has values of $0.94$ below and $0.89$ above the chiral phase transition. The discontinuity at $k_F$ has a size of $0.82-0.85$ for the lower and of $0.72-0.75$ for the higher densities. These results can be readily compared to the well established nuclear matter values, see \cite{Dickhoff:2005mb} for a compilation. At $T=0$ and normal nuclear matter density, the depletion of the nucleon momentum distribution is on the order of $15-18\%$. The discontinuity at $k_F$ has a size of $0.7-0.75$.

We can conclude that the short-range effects in quark matter in the chirally restored phase are of similar size as in normal nuclear matter. In the chirally broken phase, where the quarks reach the same particle number density $\rho_0$ as the nucleons at normal density, the correlations are slightly smaller.

It is also interesting to compare the present results to our previous calculations. The right panel of Fig.~\ref{fig:momdist} shows the momentum distributions from \cite{Froemel:2001iy} at two different densities ($\effmc=0$). The high-momentum tail at $\rho=3\rho_0$ is more than one order of magnitude smaller than the $4\rho_0$ result of the \order{1/N_c} approach. A density of $9\rho_0$ is needed to generate short-range effects on the same level as in the \order{1/N_c} approach at $0.5\rho_0-1\rho_0$. This confirms again, cf.~Section~\ref{sec:res_width_scatt}, that the loop-expansion has missed significant contributions to the short-range correlations.

\section{Summary and Outlook\label{ch:outlook}}

In this work we have explored the content of short-range correlations in cold and dense quark matter within the framework of the SU(2) NJL model. Our approach is based on techniques that haven proven to be very successful in nuclear matter. Employing a next-to-leading order expansion in $1/N_c$ for the quark self energy, we have constructed a fully self-consistent model. By summing up all \order{1/N_c} self energy diagrams, we have dynamically generated RPA mesons. Their properties are -- in contrast to the Hartree+RPA approach -- self-consistently fed back into the quark self energy. To preserve analyticity, the reals parts of the quark self energy and the RPA polarizations have been calculated from dispersion integrals -- non-dispersive terms were identified explicitly.

The RPA pions are not the Goldstone modes of our approach. Presently, a next-to-leading order approach to the NJL model that is fully self-consistent \emph{and} yields Goldstone pions does not exist. An analysis of the structure of the quark width has shown that the too large pion mass does not influence which processes contribute to the on-shell width. However, the spacelike off-shell components of the RPA pion propagator, that are important for the generation of the on-shell quark width, depend on the RPA pion mass. Thus, our results for the chirally broken phase represent a conservative estimate of the short-range effects.

In calculations at finite chemical potentials and zero temperature, we have investigated the (off-shell) structure of the quark width, the size and density dependence of the short-range effects, and their influence on the properties of the medium. The short-range correlations are not strong enough to turn the first-order phase transition into a smooth crossover at zero temperature. Nonetheless, the mass and density gaps at the chiral phase transition have dropped by $40-50\%$. Above the phase transition, the correlations quickly saturate due to Pauli blocking and the NJL cutoff.

We have found quark widths that are similar in shape but one order of magnitude larger than in the loop-expansion of \cite{Froemel:2001iy}. The larger results of the present work arise from a consistent expansion of the quark self energy in $1/N_c$. The comparison of the quark matter results to nuclear matter calculations shows that the short-range correlations are of similar magnitude in both systems. A complete description of nuclear systems is only possible when the short-range effects are taken into account. Therefore, the influence of short-range correlations on the quarks (and the RPA mesons) should also not be ignored.

Let us consider the QCD phase diagram to stress this point: Phenomenological models are used to determine the properties of the chiral phase transition. Many models suggest that a first-order phase transition turns into a crossover at a critical point. As we have seen in this work, short-range correlations change the properties of the phase transition with respect to quasiparticle calculations. Nuclear matter calculations indicate that the short-range correlations will be even larger at finite temperatures \cite{Froemel:2003dv}. Hence, the mean field models will most likely predict a wrong position for the critical point.

The present \order{1/N_c} approach can be improved in many ways. The most interesting but also most complicated extension would be to fix the chiral properties of the RPA pions. A first step into that direction would be to include the next-to-leading order corrections to the RPA polarizations -- or to generate higher order vertex corrections via Bethe--Salpeter equations -- in the self-consistent calculation. Such an approach would allow us to take a closer look at pions with more reasonable masses and at the chiral theorems in the presence of short-range correlations. Our investigations from Chapter~\ref{ch:qq_scattering} indicate that the on-shell width of the quarks could also be larger than determined here.

Most likely, the short-range effects in quark matter will -- like those in nuclear matter -- increase significantly at higher temperature. This should be checked in an explicit calculation. We have already discussed the possible consequences for the structure of the chiral phase transition and the critical point.

We have ignored the phenomenon of color superconductivity \cite{Rajagopal:2000wf,Buballa:2003qv} in the present approach. Diquark states can be dynamically generated in the same fashion as the RPA mesons. A diquark condensate may have influence on the short-range effects in the chirally restored phase \cite{Carter:1998ji}. The short-range effects will also have some influence on the diquark condensate. For the investigation of systems like neutron stars and hypernuclear matter, it would also be interesting to extend our model to asymmetric quark matter and flavor SU(3). Similar approaches exist already for nuclear matter.

\section*{Acknowledgments}

We would like to thank Ulrich Mosel for stimulating discussions, for reading the manuscript, and for continuous support throughout the preparation of this work. Parts of the calculations were done at the Center for Scientific Computing (CSC) in Frankfurt. The package CUBPACK by R. Cools and A. Haegemans \cite{Cools:2003} was used for the numerical calculation of multidimensional integrals. This work was supported by BMBF.


\appendix

\section{RPA on the Hartree(--Fock) level\label{app:hfrpa}}

For the present work it is instructive to review certain aspects of the quasiparticle RPA. An investigation at $T=\mu=0$ (chirally broken phase) provides insights about the Goldstone boson character of the RPA pions and the proper use of dispersion integrals.

It has been shown in \cite{Klevansky:1992qe} that the time-ordered RPA pion polarization (\ref{eq:msecu_start}) of the Hartree(--Fock)+RPA approach at $T=\mu=0$ has the form
\begin{equation}
	\label{eq:mf_pipol}
	- i \msec_\pi (k^2) = -4N_f N_c \int\dvp \frac 1{p^2-\effmc^2}+2N_f N_c k^2 \mathrm{I}(k^2) 
\end{equation}
where $\effmc=m_0+\qseh(+\qsef_s)$ and
\begin{align}
	\mathrm{I}(k^2)= \int\dvp \frac 1{[(p+{\textstyle\frac 12k})^2-\effmc^2][(p-{\textstyle\frac 12k})^2-\effmc^2]} \,.
	\label{eq:int_i}
\end{align}
Note that the \emph{imaginary} part of $\mathrm{I}$ corresponds to the \emph{real} part of $\mse^\chron$. The time-ordered quark propagator is given by $\qpclo^{-1}=\slashed{k}-\effmc$. Using Eq.~(\ref{eq:hartree_se}),
\begin{align}
	 - 4 N_f N_c \int\dvp \frac 1{p^2-\effmc^2} = \frac{i}{2\njlkpl}\frac{\qseh}{\effmc} \,, \label{eq:hartshift}
\end{align}
we can identify the denominator of the RPA pion propagator $\mpc_\pi$ (\ref{eq:mesprop}) with
\begin{align}
	\label{eq:hfproppi_denom}
	1+ 2\njlkpl\msec_\pi (k^2) &=   \frac{\effmc - \qseh}\effmc +4 i \njlkpl N_f N_c k^2 \mathrm{I}(k^2) \,.
\end{align}

If the pions are Goldstone modes, their mass should vanish in the chiral limit (and be small for finite $m_0$) when chiral symmetry is spontaneously broken. $\mpc_\pi$ should then have a pole at $k^2=0$. The second term on the rhs.~of (\ref{eq:hfproppi_denom}) is surely zero for $k^2=0$. For the first term we find in the Hartree approximation, $\effmc=m_0+\qseh$,
\begin{align}
	\frac{\effmc - \qseh}\effmc = \frac{m_0}{m_0+\qseh} \,. \label{eq:dshift_m0_hartree}
\end{align}
This expression is zero in the chiral limit. For finite $m_0$ we have $m_0/\qseh \approx 0.016$. Hence, the Hartree+RPA pions behave like Goldstone bosons.

The Hartree--Fock approximation is not a clean expansion in $1/N_c$. However, we can use it to estimate the next-to-leading order effects. Using $\effmc=m_0+\qseh+\qsef_s$ and $\qseh=2N_f N_c \qsef_s$ \cite{Klevansky:1992qe}, the first term of (\ref{eq:hfproppi_denom}) becomes
\begin{align}
	\frac{\effmc - \qseh}\effmc 
		\approx \frac{\qsef_s}{\qseh + \qsef_s} = \frac 1 {2 N_f N_c + 1} \approx 0.08. \label{eq:fock_shift}
\end{align}
This term is now significantly larger than $m_0/\qseh$. Thus, the Hartree--Fock+RPA pions will not behave like Goldstone bosons -- even in the chiral limit they will have a finite mass. Fig.~\ref{fig:1_re_Pi_pi} shows $1+2\njlkpl\Re\mseret_l$ from a Hartree+RPA and a Hartree--Fock+RPA calculation. The zeros determine the pole of the propagator. We can see that even small shifts have significant influence on the pion mass. The shift of $0.08$ raises $m_\pi$ by  $150\MeV$. The \order{1/N_c} approach includes the Fock self energy. If no cancellation with other \order{1/N_c} terms occurs, the \order{1/N_c} RPA pions will behave even less like Goldstone bosons.

The denominator of $\mpc_\sigma$ is found in the same way as (\ref{eq:hfproppi_denom}), cf.~\cite{Klevansky:1992qe},
\begin{align}		
	\label{eq:hfpropsigma_denom}
	1+ 2\njlkpl\msec_\sigma (k^2) &=   
		\frac{\effmc - \qseh}\effmc +4 i \njlkpl N_f N_c \left(k^2-4\effmc^2\right) \mathrm{I}(k^2) \,.
\end{align}
The first term is of minor importance here. As long as it becomes not too large, the poles of $\mpret_\sigma$ are determined approximately by the zero of the second term, i.e., $m_\sigma\approx 2\effmc$.

We turn now to the second term on the rhs.~of (\ref{eq:mf_pipol}). The real and the imaginary part of $\mathrm{I}(k^2)$ can be calculated explicitly using either the residue theorem or Cutkosky rules \cite{Peskin:1995} when a three-momentum cutoff is used to regularize the integral. Such a calculation shows that $\Re\mathrm{I}(k^2)\sgn(k_0)$ -- the main contribution to $\Im\mseret_{\sigma,\pi}$, cf.~(\ref{eq:sec_seret},\ref{eq:mf_pipol}) -- and $\Im\mathrm{I}(k^2)$ are related by a dispersion integral without additional constant terms,
\begin{align}
	\Im \mathrm{I}^{\ret} (k_0,\veq k)=
	-\frac 1\pi \mathcal{P} \int_0^\infty dq_0^2 \frac{\Re \mathrm{I}^\ret(q_0,\veq k)}{q_0^2-k_0^2} \,, \label{eq:imi_exact}
\end{align}
where $\Re \mathrm{I}^\ret(k_0,\veq k)=\Re \mathrm{I}(k^2)\sgn(k_0)$ and $\Im\mathrm{I}^\ret(k)=\Im\mathrm{I}(k^2)$, cf.~(\ref{eq:sec_seret}).

When $\Re \mseret_{\sigma,\pi}(k^2)$ shall be calculated from a dispersion integral, it must be taken into account that the $k_0$ dependence of $\Im \mseret_{\sigma,\pi}$ is not given by $\Re \mathrm{I}^\ret$ alone but by $(k^2-4\effmc^2) \Re \mathrm{I}^\ret(k_0,\vec k^2)$ and $k^2 \Re \mathrm{I}^\ret(k_0,\vec k^2)$, respectively. Calculating the dispersion integral over $\Re \mathrm{I}^\ret$ and multiplying the result with a factor $k^2$ -- i.e., inserting the dispersive result for $\Im \mathrm{I}$ into $\Re \mseret_{\sigma,\pi}$ -- is not equivalent to calculating a dispersion integral over $k^2 \Re \mathrm{I}^\ret \sim \Im \mseret_{\sigma,\pi}$. The difference is a $k_0$ independent term,
\begin{align}
	k^2 \Im \mathrm{I} (k_0,\veq k)&=
	-\frac 1\pi k^2 \mathcal{P} \int_0^\infty dq_0^2 \frac{\Re \mathrm{I}^\ret(q_0,\veq k)}{q_0^2-k_0^2} \label{eq:disp_shift} \\
	&= \frac 1\pi \int_0^\infty dq_0^2 \Re \mathrm{I}^\ret(q_0,\veq k)
			-\frac 1\pi \mathcal{P} \int_0^\infty dq_0^2 \frac{(q_0^2-\vec k^2)\Re \mathrm{I}^\ret(q_0,\veq k)}{q_0^2-k_0^2} \notag\\
	&= \underbrace{\frac 1\pi \int_0^\infty dq_0^2 \Re \mathrm{I}^\ret(q_0,\veq k)}_{\mbox{const.~in $k_0$}}
			-\frac 1{2 N_f N_c \pi} \mathcal{P} \int_0^\infty dq_0^2 \frac{\Im \mseret_\pi (q_0,\veq k)}{q_0^2-k_0^2} \,, \notag
\end{align}
and likewise for the $\sigma$ case.

Using the dispersion integral over $\Re \mathrm{I}^\ret$ and multiplying with $k^2$ afterwards, is equivalent to the direct calculation of $\Re \mseret_{\sigma,\pi}$.  Hence, the $k_0$ independent term in the last line of (\ref{eq:disp_shift}) must be calculated explicitly when $\Re \mseret_{\sigma,\pi}$ is calculated from a dispersion integral over $\Im \mseret_{\sigma,\pi}$,
\begin{align}
	\Re\mseret_{\sigma,\pi}(k) 
		= \frac 1\pi \mathcal{P} \int_0^\infty dq_0^2 \frac{\Im \mseret_{\sigma,\pi} (q_0,\veq k)}{q_0^2-k_0^2} 
	 - \frac{2 N_f N_c}\pi \int_0^\infty dq_0^2 \Re \mathrm{I}^\ret(q_0,\veq k)
				-\frac{\qseh}{2\njlkpl \effmc} \,.  \label{eq:qp_remseret_disp}
\end{align}
The first term on the rhs.~is the regular dispersion integral. The third term is the constant shift from (\ref{eq:hartshift}). A decomposition similar to (\ref{eq:qp_remseret_disp}) is found for the \order{1/N_c} approach, cf.~Appendix~\ref{app:re_mespol}. Note that $\mse_{\mathrm{n}}$ in (\ref{eq:remseret_full}) corresponds to the third term of (\ref{eq:qp_remseret_disp}) and $\mseret_{\mathrm{d}}$ in (\ref{eq:remseret_full}) corresponds to the sum of the first and the second term of (\ref{eq:qp_remseret_disp}).

\section{Real parts of the RPA meson polarizations\label{app:re_mespol}}

In the following we will examine the complex polarizations to identify the non-dispersive part of $\Re\mseret_l$. Likewise, we will investigate the quark self energy in Appendix~\ref{app:re_qse}. The Feynman rules of the real-time formalism provide no direct access to the retarded self energy and polarizations \cite{Danielewicz:1982kk,Das:1997}. Hence, we will start from the time-ordered $\qsec$ and $\msec_l$. They are related to $\qseret$ and $\mseret_l$ in a simple way \cite{Das:1997}:
\begin{equation}
	\label{eq:sec_seret}
	\begin{aligned}
		\Re \qsec(k) &= \Re\qseret(k) \,, 			\\
		\Re\msec_l(k) &= \Re\mseret_l (k) \,,
	\end{aligned}
	\qquad
	\begin{aligned}
		\Im \qsec(k) &= [1-2\nf(k_0)]\Im\qseret(k) \,,    \\
		\Im\msec_l(k) &= [1+2\nb(k_0)] \Im\mseret_l(k) \,, 	
	\end{aligned}
\end{equation}
with the zero temperature limits $1-2\nf(k_0) \rightarrow \sgn(k_0-\mu)$ and $1+2\nb(k_0) \rightarrow \sgn(k_0)$. The imaginary parts of $\qseret$ and $\mseret_l$ that are found in this approach are of course equivalent to the widths that we have found earlier (\ref{eq:qwidth_sp},\ref{eq:mbr_sigma_pi}).

\subsection{Decomposition of the RPA polarizations}

In this section, we will frequently use the relations between the different kinds of propagators ($\qpc,\qpret,\qpgk$) from Section~\ref{sec:qprop} without further reference. For the rather technical manipulations we introduce the shorthand notation $p_\pm = p \pm {\textstyle\frac 12} k$ and, cf.~(\ref{eq:eff_momentum}),
\begin{align*}
	\qpr^{\chron,\gtrless,\ret,\av}_\pm &= \qpr^{\chron,\gtrless,\ret,\av}(p_\pm)  \,,
			& \effp^{\ret,\av}_{\pm,\mu} &= p^\pm_\mu-\qse^{\ret,\av}_\mu(p_\pm)  \,, \\	
	 \nfpm &= \nf(p^\pm_0)  \,,
			& \effm^{\ret,\av}_\pm &= m_0+\qse^{\ret,\av}_s(p_\pm) \,.
\end{align*}

The time-ordered polarizations of the RPA mesons are given by
\begin{align}
	-i\msec_l (k) = -\int \dvp
				\Trace \left[ \Gamlt  \qpc_+ \Gaml \qpc_- \right] \,.  \label{eq:msecu_start}
\end{align}
We can replace the time-ordered propagators by $\qpc(k)=\qpret(k)+\qpkl(k)$, 
\begin{align}
	\msec_l (k) = -i\int \dvp 
				 \Trace &\left[ \Gamlt \qpret_+ \Gaml \qpret_- +\Gamlt \qpret_+ \Gaml \qpkl_- 
				   +\Gamlt \qpkl_+ \Gaml \qpret_- +\Gamlt \qpkl_+ \Gaml \qpkl_-	\right] \,.  \label{eq:msec_retgk}
\end{align}
The first term of the integrand is zero, cf.~(\ref{eq:sret_ints}). Using $\qpret(k)=\Re\qpret(k)-\frac i2 \qsp(k)$ and $\qpkl(k)=i\qsp(k)\nf(k_0)$, the polarizations can be split up into their real and imaginary parts. With the help of (\ref{eq:sec_seret}) and the relation $\nf(p_+)+\nf(p_-)-2\nf(p_+)\nf(p_-)=[\nf(p_-)-\nf(p_+)][1+2\nb(k)]$ we find
\begin{align}
	\mseret_l (k) = -i \int \dvp
		\Trace \left[ \Gamlt \qpret_+\Gaml \qpkl_- +\Gamlt \qpkl_+ \Gaml \qpav_- \right]	\,. \label{eq:re_mseret}
\end{align}

Replacing the non-ordered propagators in (\ref{eq:msec_retgk},\ref{eq:re_mseret}) by $\qpkl_\pm = [\qpav_\pm-\qpret_\pm]\nfpm$ yields expressions for $\mse^{\chron,\ret}$ that contain only retarded and advanced propagators,
\begin{align}
	\mseret_l (k) = -i \int \dvp
			 \Trace  &\left[ 
				 -\Gamlt \qpret_+ \Gaml \qpret_- \nfm +\Gamlt \qpav_+\Gaml \qpav_- \nfp \right.    \label{eq:mseret_retav_sh} \\
			 &\quad \left. + \Gamlt \qpret_+\Gaml \qpav_-\left\{\nfm - \nfp\right\} \right] \,,  \notag \\
	\msec_l (k) = -i\int \dvp  \Trace & \left[ \Gamlt \qpret_+ \Gaml \qpret_- (1-\nfp)(1-\nfm) 
		+\Gamlt \qpav_+ \Gaml \qpav_- \nfp\nfm \right.  \label{eq:msec_retav_sh} \\
		&\quad \left. +\Gamlt \qpret_+ \Gaml \qpav_-	(1-\nfp)\nfm 
		+ \Gamlt \qpav_+ \Gaml \qpret_- \nfp(1-\nfm)\right] \,.  \notag
\end{align}
These expressions are a good starting point for our search for the non-dispersive part of $\Re\mseret_l(k)$. The structure of $\qpr^{\ret,\av}$ resembles -- even in an interacting medium -- that of free propagators, cf.~(\ref{eq:explqpret}). Thus, an investigation similar to the Hartree+RPA approach at $T=\mu=0$, see Appendix~\ref{app:hfrpa} and \cite{Klevansky:1992qe} for details, is possible.

Instead of inserting the explicit forms of $\qpr^{\ret,\av}$ (\ref{eq:explqpret}) into (\ref{eq:mseret_retav_sh}) or (\ref{eq:msec_retav_sh}) right away, we will investigate the general expression $\Trace\Gamlt\qpr_+\Gaml\qpr_-$ first. In this trace, the indices $\pm$ do not only mark the four-momentum but implicitly also the kind ($\ret$, $\av$) of the propagators and the self energies contained in them. Later we will derive expressions for specific combinations of retarded and advanced propagators from this general form. Inserting (\ref{eq:explqpret}) into the general trace, we get
\begin{align}
	\Trace \Gamnjlt_{\sigma,\pi} \qpr_+\Gamnjl_{\sigma,\pi} \qpr_- 
		= 4N_f N_c 
			\frac{\effp^+_\mu \effp_-^\mu \pm \effm_+\effm_-}{(\effp^2_+ - \effm^2_+)(\effp^2_- - \effm^2_-)} \,.	\label{eq:trace_ss}
\end{align}
The upper sign refers to the $\sigma$ and the lower sign to the $\pi$ case, respectively. $\Gamnjl_\pi$ is used as a shorthand notation for $\Gamnjl_{0,\pm}$ (isospin symmetric matter). Note that the numerator would take the simpler form $p^2-\frac 14 k^2\pm \effmc^2$ in the Hartree approximation where $\qse^{\ret,\av}_s(k)=\qsemf$ and $\qse^{\ret,\av}_{0,v}=0$.

We can rewrite the denominator of (\ref{eq:trace_ss}) in terms of partial fractions,
\begin{align}
	&\Trace \Gamnjlt_{\sigma,\pi} \qpr_+\Gamnjl_{\sigma,\pi} \qpr_-    \notag \\
	&\quad = 2 N_f N_c \left[\frac{1}{\effp^2_+ - \effm^2_+} +\frac{1}{\effp^2_- - \effm^2_-}\right]
	- 2N_f N_c \frac{(\effp_+ - \effp_-)^2 -(\effm_+ \pm \effm_-)^2}{(\effp^2_+ - \effm^2_+)(\effp^2_- - \effm^2_-)} \,.
	\label{eq:trace_ss2}
\end{align}
The first term on the rhs.~corresponds to a constant contribution to $\Re\mseret_l$. It can generate a pion mass and may spoil the Goldstone boson character of the RPA pions. The second term is complex. Its real part can in principle be calculated from a dispersion integral. $k_0$ independent shifts that might arise from this term will be discussed later.

Inserting the first term on the rhs.~of (\ref{eq:trace_ss2}) into the retarded polarizations (\ref{eq:mseret_retav_sh}) yields for the non-dispersive, constant part of the meson polarizations
\begin{align}
	\mse_{\mathrm{n}}
 	=- 4 i N_f N_c \int\dvp 
 			(-2i) \Im\frac1{\effp^2_\ret-\effm^2_\ret}\nf(p_0) \,. \label{eq:mse_nondisp}
\end{align}
To obtain this expression, we have used ${\effp_\ret}^*=\effp_\av$,  ${\effm_\ret}^*=\effm_\av$ and have performed the substitutions $p \rightarrow p\pm \frac12 k$. The rhs.~of (\ref{eq:mse_nondisp}) resembles the Hartree self energy (\ref{eq:hartree_se}). Introducing the constant term $\qse^s_\const$ we find
\begin{align}
	\qsemf 
	= 16 \njlkpl N_f N_c (m_0+\qse^s_\const) &\int\dvp \Im \frac 1{\effp^2_\ret-\effm^2_\ret} \nf(p_0) \notag \\
		 +16 \njlkpl N_f N_c &\int\dvp \Im \frac{\qseret_s(p)-\qse^s_\const}{\effp^2_\ret-\effm^2_\ret} \nf(p_0) \,.
	\label{eq:qsemf_split}
\end{align}
The value of $\qse^s_\const$ can in principle be freely choosen to shift strength between the two terms. One choice is to collect all constant contributions to $\qse^{\chron,\ret}_s$ in $\qse^s_\const$. Inserting (\ref{eq:qsemf_split}) into (\ref{eq:mse_nondisp}) yields
\begin{align}
	\mse_{\mathrm{n}} &=	 - \frac 1{2 \njlkpl }\frac{\qsemf}{m_0+\qse^s_\const} 
	 +\frac{8 N_f N_c}{m_0+\qse^s_\const} 
			\int\dvp \Im \frac{\qseret_s(p)-\qse^s_\const}{\effp^2_\ret-\effm^2_\ret} \nf(p_0) \,. \label{eq:mse_nondisp2}
\end{align}
The first term on the rhs.~corresponds to the result of the Hartree(\mbox{--}Fock) approximation (\ref{eq:hfproppi_denom}) when $\qse^s_\const$ is set to $\qseh(+\qsef)$. The second term results from using full propagators. In the Hartree(\mbox{--}Fock) approximation, where $\qseret_s(k)=\qseh(+\qsef)$, this term will vanish.

\subsection{The dispersive part of the RPA polarizations}

We turn now to the second part of (\ref{eq:trace_ss2}). Inserting this expression into Eq.~(\ref{eq:msec_retav_sh}) we find for the dispersive part of the time-ordered polarizations
\begin{equation}
	\label{eq:msec_nfnfnfnf}
	\begin{split}
	\msec_{\mathrm{d},l} (k)
	=
	 2i N_f N_c \int\dvp & \left[ \frac{(\effp_+^\ret - \effp_-^\ret)^2 -(\effm_+^\ret \pm \effm_-^\ret)^2}
	 		{(\mbox{$\effp^\ret_+$}^2 - \mbox{$\effm^\ret_+$}^2)(\mbox{$\effp^\ret_-$}^2 - \mbox{$\effm^\ret_-$}^2)}
	 		(1-\nfp)(1-\nfm) \right. \\
	 &\quad+
	  \frac{(\effp_+^\av - \effp_-^\av)^2 -(\effm_+^\av \pm \effm_-^\av)^2}
	 		{(\mbox{$\effp^\av_+$}^2 - \mbox{$\effm^\av_+$}^2)(\mbox{$\effp^\av_-$}^2 - \mbox{$\effm^\av_-$}^2)}
		\nfp\nfm   \\
	&\quad+	 		
	 \frac{(\effp_+^\ret - \effp_-^\av)^2 -(\effm_+^\ret \pm \effm_-^\av)^2}
	 		{(\mbox{$\effp^\ret_+$}^2 - \mbox{$\effm^\ret_+$}^2)(\mbox{$\effp^\av_-$}^2 - \mbox{$\effm^\av_-$}^2)}
	 		(1-\nfp)\nfm \\
	&\quad+	 		
	 \left. \frac{(\effp_+^\av - \effp_-^\ret)^2 -(\effm_+^\av \pm \effm_-^\ret)^2}
	 		{(\mbox{$\effp^\av_+$}^2 - \mbox{$\effm^\av_+$}^2)(\mbox{$\effp^\ret_-$}^2 - \mbox{$\effm^\ret_-$}^2)}
	 		\nfp(1-\nfm)	\right] \,. 	
	\end{split}
\end{equation}
A similar expression can be obtained from (\ref{eq:mseret_retav_sh}) for $\mseret_{\mathrm{d},l}$.

For $T=0$, the integrand in (\ref{eq:msec_nfnfnfnf}) can be simplified. Using (\ref{eq:sec_seret}) and $\qseretlo^*(k)=\qseav(k)$, we find $\Im\qsec(k)=\sgn(k_0-\mu)\Im\qseret(k)=-\sgn(k_0-\mu)\Im\qseav(k)$. This means that the time-ordered imaginary parts are identical to the retarded ones for $k_0>\mu$ and identical to the advanced ones for $k_0<\mu$. The real parts of $\qse^{\chron,\ret,\av}$ are identical for all $k_0$, cf.~(\ref{eq:sec_seret}). In (\ref{eq:msec_nfnfnfnf}) every retarded quantity is connected to a factor $(1-\nf)$ while every advanced quantity comes with a factor $\nf$. It follows that all retarded and advanced quantities can be replaced by time-ordered ones. After the replacement $\ret,\av\rightarrow\chron$, the four terms in the integrand can be added up and we find
\begin{align}
	\msec_{\mathrm{d},l} (k)
	\,\stackrel {T=0} =\, 
	 2i N_f N_c \int\dvp \frac{(\effp_+^\chron - \effp_-^\chron)^2 -(\effm_+^\chron \pm \effm_-^\chron)^2}
	 		{(\mbox{$\effp^\chron_+$}^2 - \mbox{$\effm^\chron_+$}^2)(\mbox{$\effp^\chron_-$}^2 - \mbox{$\effm^\chron_-$}^2)} \,.
	 		\label{eq:qsec_simple} 
\end{align}

Before we continue, it is important to realize that energy independent terms which are not covered by dispersive integrals can be isolated in the $k_0\rightarrow \pm\infty$ limit. By inserting the effective masses and momenta explicitly, we find for the numerators on the rhs.~of (\ref{eq:qsec_simple}) in the $\sigma$ and the $\pi$ case
\begin{equation}
	\label{eq:num_pol_sigmapi}
	\begin{split}
	 (\effp_+-\effp_-)^2 - (\effm_+ + \effm_-)^2 &= 
			k^2 - 4 M^2           -2 k_\mu\Delta\qse^\mu + \Delta\qse_\mu \Delta\qse^\mu  \,,  \\
	(\effp_+-\effp_-)^2 - (\effm_+ - \effm_-)^2 &= 
			k^2 - \Delta\qse^2_s -2 k_\mu\Delta\qse^\mu + \Delta\qse_\mu \Delta\qse^\mu  \,,
	\end{split}
\end{equation}
with $\Delta \qse^{s,\mu} = \qse^{s,\mu}_+  - \qse^{s,\mu}_-$ and $M= m_0 + \frac12(\qse^s_+ +\qse^s_-)$. The numerators depend on $p_0$ and $\vec p$ only in the arguments of the self energies. Up to the $\Delta\qse$ terms, the expressions in (\ref{eq:num_pol_sigmapi}) resemble the mean field results $k^2-4\effmc^2$ and $k^2$, cf.~(\ref{eq:hfpropsigma_denom},\ref{eq:hfproppi_denom}).

At large $k_0^2$, the imaginary parts as well as the four-momentum dependent, dispersive real parts of the self energy will vanish due to the cutoff of the model. Only the mean field self energy and the three-momentum dependent, real shifts to the dispersion integrals remain finite, i.e., $\qse^s_\pm \rightarrow \qsemf+\qsef_{\eff,s}(|\vec p_\pm|)$ and $\qse^\mu_\pm \rightarrow \qsef_{\eff,\mu}(|\vec p_\pm|)$. Here $\qsef_\eff$ is the effective Fock self energy that is defined below in Eq.~(\ref{eq:eff_fock_comp}).

The $p_0$ dependence of (\ref{eq:num_pol_sigmapi}) is completely lost for $k_0^2\rightarrow \infty$: The $k_0$ dependent part of the numerators reduces to $k^2-2 k_\mu\Delta\qse^\mu$. The term $2 k_\mu\Delta\qse^\mu$ -- that is not present in the mean field approaches -- does not contribute to the integral in (\ref{eq:qsec_simple}). The $p_+$ and the $p_-$ contributions in  $\Delta\qse$ cancel each other at large $k_0^2$. Hence, the energy dependence of the numerators in (\ref{eq:num_pol_sigmapi}) is -- as in the mean field case -- given by
\begin{align}
	(\effp_+ - \effp_-)^2 -(\effm_+ \pm \effm_-)^2 \quad \xrightarrow{k_0^2\rightarrow \infty} \quad
		 k_0^2 +\ordersymb(k_0^0)  \,.
	\label{eq:numenergydep}
\end{align}
The same result can be found for the more general expression in (\ref{eq:msec_nfnfnfnf}). At large $k_0^2$, where the self energies are purely real, the distinction between $\qseret$ and $\qseav$ becomes irrelevant in the numerators. Note that the second term on the rhs.~of (\ref{eq:numenergydep}), \order{k_0^0}, is still $\vec p_\pm$ dependent. We will come back to that point in Appendix~\ref{app:re_qse}.

Let us briefly comment on the $k\rightarrow 0$ behavior of the numerators in (\ref{eq:msec_nfnfnfnf},\ref{eq:qsec_simple}).
This is important for the discussion of the pion mass in Section~\ref{sec:meson_masses}. The integrand of (\ref{eq:qsec_simple}) vanishes for $k\rightarrow 0$ in the pion case since $\effp^\chron_+=\effp^\chron_-$ and $\effm^\chron_+=\effm^\chron_-$ for $k\rightarrow 0$. In (\ref{eq:msec_nfnfnfnf}) the terms that contain either only retarded or only advanced self energies will vanish in the same way. The mixed terms become zero already for $k_0=0$, due to the distribution functions: When $p_0^+=p_0^-=p_0$, the factors $(1-\nfpm)\nfmp$ could only be finite for $p_0<\mu$ \emph{and} $p_0>\mu$ at the same time.

We turn our attention now to the denominator of (\ref{eq:qsec_simple}) and introduce the integral
\begin{align}
	 \mathrm{I}(k) = \int\dvp \frac{1}
	 		{(\mbox{$\effp^\chron_+$}^2 - \mbox{$\effm^\chron_+$}^2)(\mbox{$\effp^\chron_-$}^2 - \mbox{$\effm^\chron_-$}^2)} \,. \label{eq:full_int_i}
\end{align}
Note that the \emph{imaginary} part of $\mathrm{I}$ corresponds to the \emph{real} part of $\mse^{\chron,\ret}$. We can introduce a retarded integral via $\Re \mathrm{I}^\ret(k)=\Re \mathrm{I}(k)[1+2\nb(k_0)]$, $\Im\mathrm{I}^\ret(k)=\Im\mathrm{I}(k)$, cf.~(\ref{eq:sec_seret}). 
In the mean field approximations, $\Re\mathrm{I}(k)$ and $\Im\mathrm{I}(k)$ can be calculated analytically when a simple cutoff scheme is used. On that level, $\Im\mathrm{I}^\ret(k)$ calculated from a dispersion integral is identical to $\Im\mathrm{I}^\ret(k)$ obtained from the direct calculation, cf.~Appendix~\ref{app:hfrpa}. No constant shifts must be added to the dispersion integral (\ref{eq:imi_exact}) .

The complex self energies that enter (\ref{eq:full_int_i}) in the \order{1/N_c} approach do not introduce additional poles to the integrand. They remain small and vanish or become real and energy independent for energies a few times larger than the cutoff $\njlcut$. Thus, the full calculation does not introduce additional shifts to the dispersion relation and -- like in the mean field case -- the relation (\ref{eq:imi_exact}) should hold. The complex integral $\mathrm{I}^\ret(k)$ of the \order{1/N_c} approach is found by inserting the second term of (\ref{eq:trace_ss2}) into (\ref{eq:mseret_retav_sh}), cf.~Eq.~(\ref{eq:msec_nfnfnfnf}), and replacing the numerators of all terms in the integrand by $1$,
\begin{align}
	\label{eq:i_ret_full}
	\mathrm{I}^\ret(k)
	= 
	 \int & \dvp  \left[ \frac{1}
	 		{(\mbox{$\effp^\ret_+$}^2 - \mbox{$\effm^\ret_+$}^2)(\mbox{$\effp^\av_-$}^2 - \mbox{$\effm^\av_-$}^2)}
	 		 (\nfm -\nfp) \right. \\
	 & \left. -\frac{1}
	 		{(\mbox{$\effp^\ret_+$}^2 - \mbox{$\effm^\ret_+$}^2)(\mbox{$\effp^\ret_-$}^2 - \mbox{$\effm^\ret_-$}^2)}
	 		\nfm 
	+	 \frac{1}
	 		{(\mbox{$\effp^\av_+$}^2 - \mbox{$\effm^\av_+$}^2)(\mbox{$\effp^\av_-$}^2 - \mbox{$\effm^\av_-$}^2)}
	 		\nfp 	 \right] \,.	 \notag
\end{align}
Using $\prpl$ and $\prwd$ from (\ref{eq:denom_prop_real},\ref{eq:denom_prop_imag}) we find for the real part of $\mathrm{I}^\ret$ explicitly
\begin{align}
	\Re \mathrm{I}^\ret(k)=\int \dvp \frac{2\prwd_+\prwd_-}{(\prpl_+^2+\prwd_+^2)(\prpl_-^2+\prwd_-^2)} (\nfm - \nfp) \,.
\end{align}

Eq.~(\ref{eq:imi_exact}) does not translate directly to the polarizations. The numerators (\ref{eq:num_pol_sigmapi}) in the integral for $\mseret_{\mathrm{d},l}(k)$ are more complicated than the trivial ones in (\ref{eq:i_ret_full}). A factor $k^2$ occurs which changes the $k^2\rightarrow \infty$ behavior. This factor will generate a $k_0$ independent shift when it is included in the dispersion integral, see (\ref{eq:disp_shift}),
\begin{align}
	\label{eq:full_mshift}
	k^2 \Im \mathrm{I}^{(\ret)} (k_0,\veq k)
	&=\frac 1\pi \int_0^\infty dp_0^2 \Re \mathrm{I}^\ret(p_0,\veq k)
			+\frac 1\pi \mathcal{P} \int_0^\infty dp_0^2 
					\frac{(p_0^2-\vec k^2)\Re \mathrm{I}^\ret(p_0,\veq k)}{k_0^2-p_0^2} \,. 
\end{align}
The numerator in the last term on the rhs.~is just $p^2\Re\mathrm{I}^\ret(p)$ for $\vec p=\vec k$, i.e., this term is a dispersion for $k^2\mathrm{I}(k)$ instead of $\mathrm{I}(k)$ alone. On account of (\ref{eq:num_pol_sigmapi}) this is just one of the terms which appear in (\ref{eq:qsec_simple}).

The other terms in the numerators of (\ref{eq:qsec_simple}) should not generate additional shifts: They are either constant or -- as discussed earlier -- well behaved functions that vanish for large $k_0$. Hence, the integrals
\begin{equation}
	\label{eq:j_sigmapi}
	\begin{split}
	 \mathrm{J}_\sigma(k) &= \int\dvp \frac{4 M^2 + 2 k_\mu\Delta\qse^\mu - \Delta\qse_\mu \Delta\qse^\mu} 
	 		{(\mbox{$\effp^\chron_+$}^2 - \mbox{$\effm^\chron_+$}^2)(\mbox{$\effp^\chron_-$}^2 - \mbox{$\effm^\chron_-$}^2)} 
	 			\,, \\
	 \mathrm{J}_\pi(k) &= \int\dvp \frac{\Delta\qse^2_s + 2 k_\mu\Delta\qse^\mu - \Delta\qse_\mu \Delta\qse^\mu} 
	 		{(\mbox{$\effp^\chron_+$}^2 - \mbox{$\effm^\chron_+$}^2)(\mbox{$\effp^\chron_-$}^2 - \mbox{$\effm^\chron_-$}^2)} 
	 			\,, 
	\end{split}
\end{equation}
cf.~(\ref{eq:num_pol_sigmapi}), will satisfy corresponding dispersion relations without shifts (\ref{eq:imi_exact}) as $\mathrm{I}(k)$. The integrals $\mathrm{J}^\ret_{\sigma,\pi}(k)$ can be constructed by inserting the second term of (\ref{eq:trace_ss2}) into (\ref{eq:mseret_retav_sh}), bringing the numerators into the form (\ref{eq:num_pol_sigmapi}), and removing the $k^2$ terms.

$\mseret_{\mathrm{d},l}$ can be decomposed into $\mseret_{\mathrm{d} \sigma,\pi}(k)=2i N_f N_c [ k^2 \mathrm{I}^\ret(k)- \mathrm{J}^\ret_{\sigma,\pi}(k)]$. The dispersion relation for $\mseret_{\mathrm{d},l}$ is then given by the dispersion relations for  $k^2 \mathrm{I}^\ret(k)$ and $\mathrm{J}^\ret_{\sigma,\pi}(k)$. A shift proportional to the one in (\ref{eq:full_mshift}) will be the only correction to the dispersion integral over $\Im\mseret_{\mathrm{d},l}$. Using (\ref{eq:def_mbr}) we find
\begin{align}
	\Re\mseret_{\mathrm{d},l} (k)
		= \frac 1\pi \mathcal{P} \int_0^\infty dp_0^2 \frac{p_0\mbr_l (p_0,\veq k)}{k_0^2-p_0^2} 
				- \frac{2 N_f N_c}\pi \int_0^\infty dp_0^2 \Re \mathrm{I}^\ret(p_0,\veq k) \,. \label{eq:remseret_d}
\end{align}
Eventually, the dispersion relation for the retarded RPA polarizations -- including all $k_0$ independent shifts -- is found by combining the results from Eqs.~(\ref{eq:mse_nondisp2}) and (\ref{eq:remseret_d}). The real parts which enter our numerical calculations are
\begin{align}
	\Re\mseret_l(k)=\mse_{\mathrm{n}}+\Re\mseret_{\mathrm{d},l}(k) \label{eq:remseret_full}\,.
\end{align}

\section{Real parts of the quark self energy\label{app:re_qse}}

The \order{1/N_c} meson exchange diagram in the first line of Fig.~\ref{fig:dyson_full} may contain $k_0$ independent, non-dispersive contributions to the real part of the quark self energy. Searching these components is not as important as for the mesons. The quark mass is determined predominantly by the \order{1} mean field self energy $\qsemf$, cf. Section~\ref{sec:os_se}, that we calculate separately. In the following, we will use the relations between the different kinds of propagators ($\qpr^{\chron,\ret,\gtrless},\mpr^{\chron,\ret,\gtrless}$) from Sections~\ref{sec:qprop} and \ref{sec:mprop} without further reference.

Using Feynman rules, we find for the time-ordered quark self energy
\begin{align}
	-i\qsec(k)=-i\qsemf + \sum_l \int \dvp \Gaml \qpc(p) \Gamlt \mpc_l(p-k) \,. \label{eq:qsecu_start}
\end{align}
An investigation of the integral in the same way as in Appendix~\ref{app:re_mespol} is not possible: In Appendix~\ref{app:re_mespol} the explicit structure of the quark propagator is utilized to decompose and simplify the integrand. The structure of the RPA propagator
\begin{align}
	\mpc_l (k)=-\frac{2 \njlkpl}{1+2\njlkpl\msec_l (k)}  \label{eq:effmesprop}
\end{align}
is completely different from the quark propagator. Thus, the methods of Appendix~\ref{app:re_mespol} are not applicable here. It is not feasible to replace $\mpc_l$ by a standard meson propagator, $\mespclo^{-1}(k)=k^2-m^2-\msec(k)$, in (\ref{eq:qsecu_start}) since the properties of $\mpc_l$ and $\mespc_l$ differ -- especially in the high-energy properties. The propagator $\mespc$ will vanish in the $k_0^2\rightarrow\infty$ limit. In our effective propagators, the polarization should be -- up to $\modk$ independent terms -- suppressed by the cutoff. Therefore, $\mpc$ will approach a finite value, cf.~(\ref{eq:effmesprop}).

We split up the RPA propagator into an energy independent part $\mpinf(|\veq k|)$ that corresponds to the limit at high $k_0^2$ and an effective propagator $\empc_l$ that will vanish for $k_0^2\rightarrow \infty$,
\begin{align}
	\mpc_l(k)=\mpinf(|\veq k|) + \empc_l(k) \,. \label{eq:split_mesprop}
\end{align}
The same decomposition can be done for $\mpr^{\ret,\av}_l$. Since $\mpinf$ is real, we find $\msp_l(k)= -2\Im\empret_l(k)$ and, using (\ref{eq:mpgksp}),   $\mpgk_l(k) = \empgk_l(k)$.

Inserting the decomposed RPA propagator into the time-ordered self energy (\ref{eq:qsecu_start}) yields
\begin{align}
	\qsec(k) -\qsemf
	&= \sum_l \int \dvp \left[ i \Gaml \qpc(p) \Gaml \mpinf(|\vec p-\veq k|) 
	+ i \Gaml \qpc(p) \Gamlt \empc_l(p-k) \right]\,. \label{eq:qsec_disp_decomp}
\end{align}
The first term is a $k_0$ independent contribution to $\Re\qsec$. Thus, it is not covered by the  dispersion integral. In the simplest case, $\mpinf=-2\njlkpl$ (for $\msec\rightarrow 0$), this term becomes just the constant Fock self energy. This is not surprising since the Fock diagram is part of the effective meson exchange, cf.~Fig.~\ref{fig:mesexse}.

Because of its structure, we will refer to the first term on the rhs.~of (\ref{eq:qsec_disp_decomp}) as effective Fock self energy $\qsef_{\eff}$.
Using the decomposition $\qpc(p)=\Re\qpret(p) -\frac i2 \qsp(p) [1-2\nf(p_0)]$ and Eqs.~(\ref{eq:norm_qsp2},\ref{eq:sret_ints}), we find
\begin{equation}
	\label{eq:eff_fock_comp}
	\begin{split}
		\qsef_{\eff,s} (|\veq k|)&= 2 \int\dvp  \mpinf(|\vec p-\veq k|) \qsp_s(p) \nf(p_0) \,, \\
		\qsef_{\eff,0} (|\veq k|)&= 2 \int\dvp  \mpinf(|\vec p-\veq k|) \qsp_0(p) [1-2\nf(p_0)] \,,  \\
		\qsef_{\eff,v} (|\veq k|)&= -4 \int\dvp \mpinf(|\vec p-\veq k|) \cos\vartheta \qsp_v(p) \nf(p_0) \,.
	\end{split}
\end{equation}
Note that $\qsp_0$ is symmetric in $p_0$ in the vacuum while $[1-2\nf(p_0)]$ is antisymmetric. Thus, $\qsef_{\eff,0}$ becomes finite only in the presence of a medium. Since $\qsp_v(p)=\qsp_v(p_0,|\veq p|)$ and $\int_{-1}^{+1} d\cos\vartheta \cos\vartheta=0$, $\qsef_{\eff,v}$ will be zero when $\mpinf$ is constant.

As discussed before, a search for shifts to the dispersion integral in the second term on the rhs.~of Eq.~(\ref{eq:qsec_disp_decomp}) is complicated. We just give a plausibility argument here, why such shifts should not occur. In general, the existence of constant shifts is related to the level of divergence of the integrand in the dispersion integral. The RPA polarizations (\ref{eq:qsecu_start}) include two quark propagators $\sim p_\pm/p_\pm^2$ (for large $k^2$). This gives rise to a term $\sim k^2$ in the numerator of the integrand. We have shown in Eq.~(\ref{eq:full_mshift}) how the inclusion of the factor $k^2$ in the dispersion integral for $\Re\mseret_l$ generates a constant term. The other terms in the numerator of (\ref{eq:qsecu_start}) are of lower order in $k$ and do not generate such shifts.

In contrast to the polarizations, the self energy integral in (\ref{eq:qsec_disp_decomp}) contains only one quark propagator and one effective meson propagator $\empc_l$. Assuming that $\empc_l(p\pm k)\sim 1/(p\pm k)^2$ for large $k^2$ -- like for the standard propagator $\mespc_l$ -- the last integrand of (\ref{eq:qsec_disp_decomp}) does not contain a factor $k^2$ in the numerator. The denominator, on the other hand, resembles that of the polarizations. Consequently, the imaginary part of the quark self energy (\ref{eq:qsec_disp_decomp}) is of lower order in $k$ than the imaginary parts of the RPA polarizations. We can expect that the dispersion integral for $\Re\qseret$ does not include a shift similar to the one in (\ref{eq:full_mshift}). If we overlook any (next-to-leading order) shifts by applying this argument, they should be small compared to the leading-order Hartree self energy.

The complete dispersion relation for the real part of the retarded quark self energy is then given by
\begin{align}
	\Re\qseret (k_0,\veq k) = \qsemf+\qsef_\eff(\veq k) +\frac 1 {2\pi} \mathcal{P}\int^{+\infty}_{-\infty} d p_0
				\frac{\qbr(p_0,\veq k)}{k_0-p_0}  \,.  \label{eq:qrk_disp}
\end{align}
The Lorentz components of $\Re\qseret$ can be found in the usual way (\ref{eq:extract_lorentz}). Note that all components depend on $|\veq k|$ only, not on the full three-momentum $\vec k$.

A remarkable interplay between quarks and mesons can be observed here. For $k_0^2\rightarrow \infty$, $\Re\qseret$ approaches the $|\veq k|$ dependent, finite value of $\qsemf+\qsef_\eff(\veq k)$. The momentum dependence is introduced by the high-energy limit of the RPA propagators, cf.~(\ref{eq:split_mesprop},\ref{eq:eff_fock_comp}). The momentum dependent limit of the meson propagators, on the other hand, is generated by $\qsef_\eff(\veq k)$ that enters Eqs.~(\ref{eq:num_pol_sigmapi},\ref{eq:numenergydep}). This self-consistent effect is not present in the Hartree(\mbox{--}Fock)+RPA approximations since the polarizations are not fed back into the quark self energy. There, the $k_0^2\rightarrow \infty$ limits of $\Re\qseret$ and $\Re\mseret$ are constant in $\vec k$.


\end{document}